\documentclass[prl,twocolumn,superscriptaddress]{revtex4}

\usepackage{anysize}
\usepackage{graphicx}
\usepackage{epsfig}
\usepackage{verbatim}
\usepackage{amssymb}
\usepackage{amsfonts,amsmath}
\usepackage{bm}
\usepackage{eufrak}
\usepackage[colorlinks]{hyperref}
\usepackage{fancyhdr}
\usepackage{relsize}
\usepackage[T1]{fontenc}
\usepackage{xr}
\usepackage[final]{pdfpages}

\marginsize{1.5cm}{1.5cm}{1cm}{1cm}
\hypersetup{colorlinks=true,
	linkcolor=black,
	anchorcolor=black,
	citecolor=black,
	urlcolor=black
}

\providecommand{\be}{\begin{equation}}
\providecommand{\ee}{\end{equation}}
\providecommand{\bea}{\begin{eqnarray}}
\providecommand{\eea}{\end{eqnarray}}
\providecommand{\beas}{\begin{eqnarray*}}
\providecommand{\eeas}{\end{eqnarray*}}

\providecommand{\beni}{\begin{equation*}}
\providecommand{\eeni}{\end{equation*}}

\providecommand{\bw}{\begin{widetext}}
\providecommand{\ew}{\end{widetext}}

\newcommand{\nn}{\nonumber}

\newlength{\bilderlength}

\newlength{\figsize}
\newcommand{\st}{\scriptscriptstyle}

\begin{document}

\title{Spatial organization of bacterial transcription and translation}
\author{Michele Castellana}
\affiliation{Joseph Henry Laboratories of Physics, Princeton University, Princeton, NJ 08544}
\affiliation{Lewis-Sigler Institute for Integrative Genomics, Princeton University, Princeton, NJ 08544}
\affiliation{Laboratoire Physico-Chimie Curie, Institut Curie, CNRS UMR168, 75005 Paris, France}
\author{Sophia Hsin-Jung Li}
\affiliation{Department of Molecular Biology, Princeton University, Princeton, NJ 08544}
\author{Ned S. Wingreen}
\affiliation{Lewis-Sigler Institute for Integrative Genomics, Princeton University, Princeton, NJ 08544}

\begin{abstract}
In bacteria such as \emph{Escherichia coli}, DNA is compacted into a nucleoid near the cell center, while ribosomes---molecular complexes that translate messenger RNAs (mRNAs) into proteins---are mainly localized to the poles. We study the impact of this spatial organization using a minimal reaction-diffusion model for the cellular transcriptional-translational machinery. 
While genome-wide mRNA-nucleoid segregation still lacks experimental validation, our model predicts that $\sim90\%$ of mRNAs are segregated to the poles. In addition, our analysis reveals a ``circulation'' of ribosomes driven by the flux of mRNAs, from synthesis in the nucleoid to degradation at the poles. 
We show that our results are robust with respect to multiple, biologically relevant factors, such as mRNA degradation by RNase enzymes, different phases of the cell division cycle and growth rates, and the existence of non-specific, transient interactions between ribosomes and mRNAs.
Finally, we confirm that the observed nucleoid size stems from a balance between the forces that the chromosome and mRNAs exert on each other. This suggests a potential global feedback circuit in which gene expression feeds back on itself via nucleoid compaction. 
\end{abstract}

\maketitle

\section{Introduction}\label{intro}

\setlength{\parskip}{5pt plus 0pt minus 0pt}

The cytoplasm of many bacterial cells exhibits a striking spatial organization: rather than filling the entire cell volume, the DNA forms a condensed structure called a ``nucleoid'' that is generally localized near midcell (Fig. \ref{fig1}) \cite{lewis2004bacterial,bakshi2012superresolution}. Moreover, ribosomes---large molecular complexes that translate messenger RNAs (mRNAs) into proteins---are observed to be anti-localized from the nucleoid \cite{bakshi2012superresolution}. These observations raise two natural questions: (1) What physical processes are responsible for this subcellular organization? (2) How does this internal structure influence the basic processes of mRNA transcription and protein translation in the cell? 

In the model bacterium \textit{Escherichia coli} (\textit{E. coli}), $\sim\, \!1.5 \, \rm mm$ of  supercoiled DNA are compacted into a $\sim\,\!  1 \, \mu \rm m^3$ nucleoid volume \cite{fisher2013four}, thus forming a dense DNA mesh with average pore diameter $\sim\! 50 \, \rm nm$. As a result, free ribosomes, with diameter $\sim\! 20 \,  \rm nm$, can readily diffuse into the nucleoid \cite{sanamrad2014single}, while polysomes,   molecular complexes composed of mRNAs with multiple bound ribosomes and having an effective diameter $\gtrsim 50 \, \rm nm$, anti-localize from the nucleoid due to excluded-volume effects. \textit{In vivo} measurements of mRNA mobility suggest a typical diffusion coefficient of $D \sim 0.05 \, \mu \rm m^2 / \rm s$, implying that mRNAs formed in the nucleoid by transcription from DNA can diffuse out of the nucleoid to the ribosome-rich regions in a few seconds---a time significantly shorter than the typical mRNA lifetime of $\sim\!\, 5 \, \rm min$ \cite{bakshi2012superresolution,bernstein2004global}. These observations suggest that most mRNAs formed in the nucleoid diffuse out of the nucleoid to the ribosome-rich regions where ribosomes and mRNAs are colocalized, and where the bulk of translation occurs.   

In recent years, the  idea that mRNA localization may play a functional role in bacteria \cite{rudner2010protein,dossantos2012diviva} has inspired a variety of measurements of mRNA localization. These studies have provided evidence for multiple,  mRNA-specific localization patterns, such as localization within the cell cytoplasm \cite{valencia-burton2007rna}, to the cell membrane or to the cell poles \cite{nevo-dinur2011translation}, and at the nascent septum separating daughter cells \cite{dossantos2012diviva}. However, such experiments have proven to be challenging \cite{golding2004rna,kannaiah2014protein} and limited to specific mRNAs \cite{pilhofer2009fluorescence,buxbaum2014right}.
As a result, genome-wide ribosome-mRNA colocalization still lacks experimental validation. In this study, we employ known reaction-diffusion properties of mRNAs, ribosomes, and the nucleoid to predict the physical origin and the extent of overall, genome-wide mRNA localization.

Our approach also describes non-specific, transient interactions between ribosomes and mRNAs. In this regard, recent studies in \textit{E. coli} \cite{bakshi2012superresolution} and \textit{Caulobacter crescentus} \cite{montero2012in}  observed an increase of the diffusion coefficient of non-translating ribosomes under depletion of the pool of mRNAs, which has been interpreted as evidence for non-specific, transient bindings between ribosomes and mRNA molecules in \cite{montero2012in}.  On the other hand, another study in \textit{E. coli} reported that the diffusion coefficient of non-translating ribosomes is not affected by mRNA depletion \cite{sanamrad2014single}.
Here, we show that if transient ribosome-mRNA interactions are significantly faster than other relevant time scales, the reaction-diffusion equations can be substantially simplified by treating such interactions as a local Poisson process. 
Finally, we consider the opposing forces exerted by the compressed nucleoid and by polarly localized mRNAs and confirm that the observed nucleoid size results from the balance of these dominant forces, and we present a simple analytical formula for the degree of nucleoid compaction under different physiological conditions, e.g. for different amounts of DNA and mRNA in the cell.

\section{Results}\label{res}

\begin{figure}
\centering\includegraphics[scale=0.7]{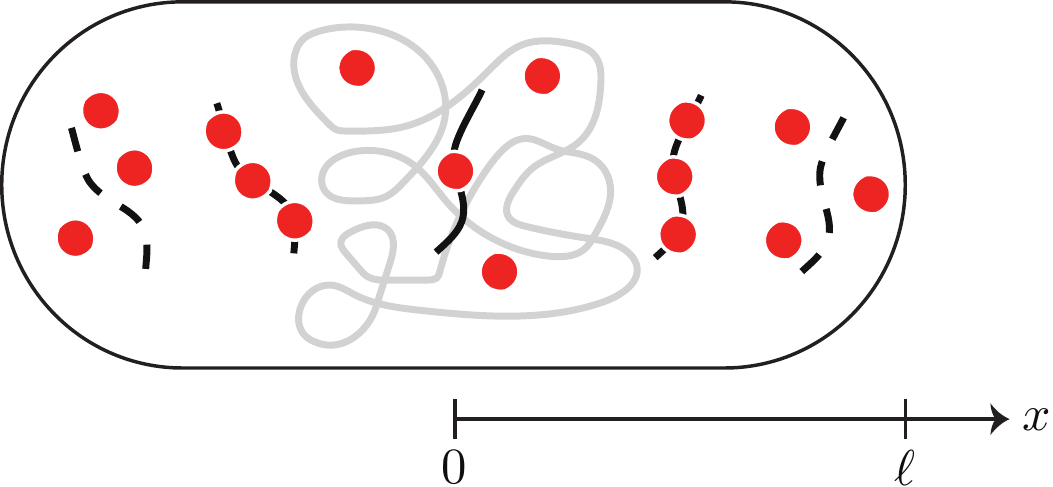}
\caption{
Schematic of the spatial organization of transcription and translation in \textit{E. coli}. mRNAs (black solid curves) are transcribed in the nucleoid---the condensed DNA chromosome  at the cell center (gray solid curve). Ribosomes (red circles) bind to mRNAs forming polysomes, i.e. mRNAs with multiple bound ribosomes. Polysomes diffuse preferentially out of the nucleoid due to excluded-volume effects. Eventually the mRNA molecules are degraded (black dashed lines). In our 1D model, the coordinate $x$ runs along the long axis of the cell and, assuming symmetry, we model only the right half of the cell ($0\leq x \leq \ell)$.  
\label{fig1}}
\end{figure}

We describe the coupled dynamics of ribosomes and mRNAs in an \textit{E. coli} cell using a minimal, 1D reaction-diffusion model. We introduce a coordinate $x$ running along the long axis of the cell and, given the approximate left-right symmetry  of a typical \textit{E. coli} cell, we consider only the right half, from $x=0$ at the cell center to $x= \ell$ at the right cell pole (Fig. \ref{fig1}). 
\textit{In vitro} measurements of the assembly dynamics of the translation-initiation complex suggest that the \textit{in vivo} binding rate of 30S ribosomal subunits to mRNAs is significantly larger than the unbinding rate, thus implying that the majority of mRNAs have a 30S subunit bound at the translation-initiation site \cite{milon2012real}. If so, translation is largely governed by the dynamics of 50S ribosomal subunits, and therefore we initially consider only the 50S units, which we refer to simply as ``ribosomes''.
We further assume that ribosomes may undergo transient, non-specific binding to mRNAs; extended versions of our model including the two ribosomal subunits and disallowing non-specific ribosome-mRNA interactions are presented later, and they confirm qualitatively the results obtained with the simple model discussed here---see Supporting Information, sections \ref{secs3} and \ref{secs6}, for details.
The 1D concentration of free (F) ribosomes, $c_{\st \rm F}(x)$, denotes the number of F ribosomes per unit length in an infinitesimal slice of the cell perpendicular to the $x$ axis. Similarly,  $\rho_{m,n}(x)$ is the 1D concentration of mRNAs with $m$ transiently bound (B) \cite{montero2012in} ribosomes and $n$ translating (T) ribosomes. 
As shown in section \ref{secs9}, the average number of ribosomes per mRNA, $m+n \sim 12$, is well below the maximum total number of ribosomes, $m+n \sim 100$, that could be linearly packed onto a typical mRNA. Thus, we consider only mRNA species with $m\leq m_{\rm max}$, $n\leq n_{\rm max}$, where $m_{\rm max}$ and $n_{\rm max}$ are some maximal numbers of allowed ribosomes per mRNA, chosen large enough to account for all typical mRNA species present in the cell. Importantly, this choice reduces the number of mRNA species present in our model, thus making it computationally tractable.

The resulting reaction-diffusion equation for the F-ribosome concentration is 
\bea\label{eqF}\nn
\frac{\partial c_{\st \rm F}(x,t)}{\partial t} =  
D_{\st \rm F} \left[ \frac{\partial^2 c_{\st \rm F}(x,t)}{\partial x^2} v_{\st \rm F}(x) -  c_{\st \rm F}(x,t) \frac{d^2 v_{\st \rm F}(x)}{d x^2} \right]  && \\ \nn
 - k^{\st \rm B}_{\rm on} c_{\st \rm F}(x,t) \hspace{-1mm}  \sum_{m}  \sum_{n} \rho_{m , n}(x,t)  \hspace{-0.5mm}+\hspace{-0.5mm}  k^{\st \rm B}_{\rm off} \hspace{-1mm}  \sum_{m} \sum_{n}    m  \, \rho_{m, n}(x,t)  && \\ \nn
 - k^{\st \rm T}_{\rm on} c_{\st \rm F}(x,t) \hspace{-1mm}  \sum_{m}  \sum_{n} \rho_{m , n}(x,t)  \hspace{-0.5mm}+\hspace{-0.5mm} k^{\st \rm T}_{\rm off} \hspace{-1mm} \sum_{m} \sum_{n}    n  \, \rho_{m, n}(x,t) && \\ 
 + \beta \sum_{m} \sum_{n}  \,(m+n) \,   \rho_{m , n}(x,t), &&
\eea
where only mRNA species with allowed values of $0 \leq m \leq {m_{\rm max}}$ and  $0 \leq n \leq {n_{\rm max}}$ are considered. The first term on the right-hand side (RHS) represents diffusion including excluded-volume effects due to the condensed DNA (Fig. \ref{fig1}). In this term, $D_{\st \rm F}$ is the diffusion coefficient for F ribosomes,  which  incorporates crowding effects due to ribosomes, mRNAs and other macromolecules, while $v_{\st \rm F}(x)$ is the fractional volume available to an F ribosome within the DNA mesh at position $x$ (Fig. \ref{figs20}), see sections \ref{secs1} and \ref{secs2} for details. The first term in the second line represents F ribosomes binding to all possible mRNA species with $m$ B ribosomes and $n$ T ribosomes, and thus becoming B ribosomes. This term is proportional to the ${\rm F}\rightarrow {\rm B}$ transition rate $k^{\st \rm B}_{\rm on}$ and to the total  density of mRNA. 
In principle, $k^{\st \rm B}_{\rm on}$ should decrease with the ribosome occupancy number $m+n$: however, here $m+n$ is much smaller than the maximum packing density, thus this effect is small, see above.
The next term in the second line describes a B ribosome unbinding from an mRNA, where $k^{\st \rm B}_{\rm off}$ denotes the unbinding rate and the multiplicity factor $m$ accounts for multiple B ribosomes on the mRNA. Similarly, the third line represents transitions between F and T ribosomes, where $k^{\st \rm T}_{\rm on}$ is assumed independent of $m$, $n$ (see above).  Since  measurements suggest  that the lifetime of the B state is significantly shorter than that of the T state \cite{montero2012in}, here the $\rm F \rightarrow \rm B \rightarrow T$ transition is incorporated into the $\rm F \rightarrow \rm T$ transition, with an effective rate $k^{\st \rm T}_{\rm on}$.
Finally, the last line represents B and T ribosomes being freed from mRNA molecules as these are degraded at rate $\beta$ \cite{chen2015genome}.

The 1D reaction-diffusion equations for the mRNA densities are
\bea\label{eqR}\nn
\frac{\partial \rho_{m,n} (x,t)}{\partial t} &=&  \\ \nn
D \left[ \frac{\partial^2 \rho_{m,n} (x,t)  }{\partial x^2}  v_{m+n}(x) - \rho_{m , n}(x,t) \frac{d^2 v_{m+n}(x)}{d x^2} \right]  &&\\ \nn
-  \, k^{\st \rm B}_{\rm on} c_{\st \rm F}(x,t) \rho_{m , n}(x,t)  -  k^{\st \rm B}_{\rm off} \, m \, \rho_{m , n}(x,t)  && \\ \nn
 - \, k^{\st \rm T}_{\rm on} c_{\st \rm F}(x,t) \rho_{m , n}(x,t)  -  k^{\st \rm T}_{\rm off} \, n \,  \rho_{m , n}(x,t)   &&\\ \nn
 +  \, k^{\st \rm B}_{\rm on} c_{\st \rm F}(x,t) \rho_{m-1 , n}(x,t)  +  \, k^{\st \rm B}_{\rm off} \,  (m+1)  \, \rho_{m+1 , n}(x,t)  && \\ \nn
 +  \, k^{\st \rm T}_{\rm on} c_{\st \rm F}(x,t) \rho_{m , n-1}(x,t)  + \, k^{\rm T}_{\rm off} \, (n+1)  \,  \rho_{m , n+1}(x,t)  && \\ 
 + \delta_{m,0}\, \delta_{n,0} \, \alpha(x) - \beta \, \rho_{m , n}(x,t).
\eea
Here $D$ is the average mRNA diffusion coefficient in the cytoplasm, and $v_{m+n}(x)$ is the fractional available volume within the nucleoid for an mRNA with $m+n$ attached ribosomes (Fig. \ref{figs20}), see section \ref{secs2} for details.  The third and fourth lines represent binding and unbinding of B and T ribosomes from mRNAs of species $m$, $n$, while the fifth line represents B ribosomes binding to mRNAs of species $m-1$, $n$ or unbinding from mRNAs of species $m+1$, $n$. 
Similarly, the terms in the sixth line represent T ribosomes binding and unbinding from mRNAs of species $m, n-1$ and $m,n+1$. Finally, the last line represents transcription of initially ribosome-free mRNAs according to the nucleoid profile $\alpha(x)$, and mRNA degradation at rate $\beta$. 

As transcriptional and translational time scales (${\lesssim 1 \, \rm{min}}$) are fast compared to cell doubling times  (${\gtrsim 20 \, \rm{min}}$), we focus on steady-state conditions. At steady state,  in Eq. \eqref{eqF} we enforce a constraint  on the  total number of ribosomes, $2\int_0^\ell dx [c_{\st \rm F}(x) + \sum_{m=0}^{m_{\rm max}} \sum_{n=0}^{n_{\rm max}} (m+n) \rho_{m,n}(x)] = N_{\rm tot}$, and we set a no-flux boundary condition at the cell pole, $\left. [d c_{\st \rm F}(x)/dx \, v_{\st \rm F}(x)  - c_{\st \rm F}(x) \, d v_{\st \rm F}(x)/dx ]\right|_{x=l} = 0$, see section \ref{secs1}. Similarly, in Eq. \eqref{eqR} we impose no-flux boundary conditions at the cell pole and at the cell center,   $ [ d \rho_{m,n}(x)/dx \, v_{m+n}(x)   - \rho_{m,n}(x) \, d v_{m+n}(x)/dx ]|_{x=0,l} = 0$, the latter reflecting the left-right symmetry  of the cell. According to this symmetry, the flux of F ribosomes at midcell must also vanish, and this follows directly from the boundary conditions above---see section \ref{secs11}.

\begin{figure}
\centering\includegraphics[scale=1.8]{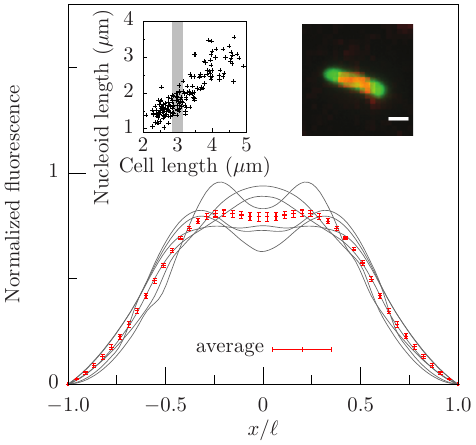}
\caption{
DNA fluorescence along the long cell axis for $3 \, \mu \rm m$-long \textit{E. coli} cells grown in glucose minimal media. Cells were stained with SYTOX Orange and imaged at exponential phase. Fluorescence for a few representative cells (gray), and resulting average fluorescence over $35$ cells with  standard error of the mean (red), both symmetrized and normalized to unit area. Left inset: nucleoid vs. cell length, where the gray area includes cells with length with $5 \%$ of $3 \, \mu \rm m$. Right inset: ribosomal protein S2-YFP (green) and nucleoid (red). Scale bar: $1 \, \mu \rm m$. 
\label{fig4}}
\end{figure}

We fix the model parameters from experimental data as follows.
We considered \textit{E. coli} cells in glucose minimal media with a $\sim 1/ \rm h$ growth rate, see section \ref{secs12}, and focused on the mid-phase of the division cycle by selecting cells with length  within $5 \%$ of a typical, medium length of $2 \, \ell = 3 \, \mu \rm m$, compare left inset in Fig. \ref{fig4}. We then rescaled the resulting DNA fluorescence profiles to a cell length of $2 \, \ell$, and we estimated the nucleoid profile  along the long cell axis by averaging over multiple cells, see main panel in Fig. \ref{fig4} and section \ref{secs2}. 
The F ribosome diffusion coefficient was taken to be $D_{\st \rm F} = 0.4 \, \mu {\rm m}^2/{\rm s}$, while the diffusion coefficient of mRNAs was set at the average diffusion coefficient of polysomes, $D = 0.05 \, \mu {\rm m}^2/{\rm s}$ \cite{bakshi2012superresolution,sanamrad2014single}. 
These diffusion coefficients were obtained from measurements of  mean square displacements of free and mRNA-bound ribosomal subunits in living \textit{E. coli} cells \cite{bakshi2012superresolution}, and thus include the effects of cytoplasmic crowding.
The F-ribosome available volume $v_{\st \rm F}(x)$ was estimated from the concentration profiles of free ribosomal subunits  \cite{sanamrad2014single}, and similarly for the mRNA available-volume profiles $v_{m+n}(x)$ (section \ref{secs2}). 
The mRNA degradation rate was taken to be $\beta =3 \times 10^{-3} / \rm s$, corresponding to a mean mRNA lifetime of $\sim\! 5 \, \rm min$  \cite{bernstein2004global}. The total mRNA production rate $\alpha_{\rm tot} = 2\int_0^\ell dx \, \alpha(x)$ was obtained from the total number of mRNAs per cell via the relation $ \alpha_{\rm tot}/\beta = \, N_{\rm mRNA} = 5 \times 10^3$ \cite{taniguchi2010quantifying}, while the profile of mRNA production $\alpha(x)$ was  chosen to be proportional to the DNA density  $\varphi(x)$, see section \ref{secs2} for details. 
The average time for a ribosome to complete protein translation is estimated to be $\tau_{\st \rm T} = 40 \, \rm s$ \cite{siwiak2013transimulation}, so we took the ${\rm T}\rightarrow {\rm F}$ transition rate to be $k^{\st \rm T}_{\rm off} = 1/\tau_{\st \rm T}  = 2.5 \times 10^{-2} / \rm s$. To set the other transition rates we used the observation that $\sim \, \!\!80 \%$ of ribosomes are T ribosomes \cite{bakshi2012superresolution}, with non-T ribosomes estimated to spend $\sim\! 90 \%$ of their time as B ribosomes and $\sim\! 10 \%$ of their time as F ribosomes, where the $90\% \!:\!10\%$ division of non-T ribosomes between B and F ribosomes is inferred from ribosome diffusion rates in \textit{Caulobacter crescentus} \cite{montero2012in}. 
Global equilibrium between T and F ribosomes at steady state requires the ${\rm F}\rightarrow {\rm T}$ transition rate $k^{\st \rm T}_{\rm on}$ to satisfy  $N_{\st \rm F} \, \overline{\rho}\, k^{\st \rm T}_{\rm on} = N_{\st \rm T} \, k^{\st \rm T}_{\rm off}$, where $N_{\st \rm F}$ and $N_{\st \rm T}$ are the total number of F and T ribosomes, respectively,  and $\overline{\rho} \simeq 1.7 \times 10^3 /\mu {\rm m}$ is the average total mRNA axial density \cite{taniguchi2010quantifying}. Also, in the equilibrium condition above we neglected the rate $\beta N_{\rm \st T}$ at which mRNA-bound T ribosomes are freed by mRNAs that are being degraded, because this rate is much smaller than $N_{\st \rm T} k^{\rm \st T}_{\rm off}$. According to the estimate above for the average number of F, T, and B ribosomes, we have $N_{\st \rm T}/N_{\st \rm F} = 80\%/(20\% \times 10\%)$,  which yields $k^{\st \rm T}_{\rm on} = 6 \times 10^{-4} \mu {\rm m} / {\rm s}$.  
The total number of ribosome in the cell was taken to be $N_{\rm tot} = 6 \times 10^4$ \cite{bakshi2012superresolution}.  As for the the binding-unbinding rates $k^{\st \rm B}_{\rm on}$, $k^{\st \rm B}_{\rm off}$ of the ${\rm F}\leftrightarrow {\rm B}$ transition, the equilibrium condition  reads $N_{\st\rm F} \, \overline{\rho}\, k^{\st \rm B}_{\rm on} = N_{\st \rm B} \, k^{\st \rm B}_{\rm off}$ where,
since the ${\rm F}\leftrightarrow {\rm B}$ transition  occurs on timescales not longer than $\sim 1 \, \rm s$ \cite{montero2012in}, we neglected the mRNA-decay term $\beta N_{\rm \st B} \ll N_{\st \rm B} \, k^{\st \rm B}_{\rm off}$.  Together with the estimation above for the average number of F, T, and B ribosomes $N_{\st \rm F} /N_{\st \rm B} = 2\%/18\%$, this equilibrium  condition provides an estimate for the ratio $k^{\st \rm B}_{\rm on}/k^{\st \rm B}_{\rm off} \simeq 5.4 \times 10^{-3} \, \mu \rm m$, but it does not specify the individual values of $k^{\st \rm B}_{\rm on}$, $k^{\st \rm B}_{\rm off}$. However, because the timescale of the ${\rm F}\leftrightarrow {\rm B}$ transition  is significantly shorter than other relevant timescales \cite{montero2012in}, these processes can be treated in the limit where they are at rapid equilibrium. In this limit, the problem can be significantly simplified, and the set of $1+({m_{\rm max}}+1)({n_{\rm max}}+1)$ reaction-diffusion Eqs. \eqref{eqF}, \eqref{eqR}  reduces to a set of ${n_{\rm max}}+2$ equations which completely characterize the solution $c_{\rm \st F}(x)$ and $\rho_{m,n}(x)$ for any $m$ and $0 \leq n \leq {n_{\rm max}}$. Importantly, these rapid-equilibrium equations depend on $k^{\st \rm B}_{\rm on}$, $k^{\st \rm B}_{\rm off}$ only through their ratio---see section \ref{secs7} for details.

\begin{figure}
\centering\includegraphics[scale=1.6]{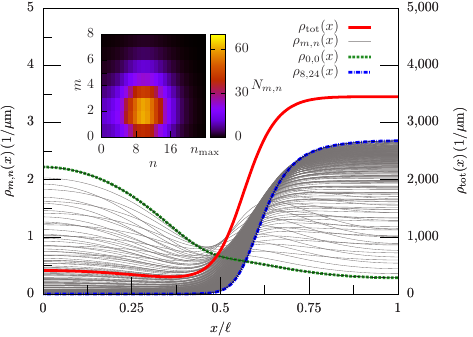}
\caption{
Steady-state mRNA and polysome distributions.  Total mRNA density $\rho_{\rm tot}(x)$ (red) and density $\rho_{m,n}(x)$ of mRNAs with $m$ transiently bound (B) ribosomes and $n$ translating  (T) ribosomes for  $0\leq m\leq m_{\rm max} = 8$ and $0\leq n \leq n_{\rm max} = 24$   (gray). The density $\rho_{0,0}(x)$ of ribosome-free mRNAs  (green) and the density $\rho_{8,24}(x)$ of mRNAs with the largest number  of T and B ribosomes considered  (blue) are also shown. The profiles $\rho_{m,n}(x)$,  $\rho_{0,0}(x)$, and $\rho_{8,24}(x)$ are normalized to unit area. Inset: distribution of mRNA species, shown as a heat map of the number $N_{m,n}$ of mRNAs with $m$ B ribosomes and $n$ T ribosomes in the right half of the cell. The maximal number of T ribosomes per mRNA used in our model, $n_{\rm max}=24$, is indicated.
\label{fig2}}
\end{figure}

\begin{figure}
\centering\includegraphics[scale=1.8]{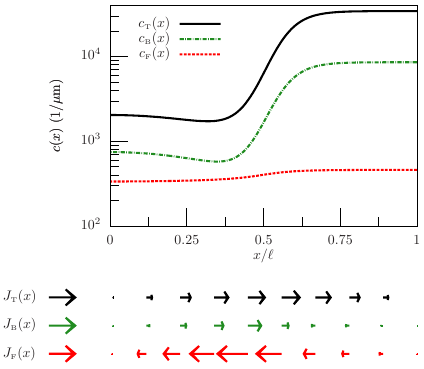}
\caption{Steady-state ribosome concentrations and ribosome fluxes. Top: Concentrations $c_{\rm \st T}(x)$, $c_{\rm \st B}(x)$, and $c_{\rm \st F}(x)$ of translating (T), transiently bound (B), and free (F) ribosomes in the right half of the cell (compare Fig. \ref{fig1}). Bottom: Fluxes of T, B, and F ribosomes along the cell's long axis depicted in the top panel. The arrow length is proportional to local ribosome flux, and the arrows in the legends correspond to a flux of $20/\rm s$. 
\label{fig3}}
\end{figure}

We numerically solved Eqs. \eqref{eqF}, \eqref{eqR} at steady state in the rapid-equilibrium limit for B ribosomes, by fixing the maximal number of allowed T ribosomes per mRNA at  ${n_{\rm max}}=24$, see section \ref{secs7} for details. The resulting mRNA profiles and ribosome concentrations are shown in Figs. \ref{fig2} and \ref{fig3}, respectively. As shown in Fig. \ref{fig2}, the total mRNA profile $\rho_{\rm tot}(x) = \sum_{m=0}^\infty \sum_{n=0}^{n_{\rm max}} \rho_{m,n}(x)$ is markedly localized outside the nucleoid region---most of the mRNAs are segregated from the nucleoid because of excluded volume. Indeed, the density profiles $\rho_{m,n}(x)$ show that the more ribosomes an mRNA has bound, the more segregated the mRNA is from the nucleoid, see Fig. \ref{figs3} for details. 
Since mRNAs are generated at the nucleoid, the strong segregation of  mRNAs away from the nucleoid at steady state must result from the majority of mRNAs being loaded with ribosomes so that excluded volume biases their diffusion away from the nucleoid. This conclusion is confirmed in the inset of Fig. \ref{fig2}, which shows a heat map of the total number $N_{m,n} = \int_0^\ell dx \rho_{m,n}(x)$ of mRNAs of species $m,n$. Most mRNAs are loaded with $\sim\! 10$ T ribosomes and $\sim\! 2$ B ribosomes. 
These average loading numbers are in agreement with the  experimental estimates above for the ribosome numbers \cite{bakshi2012superresolution,taniguchi2010quantifying}:  The number of T ribosomes per mRNA is $N_{\st \rm T}/N_{\rm mRNA} \sim \! (80\% \, N_{\rm tot})/N_{\rm mRNA} \sim 10$, and a similar estimate yields $\sim\! 2$ B ribosomes per mRNA.  In addition, the inset shows that the chosen value ${n_{\rm max}}=24$ is large enough to encompass all  typical mRNA species that are present.
 
Since each ribosome has a linear size of $\sim\! 20 \, {\rm nm}$ \cite{kaczanowska2007ribosome}, the effective size of an mRNA molecule with $\sim\! 10$ bound  ribosomes is significantly larger than the pore size of the DNA mesh in the nucleoid, which we estimate to be $\sim\! 50 \,  \rm nm$, see section \ref{secs2} for details. Thus, the majority of mRNAs experience strong excluded-volume effects which push them out of the nucleoid region.

Since mRNAs are created by transcription in the nucleoid but end up segregated away from the nucleoid, there must be a flux of mRNAs toward the cell poles. Given that new mRNAs are rapidly loaded with T ribosomes at a rate $k^{\st \rm T}_{\rm on} N_{\st \rm F}/(2\,  \ell)= 6 \times 10^{-4} \mu {\rm m} / {\rm s} \times 1200/ 3 \mu {\rm m} \simeq 0.24 / {\rm s}$, implying full occupation by $\sim\! 10$ T ribosomes in $\sim\! 3\, \rm s$,
 the poleward flux of mRNAs carries with it a poleward flux of ribosomes. Since ribosomes are conserved in our model, reflecting the long half-life of ribosomal components \cite{piir2011ribosome}, there must be a compensating flux of F ribosomes from the poles toward the nucleoid. In Fig. \ref{fig3} we show the steady-state concentration of F ribosomes $c_{\st \rm F}(x)$, the concentrations of T and B ribosomes, $c_{\st \rm T}(x) = \sum_{m=0}^\infty \sum_{n=1}^{n_{\rm max}} n \, \rho_{m,n}(x)$, $c_{\st \rm B}(x) = \sum_{m=1}^\infty \sum_{n=0}^{n_{\rm max}} m \, \rho_{m,n}(x)$, the flux $J_{\st \rm F} = -D_{\st \rm F} [d c_{\rm \st F}(x) / dx \, v_{\st \rm F}(x) - c_{\st \rm F}(x) \, d v_{\st \rm F}(x)/dx]$ of F ribosomes, and the fluxes $J_{\st \rm T}$ and $J_{\st \rm B}$ of T and B ribosomes, compare Eqs. (\ref{eqs42}), (\ref{eqs41}). As expected from the observed segregation of mRNAs,  the T and B ribosomes are markedly excluded from the nucleoid, and there is a net poleward flux of T and B ribosomes. Notably, the effect of excluded volume  in the nucleoid is so strong that mRNAs and their associated ribosomes flow from a low- to a high-concentration region. 
By contrast, F ribosomes are small enough to freely penetrate the nucleoid, and a flux of F ribosomes is driven by the gradient of these ribosomes from the poles to the nucleoid. Overall, these results illustrate and quantify a ``circular'' flux for the ensemble of T, B, and F ribosomes, compare Fig. \ref{fig1}: First, multiple F ribosomes bind to mRNAs made in the cell nucleoid. Each mRNA is thus loaded with $\sim\! 10$ T ribosomes and $\sim\! 2$  B ribosomes to become a polysome. Second, the effects of excluded volume in the nucleoid result in a net flow of these polysomes to the cell poles. Once polysomes reach the poles, they ultimately decay and free their ribosomes. This ``pumping'' of T and B ribosomes from the nucleoid to the poles results in an excess of F ribosomes at the poles, and thus in a diffusive return flux of F ribosomes to the nucleoid. 

The existence of a steady ribosome circulation implies that there must be an external source of energy driving these circular fluxes. There are two possible candidates within our model: Process (A) is the non-equilibrium creation and degradation of mRNAs, and Process (B) is mRNA and F-ribosome binding in the nucleoid and subsequent expulsion from the nucleoid by excluded-volume effects. 
Process (A) should be strictly dependent on new mRNA production, whereas Process (B) should persist even in the limit where the mRNA production and degradation rates are both low, with the total number of mRNAs fixed and equal to $N_{\rm mRNA}$. Therefore, we varied the mRNA production and degradation rates together, keeping the total mRNA number constant: 
The circulation vanished as the mRNA rates slowed (Fig. \ref{figs4}), thus identifying Process (A), the flux of new mRNAs from nucleoid to pole, as the driver of ribosome circulation.  This conclusion is confirmed by an analytical estimate for the poleward flux of T and B ribosomes, which is shown to be proportional to the mRNA production rate, see section \ref{secs4} for details. 

Before discussing  other implications of our results, it is worth considering that mRNA transcription takes a finite amount of time. For the average mRNA length of $\sim 3 \times 10^3 \, \rm nt$ discussed in section \ref{secs9} and an average mRNA elongation speed of $\sim 50 \, \rm nt/s$ \cite{vogel1994the}, we obtain a typical transcription time of $\sim 1 \,{\rm min}$, during which nascent mRNAs are bound to DNA while being elongated. We therefore extended our model to include these nucleoid-bound mRNAs, whose axial densities we denote by $\rho^{\ast}_{m,n}(x)$: these mRNAs turn into free mRNAs at a rate $\gamma = 1/(1 \, \rm min)$, see section \ref{secs10} for details. 
Besides confirming the picture in the model with one mRNA species, this extended model gives novel insights into the mechanism of co-transcriptional translation, namely the observation that ribosomes translate nascent mRNAs while these are being transcribed in the nucleoid \cite{miller1970visualization}. While it has been previously hypothesized that most of the ribosomes in the dense nucleoid region  translate co-transcriptionally \cite{bakshi2012superresolution}, our analysis shows that only $\sim \, 34 \, \%$ of these ribosomes carry out co-transcriptional translation, while a comparable fraction of $\sim \, 37 \, \%$  translates post-transcriptionally, mostly on polysomes loaded with a relatively small number of ribosomes.
The extended model also allows us to address the effects of co-transcriptional translation on the protein-synthesis rate: introducing the  efficiency $\varepsilon = 2\, k^{\st \rm T}_{\rm off}\, \hspace{-1mm} \sum_{m=0}^{\infty} \sum_{n=1}^{{n_{\rm max}}}    n  \, \int_0^\ell  dx  \rho_{m, n}(x)/N_{\rm tot}$, i.e. the average number of proteins translated per unit time per ribosome, we find that co-transcriptional translation implies a $\sim\! 3 \%$ increase in ribosome efficiency, under the assumption of B ribosome binding to all transcripts, see section \ref{secs10} for details. 

\begin{figure}
\centering\includegraphics[scale=1.97]{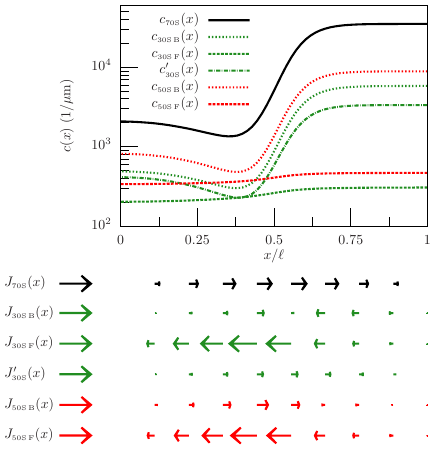}
\caption{
Steady-state ribosomal-subunit concentrations and fluxes for the  model  including 30S and 50S ribosomal subunits. Top: Concentrations $c_{\st \rm 70S}(x)$, $c_{\st \rm 30S \, B}(x)$, $c_{\st \rm 30S \, F}(x)$, $c'_{\rm \st 30S}(x)$ of 70S, transiently bound (B) 30S, free (F)  30S subunits,  and of 30S subunits bound to the translation-initiation site, respectively. We also show the concentrations $c_{\st \rm 50S \, B}(x)$ and $c_{\st \rm 50S \, F}(x)$ of B and F 50S subunits, respectively. 
Bottom: Fluxes  $J_{\st \rm 70S}(x)$, $J_{\st \rm  30S \, B}(x)$, $J_{\st \rm 30S \, F}(x)$  of 70S, B and F 30S subunits, and flux $J'_{\rm \st 30S}(x)$ of 30S subunits bound to the translation-initiation site. 
The fluxes  $J_{\st \rm  50S \, B}(x)$, $J_{\st \rm 50S \, F}(x)$ of B and F 50S subunits are also shown. Fluxes are represented along the cell's long axis depicted in the top panel, the arrow length is proportional to local ribosome flux, and the arrows in the legends correspond to a flux of $30/\rm s$. 
\label{figs8}} 
\end{figure}

We then extended our model to take account of both ribosomal subunits. During translation initiation, a 30S subunit binds to the mRNA initiation site first, and then a 50S subunit binds to the 30S subunit forming a translating 30S-50S (70S) pair \cite{laursen2005initiation}. To model this process and the spatiotemporal dynamics of the two ribosomal subunits, we introduce mRNA species with $l$ B 30S subunits, $m$ B 50S subunits, $n$ 70S pairs, and no 30S subunit at the initiation site, and denote their density by $\rho_{l,m,n}(x,t)$. Similarly, we denote by $\rho'_{l,m,n}(x,t)$ the density of mRNAs with $l$ B 30S subunits, $m$ B 50S subunits, $n$ 70S pairs, and a 30S subunit at the initiation site. 
Assuming rapid equilibrium for B 30S and 50S subunits, we solved the reaction-diffusion equations for the mRNA densities $\rho_{l,m,n}(x)$, $\rho'_{l,m,n}(x)$, and for the concentrations $c_{\st \rm 30S \, F}(x,t)$, $c_{\st \rm 50S \, F}(x,t)$ of free 30S and 50S subunits---see section \ref{secs3} for details. 
The results shown in Figs. \ref{figs8}, \ref{figs7}, and \ref{figs10}   confirm the picture obtained with the simpler one-subunit model. In particular, the mRNAs, both with and without a 30S subunit on the translation-initiation site, are strongly segregated from the nucleoid, and the larger $l$, $m$, $n$, the stronger the segregation. 
In Fig. \ref{figs8} we show the concentrations and fluxes of the ribosomal subunits: the  70S subunits, the B 30S and 50S subunits, and the 30S subunits bound to translation-initiation sites are all strongly segregated from the nucleoid, compare Fig. \ref{fig3}, while F 30S and 50S subunits are only slightly excluded from the nucleoid. Note that the concentrations of F and B 30S subunits are lower than the corresponding concentrations of F and B 50S subunits because for our choice of parameters the majority of non-translating 30S subunits are bound to translation-initiation sites, thus reducing the number of F and B 30S subunits. 
The fluxes of 70S,  B 30S and 50S subunits, and of 30S subunits bound to translation-initiation sites are directed toward the cell poles, while the compensating fluxes of F 30S and 50S subunits are directed toward the cell center. 

The two-subunit model  was then extended to include additional, biologically relevant features: In section \ref{secs8} we incorporated in the model  the mechanism of mRNA degradation by RNase enzymes \cite{chen2015genome,mackie2013rnase}, see Figs.  \ref{figs15}-\ref{figs14}, 
and in section \ref{secs3e} we extended the model to cells in the late phase of their division cycle, see Figs. \ref{figs17}-\ref{figs19}. We then extended the analysis above for glucose minimal media   to different growth rates: We imaged \textit{E. coli} cells in glycerol minimal and defined rich media, with growth rates $\sim 0.5/ \rm h$ and $\sim 2 / \rm h$, respectively, extracted the nucleoid profiles, and present the resulting model predictions in section \ref{secs13}, compare Figs. \ref{figs22}-\ref{figs28}.  Finally, as the existence of B ribosomes is an open question \cite{bakshi2012superresolution,montero2012in,sanamrad2014single}, in section \ref{secs6} we considered a version of the model with two ribosomal subunits and no B subunits. Overall, these results confirm all the qualitative features of the simpler, two-subunit model discussed above.

So far, our analysis has shown that the excluded volume due to DNA localization at midcell segregates the majority of mRNAs to the cell poles. In what follows, we will extend this analysis and show that the converse is also true, i.e. mRNA segregation to the poles causes nucleoid compaction at midcell. Specifically, in  section \ref{secs14} we show that mRNAs, like particles in a gas, exert an entropic force (pressure) on the nucleoid directed towards the cell center, and that this force can be computed directly from the reaction-diffusion Eqs. \eqref{eqF} and \eqref{eqR}.  On the other hand, the natural tendency of the compressed DNA polymer to increase its configurational entropy results in a an effective ``spring'' force pushing outward on the mRNAs. Exploiting the condition that these two forces must balance at mechanical equilibrium, we self-consistently determined the mRNA and F ribosome profiles, as well as the DNA density profile $\varphi(x)$, and the results are shown in Figs. \ref{figs25} and \ref{figs26}. In particular, denoting by $2 x_0$ the width of the nucleoid region centered at midcell, the resulting value of $2 x_0 \approx 1.43 \, \rm \mu m$ matches well the experimentally observed nucleoid size in  Fig. \ref{fig4}. 
Finally, a simplified version of our analysis provides a straightforward prediction for the nucleoid width under different physiological conditions: We assume that mRNAs are uniformly distributed outside the nucleoid, which they cannot penetrate, that the nucleoid is confined in a region of width $2 x_0$ centered at midcell, and we neglect the small force exerted by F ribosomes on the nucleoid (see section \ref{secs14} for details). As a result, the nucleoid size $x_0$ can be determined by solving the following force-balance equation:
\be\label{eq3}
\frac{N_{\rm mRNA}}{2(\ell-x_0)} = \frac{\pi^2}{6} \frac{\xi^2 N^{2 \nu}}{(2 \pi R^2 )^{2/3}x_0^{5/3}},
\ee

\noindent where $\xi = 200 \, \rm nm$ is twice the estimated persistence length of a segment of supercoiled DNA,  $N = 7.5 \times 10^3$  is the total number of such segments in the chromosome \cite{mondal2011entropy}, and $R = 0.5 \, \mu \rm m$ is the radius of a circular cell slice. 
In Fig. \ref{figs24} we show the predicted nucleoid size for different values of the total number of mRNAs---the larger the number of mRNAs, the more the nucleoid shrinks towards midcell due to the entropic force exerted by the polysomes.  

\section{Discussion}\label{disc}

The study of intracellular mRNA localization has attracted growing interest in recent years \cite{golding2004rna,kannaiah2014protein,pilhofer2009fluorescence,buxbaum2014right}. In eukaryotic systems, mRNA localization is a well-established mechanism for achieving a variety of functions, such as the targeted expression of  proteins to specific regions of the cell, the control of intracellular signaling, or the partition of mRNAs into daughter cells for cell-fate differentiation \cite{palacios2001getting,shal-tav2005rna}. Some functional, mRNA-specific localization patterns has also been observed in bacteria \cite{rudner2010protein}: Two examples are the membrane localization of mRNAs which code for proteins transporting lactose into the cell \cite{nevo-dinur2011translation} and mRNA localization at the cell poles, which has been shown to play a functional role in controlling sugar-utilization genes \cite{nevo-dinur2011translation,lopian2010spatial}.

In this study, we analyze the extent of bacterial genome-wide mRNA localization by means of a minimal reaction-diffusion model for the transcriptional and translational machinery in \textit{E. coli}. The experimental observation that ribosomes in \textit{E. coli}  are markedly localized to the  cell poles, and thus segregated from the nucleoid located at the cell center \cite{bakshi2012superresolution}, suggests that the majority of mRNAs are also likely to be segregated from the nucleoid.
While experiments on mRNA localization in \textit{E. coli} have proven to be challenging \cite{golding2004rna,kannaiah2014protein}, and are limited to specific mRNAs \cite{pilhofer2009fluorescence,buxbaum2014right}, our model makes a novel prediction for strong, genome-wide mRNA localization away from the nucleoid, indicating that $\sim 90 \%$ of mRNAs are typically located outside the nucleoid, and demonstrating that the total mRNA profile  resembles that of translating ribosomes (compare Figs. \ref{fig2}, \ref{fig3}).
A specific prediction of our model is that mRNA segregation is due to excluded-volume effects resulting from the condensed nucleoid DNA. Overall, this result provides novel insights into the mechanisms of mRNA segregation: While other studies for specific mRNAs raised the possibility that mRNA segregation may be associated with dynamical rearrangements of the nucleoid  \cite{stylianidou2014cytoplasmic}, our analysis indicates that genome-wide mRNA segregation can arise entirely from excluded-volume effects. Also, our result that  segregation  primarily affects mRNAs loaded with multiple ribosomes is in line with the recent experimental observation that mRNAs with multiple bound fluorescent proteins localize to the cell poles in live \textit{E. coli} cells \cite{golding2004rna}. 

Our model also reveals a ``circulation'' of ribosomes within the cell driven by the flux of newly synthesized mRNAs from the nucleoid to the poles: mRNA-bound ribosomes flow from the nucleoid to the cell poles, where they unbind from mRNAs and then diffuse back to the nucleoid to bind newly synthesized mRNA molecules. Using our model, we also  analyze the extent of co-transcriptional translation, namely the observation that ribosomes translate mRNAs that are being transcribed in the nucleoid \cite{miller1970visualization}. 
Although it has been recently hypothesized that most of the ribosomes in the DNA-rich region translate co-transcriptionally \cite{bakshi2012superresolution}, we 
find that only about a third of these ribosomes carry out co-transcriptional translation, whereas a slightly higher fraction  translates post-transcriptionally in polysomes with a relatively small loading number.

We incorporated in our model the mechanism of mRNA degradation by RNase enzymes, different phases of the cell division cycle, different growth rates, and the effect of non-specific, transient interactions between ribosomes and mRNAs \cite{bakshi2012superresolution,montero2012in,sanamrad2014single}, showing that our results are stable with respect to such variety of conditions. 

Finally, we extended our analysis to study the consequences of mRNA localization on nucleoid compaction. Using our calculated ribosome and mRNA densities, we confirmed that mRNA segregation to the poles quantitatively accounts for nucleoid compaction. Physically, the observed nucleoid size reflects the balance of two competing entropic forces---the compressive force that mRNAs (polysomes) at the poles exert on the DNA, and the expansive force exerted by the DNA on these mRNAs. Our detailed analysis supports a simplified analytical formula \eqref{eq3} that predicts nucleoid size for different physiological conditions, compare Fig. \ref{figs24}. Biologically, the compaction of the nucleoid by mRNAs creates a potential global feedback circuit: gene expression drives mRNA levels, which, by compacting the nucleoid, impact transcription-factor access and hence gene expression \cite{kuhlman2012gene}. 

To summarize, while localization of the transcriptional-translational machinery is a well-known, functional mechanism in eukaryotes, the function of such localization in bacteria is not yet well established. In this regard, our model provides novel insights into the mechanisms governing the spatial structure of transcription and translation in bacteria, and can help guide the molecular manipulation of these functions, with potentially broad applications in molecular synthetic biology and biotechnology. 

\textit{Acknowledgments}---We thank  E. Bonomi, B. Bratton, C. Broedersz, Z. Gitai, I. Golding,  J.-F. Joanny, P. Sens, J.-C. Walter, J. C. Weisshaar, and M. Z. Wilson for valuable conversations and suggestions. Research supported by National Science Foundation grant PHY-1305525, PHY-1066293, GRFP DGE-1148900, by US National Institutes of Health grants NIDA DP1DA026192 and NIAID R21AI102187, and by the hospitality of the Aspen Center for Physics.
The numerical computations presented in this article were performed on computational resources supported by the Lewis-Sigler Institute for Integrative Genomics at Princeton University.

\clearpage


\section*{Supplementary material}

# Supporting Information for "Spatial organization of bacterial transcription and translation"

Michele Castellana,[1,2,3] Sophia Hsin-Jung Li,[4] and Ned S. Wingreen[2]

[1]*Joseph Henry Laboratories of Physics, Princeton University, Princeton, NJ 08544*
[2]*Lewis-Sigler Institute for Integrative Genomics, Princeton University, Princeton, NJ 08544*
[3]*Laboratoire Physico-Chimie Curie, Institut Curie, CNRS UMR168, 75005 Paris, France*
[4]*Department of Molecular Biology, Princeton University, Princeton, NJ 08544*

## Contents

## S1  Fraction of available ribosome binding sites

In this section, we will estimate the fraction of available ribosome binding sites on an mRNA chain by making use of experimental data on mRNA synthesis and degradation in *E. coli*. To begin with, we estimate the average number

of nucleotides per mRNA, which we quantify as the ratio between the total number of nucleotides synthesized per unit time, and the total number of mRNAs synthesized per unit time. First, the number of nucleotides synthesized per unit time is given by the number of active RNA polymerases (RNAPs), i.e. $\sim 1000\,\mathrm{RNAP/cell}$, times the chain-elongation rate at which a single RNAP synthesizes an mRNA, i.e. $\sim 50\,\mathrm{nt/s}$ [1], resulting in a rate of $\sim 5 \times 10^4\,\mathrm{nt/s}$. Second, the total number of mRNAs synthesized per unit time is given by the total number of mRNAs $N_{\mathrm{mRNA}} = 5 \times 10^3$ [2] times the mRNA degradation rate $\beta = 3 \times 10^{-3}/\mathrm{s}$ [3], and it is given by $\sim 15\,\mathrm{mRNA/s}$. The resulting average number of nucleotides per mRNA is $\sim 3 \times 10^3\,\mathrm{nt}$. Given that each ribosome covers $\sim 30\,\mathrm{nt}$ [4], we obtain that there are on average $\sim 3 \times 10^3\,\mathrm{nt}/(30\,\mathrm{nt}) = 100$ available ribosome binding spots per mRNA. Finally, we observe that there is a total number of $N_{\mathrm{tot}} \sim 6 \times 10^4$ ribosomes per cell [5], i.e. there are on average $N_{\mathrm{tot}}/N_{\mathrm{mRNA}} \sim 12$ ribosomes per mRNA, which is only $\sim 12\%$ of the number of available binding spots per mRNA, thus implying that the vast majority of the ribosome binding sites are free.

## S2 Diffusion equations in the presence of excluded volume

In this section we derive the diffusion equation for a population of F ribosomes, taking into account the effects of excluded volume due to the nucleoid. The continuum limit of this equation can be derived starting from a discrete version [6]. We divide the interval $0 \le x \le \ell$, from midcell to the cell pole, into bins of width $\Delta x$, where $x_i$ denotes the position of bin $i$: Here, each bin represents a projection on the one-dimensional $x$ axis of a three-dimensional section of the cell obtained by slicing the cell perpendicularly to its long axis. The master equation for the F ribosome density in bin $i$, $c_{\mathrm{F}}(x_i, t)$, can then be obtained by considering hopping of ribosomes between adjacent bins:

$$\frac{\partial c_{\mathrm{F}}(x_i, t)}{\partial t} = d_{\mathrm{F}} \left\{ [c_{\mathrm{F}}(x_{i-1}, t) + c_{\mathrm{F}}(x_{i+1}, t)] v_{\mathrm{F}}(x_i) - c_{\mathrm{F}}(x_i, t)[v_{\mathrm{F}}(x_{i-1}) + v_{\mathrm{F}}(x_{i+1})] \right\}, \tag{S1}$$

where $d_{\mathrm{F}}$ is the bare rate at which a ribosome hops from one site to another, and $v_{\mathrm{F}}(x)$ is the fraction of available volume at position $x$, i.e. the probability that a ribosome finds enough free volume to hop into the DNA mesh at position $x$. The continuum limit of Eq. (S1) can be obtained by expanding $c_{\mathrm{F}}(x, t)$ and $v_{\mathrm{F}}(x)$ around $x = x_i$, which yields

$$\frac{\partial c_{\mathrm{F}}(x, t)}{\partial t} = D_{\mathrm{F}} \left[ \frac{\partial^2 c_{\mathrm{F}}(x, t)}{\partial x^2} v_{\mathrm{F}}(x) - c_{\mathrm{F}}(x, t) \frac{d^2 v_{\mathrm{F}}(x)}{dx^2} \right], \tag{S2}$$

where $D_{\mathrm{F}} = d_{\mathrm{F}} \Delta x^2$ is the bare diffusion coefficient. The 1D F ribosome flux at position $x$ is given by

$$J_{\mathrm{F}}(x) = -D_{\mathrm{F}} \left[ \frac{dc_{\mathrm{F}}(x)}{dx} v_{\mathrm{F}}(x) - c_{\mathrm{F}}(x) \frac{dv_{\mathrm{F}}(x)}{dx} \right]. \tag{S3}$$

Note that in the absence of reaction terms, the steady-state solution of Eq. (S1) is

$$c_{\mathrm{F}}(x) = C\, v_{\mathrm{F}}(x), \tag{S4}$$

where the constant $C$ depends on the total number of F ribosomes.

## S3 *E. coli* nucleoid imaging analysis

*E. coli* strain AFS55 [5] which contains ribosomal protein S2-YFP was inoculated overnight in the media including MOPS defined rich, glucose and glycerol minimal media (Teknova). Next day the overnight cultures were diluted by a hundred fold in the same fresh media and grew in $37\,^\circ\mathrm{C}$.

The staining procedure is adopted from a previous publication [7]. Briefly, $100\,\mu\mathrm{L}$ was taken when the OD reached 0.3 and stained with $1\,\mu\mathrm{L}$ $50\mu\mathrm{M}$ SYTOX Orange (Molecular Probes) for 10 min at $37\,^\circ\mathrm{C}$ in darkness. Cells were then washed twice with $1\,\mathrm{mL}$ fresh media and resuspended in $1\,\mathrm{mL}$ fresh media. $1\,\mu\mathrm{L}$ of cells were placed on $1\%$ low-melting agar pad (Calbiochem) made from the same media and imaged with inverted Nikon90i epifluorescent microscope equipped with a $100 \times 1.4\,\mathrm{NA}$ objective (Nikon) and Hamamazu Orca R2 CCD camera. NIS Elements software (Nikon) was used to automate image acquisition for phase contrast, YFP and mCherry fluorescent channels. Segmentation, quantification of fluorescence intensity, and cell-length measurements were further analyzed in MATLAB using customized programs.

## S4   Available-volume profiles

In this section we estimate the available volume for ribosomes, mRNAs, and polysomes inside the nucleoid. To estimate the DNA pore size inside the nucleoid, we recognize that the *E. coli* chromosome is characterized by a branched, plectonemic structure of supercoiled DNA [8]: We denote by $L$ the total length of the plectonemic DNA polymer, and by $V$ the volume wherein this polymer is confined. For simplicity, suppose that the DNA is arranged to occupy the edges of a three-dimensional cubic lattice of total volume $V$. The volume is divided into $M$ cubic pores each with edge length $a$, so that $V = M a^3$. In addition, the DNA length is related to the pore size by $L = 3 a M$, thus providing the estimate $a = \sqrt{3V/L}$. For an estimated total plectoneme length $L = 1.5$ mm [8] and a nucleoid volume $V = 1.2 \, \mu m^3$ [5], the estimated pore size is $a \simeq 50$ nm.

We now provide a rough, dimensional estimate of the available volume for an F ribosome inside the nucleoid. Continuing to use the simple cubic lattice model for the nucleoid, we estimate the excluded volume inside each cubic pore of DNA: as the ribosome approaches one of the lattice edges, the excluded volume is given by the edge length $a$ times the area within the pore from which the center of mass of the ribosome is excluded when it approaches the edge, i.e. $1/4$ of a circle with the ribosome's radius $r \simeq 10$ nm [9]. Multiplying by the total number of edges (12), we obtain an estimate for the excluded volume, $v_{\mathrm{excl}} = 12 \, a \, \pi r^2/4 = 3 \, a \, \pi r^2$. We then express the available-volume fraction in the nucleoid, which we will denote by $v_{\mathrm{F}}^{\mathrm{in}}$, in terms of the ratio between the excluded volume and the pore volume as

$$v_{\mathrm{F}}^{\mathrm{in}} = \exp\left(-\kappa \frac{v_{\mathrm{excl}}}{a^3}\right) = \exp\left(-\kappa \, \pi r^2 \frac{L}{V}\right). \tag{S5}$$

The numerical correction factor $\kappa$ is introduced to improve our dimensional estimate, and is set by fitting our model to the experimentally observed concentration profile of freely diffusing ribosomes. If we imagine dividing the cell into two regions along the axial direction, the nucleoid region (in) and the polar region outside the nucleoid (out), then according to section S2 the concentration of freely diffusing ribosomes satisfies $c_{\mathrm{F}}^{\mathrm{in}}/c_{\mathrm{F}}^{\mathrm{out}} = v_{\mathrm{F}}^{\mathrm{in}}/v_{\mathrm{F}}^{\mathrm{out}}$. Since the concentration of freely diffusing ribosomes is, roughly speaking, about 10% larger outside the nucleoid than inside [10], we have $v_{\mathrm{F}}^{\mathrm{in}}/v_{\mathrm{F}}^{\mathrm{out}} = v_{\mathrm{F}}^{\mathrm{in}} \sim 90\%$, where $v_{\mathrm{F}}^{\mathrm{out}} = 1$ because there is no DNA in the out region. As a result, we obtain $\kappa = 0.25$ in Eq. (S5).

The available-volume fraction for mRNAs loaded with $m$ B ribosomes and $n$ T ribosomes can be estimated along the same lines. Since the excluded volume for a single ribosome depends on its two-dimensional cross section $\pi r^2$ rather than on its volume, we introduce an effective radius $r_{m+n}$ for an mRNA of species $m$, $n$, where $r_{m+n}$ is chosen to reproduce the overall cross section of the mRNA along with its ribosomes:

$$\pi r_{m+n}^2 = \pi r_{\mathrm{R}}^2 + (m+n)\pi r^2, \tag{S6}$$

and $r_{\mathrm{R}} = 20$ nm is a typical mRNA radius of gyration [11]. The mRNA excluded volume in the nucleoid $v_{m+n}^{\mathrm{in}}$ is then determined by replacing $r$ with $r_{m+n}$ in Eq. (S5).

Finally, the smooth excluded-volume profiles $v_{\mathrm{F}}(x)$, $v_{m+n}(x)$ shown in Fig. S1 were obtained from $v_{\mathrm{F}}^{\mathrm{in}}$, $v_{m+n}^{\mathrm{in}}$ as follows. First, we introduce a local density of DNA length

$$\varphi(x) \propto \frac{1}{1 + \exp[\zeta(x/\ell - 1/2)]}, \tag{S7}$$

with $\zeta = 20$: this density profile is shown in Fig. S1, it was chosen to reproduce the experimental axial DNA fluorescence in Fig. 2, and normalized to the total DNA length per volume [5] by setting

$$\frac{1}{\ell} \int_0^\ell dx \, \varphi(x) = \frac{L}{V}. \tag{S8}$$

Second, we replaced $L/V$ in Eq. (S5) with $\varphi(x)$, and we obtained

$$v_{\mathrm{F}}(x) = \exp[-\kappa \, \pi r^2 \varphi(x)], \tag{S9}$$

and similarly

$$v_{m+n}(x) = \exp[-\kappa \, \pi r_{m+n}^2 \varphi(x)]. \tag{S10}$$

Denoting by

$$v_{\mathrm{R}}(x) = \exp[-\kappa \, \pi r_{\mathrm{R}}^2 \varphi(x)] \tag{S11}$$

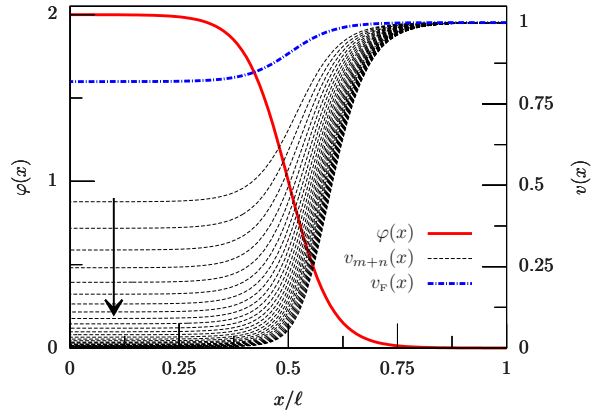

FIG. S1: Density of DNA length and available-volume profiles in 1D model of the *E. coli* transcriptional-translational machinery. Density of DNA length $\varphi(x)$ normalized to unit area, available-volume profile for free ribosomes $v_{\mathrm{F}}(x) = \exp[-\kappa \pi r^2 \varphi(x)]$, and available-volume profiles for polysomes with $m$ transiently bound and $n$ translating ribosomes $v_{m+n}(x) = \exp[-\kappa \pi r_{m+n}^2 \varphi(x)]$ for right half of cell (compare Fig. 1), where the arrow indicates the direction of increasing $m+n$. Here the constant $\kappa = 0.25$ has been estimated by fitting the observed concentration profile of freely diffusing ribosomes [10], $r = 10\,\mathrm{nm}$ is the ribosome radius [9], and $r_{m+n} = \sqrt{r_{\mathrm{R}}^2 + (m+n)r^2}$, $r_{\mathrm{R}} = 20\,\mathrm{nm}$ are the polysome effective radius and the mRNA radius of gyration [11], respectively.

the available volume for a ribosome-free mRNA, Eqs. (S6), (S9), (S10) imply that the polysome available volume can be rewritten as

$$v_{m+n}(x) = v_{\mathrm{R}}(x)[v_{\mathrm{F}}(x)]^{m+n}. \tag{S12}$$

The factorized form of Eq. (S12) has a simple intuitive interpretation if we recall that available volumes $v_{m+n}(x)$ represents the probability that an mRNA of species $m$, $n$ finds enough free volume to hop into the DNA mesh at position $x$, compare section S2. In this regard, Eq. (S12) treats the components of an mRNA of species $m$, $n$ independently: the hopping probability $v_{m+n}(x)$ is equal to the product of the hopping probability $v_{\mathrm{R}}(x)$ for the bare mRNA and of the hopping probability $v_{\mathrm{F}}(x)$ for each of the $m+n$ ribosomes carried by the mRNA.

## S5   Flux of free ribosomes at midcell

In this section we show that at steady state the no-flux boundary conditions at the cell poles for F ribosomes and at the cell poles and midcell for mRNAs in Eqs. (1) and (2) imply that the flux of F ribosomes at midcell is also zero. This absence of F-ribosome flux at midcell is a necessary physical condition due to the assumptions of ribosome conservation and cell symmetry in our model.

First, we use Eq. (S3) to rewrite the diffusive term in Eq. (1) at steady state as

$$D_{\mathrm{F}}\left[\frac{d^2 c_{\mathrm{F}}(x)}{dx^2}v_{\mathrm{F}}(x) - c_{\mathrm{F}}(x)\frac{d^2 v_{\mathrm{F}}(x)}{dx^2}\right] = -\frac{dJ_{\mathrm{F}}(x)}{dx}, \tag{S13}$$

we then integrate Eq. (1) at steady state with respect to $x$ between $x = 0$ and $x = \ell$, and we obtain

$$J_{\mathrm{F}}(0) + \int_0^\ell dx\left[-k_{\mathrm{on}}^{\mathrm{B}}c_{\mathrm{F}}(x)\sum_m\sum_n \rho_{m,n}(x) + k_{\mathrm{off}}^{\mathrm{B}}\sum_m\sum_n m\,\rho_{m,n}(x) - k_{\mathrm{on}}^{\mathrm{T}}c_{\mathrm{F}}(x)\sum_m\sum_n \rho_{m,n}(x) + k_{\mathrm{off}}^{\mathrm{T}}\sum_m\sum_n n\,\rho_{m,n}(x)\right. \tag{S14}$$
$$\left. +\beta \sum_m\sum_n (m+n)\,\rho_{m,n}(x)\right] = 0,$$

where we used Eq. (S13) and the no-flux boundary condition at the cell pole, $J_{\mathrm{F}}(\ell) = 0$, enforced in Eq. (1). We now consider Eq. (2) at steady state. We rewrite the diffusion term as $D\frac{d}{dx}\left[\frac{d\rho_{m,n}(x)}{dx}v_{m+n}(x) - \rho_{m,n}(x)\frac{dv_{m+n}(x)}{dx}\right]$,

multiply both sides by $m$, sum with respect to $m$ and $n$, and integrate with respect to $x$. We obtain

$$
\begin{aligned}
0 &= \int_0^\ell dx \sum_n \left[ -k_{\text{on}}^{\text{B}} c_{\text{F}}(x) \sum_{m=0}^{m_{\max}-1} m\,\rho_{m,n}(x) - k_{\text{off}}^{\text{B}} \sum_{m=1}^{m_{\max}} m^2\,\rho_{m,n}(x) \right. \\
&\qquad\qquad \left. + k_{\text{on}}^{\text{B}} c_{\text{F}}(x) \sum_{m=1}^{m_{\max}} m\,\rho_{m-1,n}(x) + k_{\text{off}}^{\text{B}} \sum_{m=0}^{m_{\max}-1} m(m+1)\,\rho_{m+1,n}(x) - \beta \sum_{m=0}^{m_{\max}} m\,\rho_{m,n}(x) \right] \\
&= \int_0^\ell dx \sum_n \left[ k_{\text{on}}^{\text{B}} c_{\text{F}}(x) \sum_{m=0}^{m_{\max}-1} \rho_{m,n}(x) - k_{\text{off}}^{\text{B}} \sum_{m=1}^{m_{\max}} m\,\rho_{m,n}(x) - \beta \sum_{m=0}^{m_{\max}} m\,\rho_{m,n}(x) \right],
\end{aligned}
\tag{S15}
$$

where we wrote explicitly the summation bound in terms of $m_{\max}$, and we obtained the third line by changing summation variables in the first and second terms in the second line. Multiplying Eq. (2) by $n$, summing with respect to $m$ and $n$, integrating with respect to $x$ and proceeding along the same lines, we obtain

$$
\int_0^\ell dx \sum_m \left[ k_{\text{on}}^{\text{T}} c_{\text{F}}(x) \sum_{n=0}^{n_{\max}-1} \rho_{m,n}(x) - k_{\text{off}}^{\text{T}} \sum_{n=1}^{n_{\max}} n\,\rho_{m,n}(x) - \beta \sum_{n=0}^{n_{\max}} n\,\rho_{m,n}(x) \right] = 0.
\tag{S16}
$$

Combining Eqs. (S14), (S15), and (S16), we obtain $J_{\text{F}}(0) = 0$, i.e. there is no F-ribosome flux at midcell.

## S6  Rapid equilibrium for transiently bound ribosomes

In this section, we consider the reaction-diffusion processes of ribosomes and mRNAs in the limit that transient ribosome binding and unbinding is significantly faster than all other relevant processes. As a result, the statistics of B ribosomes can be described entirely as a local Poisson process, without the need for a separate kinetic equation for each mRNA species with a different number of B ribosomes. This substantially simplifies the reaction-diffusion equations, with a major impact on computational tractability.

Let us consider Eq. (2), the reaction-diffusion equation for mRNAs from the main text, at steady state:

$$
\begin{aligned}
D \left[ \frac{d^2\rho_{m,n}(x)}{dx^2} v_{m+n}(x) - \rho_{m,n}(x) \frac{d^2 v_{m+n}(x)}{dx^2} \right] & \\
- k_{\text{on}}^{\text{B}} c_{\text{F}}(x)\rho_{m,n}(x) - k_{\text{off}}^{\text{B}} m\,\rho_{m,n}(x) - k_{\text{on}}^{\text{T}} c_{\text{F}}(x)\rho_{m,n}(x) - k_{\text{off}}^{\text{T}} n\,\rho_{m,n}(x) & \\
+ k_{\text{on}}^{\text{B}} c_{\text{F}}(x)\rho_{m-1,n}(x) + k_{\text{off}}^{\text{B}}(m+1)\,\rho_{m+1,n}(x) + k_{\text{on}}^{\text{T}} c_{\text{F}}(x)\rho_{m,n-1}(x) + k_{\text{off}}^{\text{T}}(n+1)\,\rho_{m,n+1}(x) & \\
+ \delta_{m,0}\,\delta_{n,0}\,\alpha(x) - \beta\,\rho_{m,n}(x) &= 0.
\end{aligned}
\tag{S17}
$$

In the rapid-equilibrium limit, where $k_{\text{on}}^{\text{B}}$, $k_{\text{off}}^{\text{B}}$ are both large, we set

$$
k_{\text{on}}^{\text{B}} = \lambda\,\boldsymbol{k}_{\text{on}}^{\text{B}},\ k_{\text{off}}^{\text{B}} = \lambda\,\boldsymbol{k}_{\text{off}}^{\text{B}},
\tag{S18}
$$

where $\lambda \gg 1$ and $\boldsymbol{k}_{\text{on}}^{\text{B}}$, $\boldsymbol{k}_{\text{off}}^{\text{B}}$ are of order unity. We now consider Eq. (2) to leading order in $\lambda$, and we have

$$
k_{\text{off}}^{\text{B}}(m+1)\rho_{m+1,n}(x) - k_{\text{on}}^{\text{B}} c_{\text{F}}(x)\rho_{m,n}(x) = 0,\ 0 \le m \le m_{\max}-1,
\tag{S19}
$$

which implies

$$
\rho_{m,n}(x) = \frac{1}{m!} \left( \frac{k_{\text{on}}^{\text{B}} c_{\text{F}}(x)}{k_{\text{off}}^{\text{B}}} \right)^m \rho_{0,n}(x) \equiv \boldsymbol{\rho}_{m,n}(x).
\tag{S20}
$$

Equation (S20) implies that the distribution of B ribosomes at position $x$ is Poissonian: this is a direct consequence of the rapid-equilibrium limit, where at any position $x$ binding and unbinding of B ribosomes dominates over all other processes, such as binding-unbinding of T ribosomes and diffusion.

Now, we sum Eq. (S17) with respect to $m = 0, \ldots, m_{\max}$, and we obtain the following conservation equation for the density of mRNAs at position $x$

$$
\begin{aligned}
\sum_{m=0}^{m_{\max}} \left\{ D\frac{d}{dx} \left[ \frac{d\rho_{m,n}(x)}{dx} v_{m+n}(x) - \rho_{m,n}(x)\frac{dv_{m+n}(x)}{dx} \right] - k_{\text{on}}^{\text{T}} c_{\text{F}}(x)\rho_{m,n}(x) - k_{\text{off}}^{\text{T}} n\,\rho_{m,n}(x) \right. & \\
\left. + k_{\text{on}}^{\text{T}} c_{\text{F}}(x)\rho_{m,n-1}(x) + k_{\text{off}}^{\text{T}}(n+1)\,\rho_{m,n+1}(x) \right\} + \delta_{n,0}\,\alpha(x) - \beta \sum_{m=0}^{m_{\max}} \rho_{m,n}(x) &= 0.
\end{aligned}
\tag{S21}
$$

Equations (S20) and (S21) to leading order in $\lambda$ imply the following equation involving only $\rho_{0,n}$:

$$\sum_{m=0}^{m_{\max}} \left[ -\frac{dJ_{m,n}(x)}{dx} + \omega_{m,n}(x) \right] + \delta_{n,0}\alpha(x) = 0, \tag{S22}$$

where we simplified the notation introducing the local flux of mRNAs of species $m$, $n$

$$J_{m,n}(x) = -D \left( \frac{d\boldsymbol{\rho}_{m,n}(x)}{dx} v_{m+n}(x) - \boldsymbol{\rho}_{m,n}(x) \frac{dv_{m+n}(x)}{dx} \right), \tag{S23}$$

and defining

$$\omega_{m,n}(x) = -k_{\mathrm{on}}^{\mathrm{T}} c_{\mathrm{F}}(x)\boldsymbol{\rho}_{m,n}(x) - k_{\mathrm{off}}^{\mathrm{T}}\, n\, \boldsymbol{\rho}_{m,n}(x) + k_{\mathrm{on}}^{\mathrm{T}} c_{\mathrm{F}}(x)\boldsymbol{\rho}_{m,n-1}(x) + k_{\mathrm{off}}^{\mathrm{T}}\,(n+1)\,\boldsymbol{\rho}_{m,n+1}(x) - \beta\boldsymbol{\rho}_{m,n}(x). \tag{S24}$$

Using Eqs. (S12), (S20), and (S23), and setting

$$\rho_n(x) \equiv \rho_{0,n}(x), \tag{S25}$$

we compute explicitly the first term in Eq. (S22) for large $m_{\max}$:

$$\sum_{m=0}^{\infty} \frac{dJ_{m,n}(x)}{dx} =$$
$$-D\frac{d}{dx}\sum_{m=0}^{\infty} \left( \frac{d\boldsymbol{\rho}_{m,n}(x)}{dx} v_{m+n}(x) - \boldsymbol{\rho}_{m,n}(x)\frac{dv_{m+n}(x)}{dx} \right) =$$
$$-D\frac{d}{dx}\sum_{m=0}^{\infty} \left\{ \left[ \frac{1}{m!}m \left( \frac{k_{\mathrm{on}}^{\mathrm{B}} c_{\mathrm{F}}(x)}{k_{\mathrm{off}}^{\mathrm{B}}} \right)^{m-1} \frac{k_{\mathrm{on}}^{\mathrm{B}}}{k_{\mathrm{off}}^{\mathrm{B}}} \frac{dc_{\mathrm{F}}(x)}{dx}\rho_n(x) + \frac{1}{m!} \left( \frac{k_{\mathrm{on}}^{\mathrm{B}} c_{\mathrm{F}}(x)}{k_{\mathrm{off}}^{\mathrm{B}}} \right)^{m} \frac{d\rho_n(x)}{dx} \right] v_{\mathrm{R}}(x)[v_{\mathrm{F}}(x)]^{m+n} \right.$$
$$\left. -\frac{1}{m!} \left( \frac{k_{\mathrm{on}}^{\mathrm{B}} c_{\mathrm{F}}(x)}{k_{\mathrm{off}}^{\mathrm{B}}} \right)^{m} \rho_n(x) \left( \frac{dv_{\mathrm{R}}(x)}{dx}[v_{\mathrm{F}}(x)]^{m+n} + v_{\mathrm{R}}(x)(m+n)[v_{\mathrm{F}}(x)]^{m+n-1}\frac{dv_{\mathrm{F}}(x)}{dx} \right) \right\} =$$
$$-D\frac{d}{dx}\left\{ \exp\left( \frac{k_{\mathrm{on}}^{\mathrm{B}} c_{\mathrm{F}}(x)}{k_{\mathrm{off}}^{\mathrm{B}}} v_{\mathrm{F}}(x) \right) [v_{\mathrm{F}}(x)]^{n} \left[ \frac{k_{\mathrm{on}}^{\mathrm{B}}}{k_{\mathrm{off}}^{\mathrm{B}}}\rho_n(x)v_{\mathrm{R}}(x) \left( \frac{dc_{\mathrm{F}}(x)}{dx}v_{\mathrm{F}}(x) - c_{\mathrm{F}}(x)\frac{dv_{\mathrm{F}}(x)}{dx} \right) \right.\right.$$
$$\left.\left. +\frac{d\rho_n(x)}{dx}v_{\mathrm{R}}(x) - \rho_n(x)\frac{dv_{\mathrm{R}}(x)}{dx} - n\,\rho_n(x)v_{\mathrm{R}}(x)\frac{1}{v_{\mathrm{F}}(x)}\frac{dv_{\mathrm{F}}(x)}{dx} \right] \right\}. \tag{S26}$$

The sums with respect to $m$ in the second line of Eq. (S26) are all of the form $\sum_{m=0}^{\infty} u^m/m!$, and we calculated them explicitly with a change of summation variables in some terms. The sum of $\omega_{m,n}$ in Eq. (S22) can be computed along the same lines as in Eq. (S26). We substitute Eq. (S26) in Eq. (S22), and we obtain the final set of $n_{\max}+1$ reaction-diffusion equations for $\rho_n$:

$$D\frac{d}{dx}\left\{ \exp\left( \frac{k_{\mathrm{on}}^{\mathrm{B}} c_{\mathrm{F}}(x)}{k_{\mathrm{off}}^{\mathrm{B}}} v_{\mathrm{F}}(x) \right) [v_{\mathrm{F}}(x)]^{n} \left[ \frac{k_{\mathrm{on}}^{\mathrm{B}}}{k_{\mathrm{off}}^{\mathrm{B}}}\rho_n(x)v_{\mathrm{R}}(x) \left( \frac{dc_{\mathrm{F}}(x)}{dx}v_{\mathrm{F}}(x) - c_{\mathrm{F}}(x)\frac{dv_{\mathrm{F}}(x)}{dx} \right) + \frac{d\rho_n(x)}{dx}v_{\mathrm{R}}(x) \right.\right.$$
$$\left.\left. -\rho_n(x)\frac{dv_{\mathrm{R}}(x)}{dx} - n\,\rho_n(x)v_{\mathrm{R}}(x)\frac{1}{v_{\mathrm{F}}(x)}\frac{dv_{\mathrm{F}}(x)}{dx} \right] \right\} + \exp\left( \frac{k_{\mathrm{on}}^{\mathrm{B}} c_{\mathrm{F}}(x)}{k_{\mathrm{off}}^{\mathrm{B}}} \right) [-k_{\mathrm{on}}^{\mathrm{T}} c_{\mathrm{F}}(x)\rho_n(x) - k_{\mathrm{off}}^{\mathrm{T}}\, n\,\rho_n(x)$$
$$+k_{\mathrm{on}}^{\mathrm{T}} c_{\mathrm{F}}(x)\rho_{n-1}(x) + k_{\mathrm{off}}^{\mathrm{T}}\,(n+1)\,\rho_{n+1}(x) - \beta\rho_n(x)] + \delta_{n,0}\alpha(x) = 0. \tag{S27}$$

Next, we derive the rapid-equilibrium equation for the concentration of F ribosomes starting from the reaction-diffusion equation (1) in the main text. To achieve this, we introduce the subleading correction to $\rho_{m,n}$, setting

$$\rho_{m,n}(x) = \boldsymbol{\rho}_{m,n}(x) + \frac{1}{\lambda}\Delta\rho_{m,n}(x). \tag{S28}$$

We then substitute Eq. (S28) into Eq. (S17), and we obtain:

$$k_{\mathrm{off}}^{\mathrm{B}}(m+1)\rho_{m+1,n}(x) - k_{\mathrm{on}}^{\mathrm{B}} c_{\mathrm{F}}(x)\rho_{m,n}(x) =$$
$$k_{\mathrm{off}}^{\mathrm{B}}(m+1)\left[ \boldsymbol{\rho}_{m+1,n}(x) + \frac{1}{\lambda}\Delta\rho_{m+1,n}(x) \right] - k_{\mathrm{on}}^{\mathrm{B}} c_{\mathrm{F}}(x)\left[ \boldsymbol{\rho}_{m,n}(x) + \frac{1}{\lambda}\Delta\rho_{m,n}(x) \right] =$$
$$\boldsymbol{k}_{\mathrm{off}}^{\mathrm{B}}(m+1)\Delta\rho_{m+1,n}(x) - \boldsymbol{k}_{\mathrm{on}}^{\mathrm{B}} c_{\mathrm{F}}(x)\Delta\rho_{m,n}(x) \tag{S29}$$

where in the third line we observed that the terms of order $\lambda$ cancel out because of Eq. (S20). Importantly, Eq. (S29) implies that the term in the first line, which represents the current between mRNA species $(m, n)$ and $(m+1, n)$, can be expressed in terms of the combination of subleading corrections to $\rho_{m,n}$ in the last line.

Using Eqs. (S23), (S24), (S29) in Eq. (S17), we obtain the following set of equations

$$[\boldsymbol{k}_{\mathrm{off}}^{\mathrm{B}} m\, \Delta\rho_{m,n}(x) - \boldsymbol{k}_{\mathrm{on}}^{\mathrm{B}} c_{\mathrm{F}}(x)\Delta\rho_{m-1,n}(x)] - [\boldsymbol{k}_{\mathrm{off}}^{\mathrm{B}}(m+1)\Delta\rho_{m+1,n}(x) - \boldsymbol{k}_{\mathrm{on}}^{\mathrm{B}} c_{\mathrm{F}}(x)\Delta\rho_{m,n}(x)] =$$
$$-\frac{dJ_{m,n}(x)}{dx} + \omega_{m,n}(x) + \delta_{m,0}\delta_{n,0}\alpha(x). \tag{S30}$$

Exploiting the recursive structure of Eqs. (S30) [12], we solve them iteratively for the quantities $\boldsymbol{k}_{\mathrm{off}}^{\mathrm{B}} m\, \Delta\rho_{m,n}(x) - \boldsymbol{k}_{\mathrm{on}}^{\mathrm{B}} c_{\mathrm{F}}(x)\Delta\rho_{m-1,n}(x)$, and we obtain

$$\boldsymbol{k}_{\mathrm{off}}^{\mathrm{B}} m_{\max}\Delta\rho_{m_{\max},n}(x) - \boldsymbol{k}_{\mathrm{on}}^{\mathrm{B}} c_{\mathrm{F}}(x)\Delta\rho_{m_{\max}-1,n}(x) = -\frac{dJ_{m_{\max},n}(x)}{dx} + \omega_{m_{\max},n}(x), \tag{S31}$$

$$\boldsymbol{k}_{\mathrm{off}}^{\mathrm{B}}(m_{\max}-1)\Delta\rho_{m_{\max}-1,n}(x) - \boldsymbol{k}_{\mathrm{on}}^{\mathrm{B}} c_{\mathrm{F}}(x)\Delta\rho_{m_{\max}-2,n}(x) = -\frac{dJ_{m_{\max}-1,n}(x)}{dx} + \omega_{m_{\max}-1,n}(x)$$
$$-\frac{dJ_{m_{\max},n}(x)}{dx} + \omega_{m_{\max},n}(x),$$

$$\vdots$$

$$\boldsymbol{k}_{\mathrm{off}}^{\mathrm{B}} m\, \Delta\rho_{m,n}(x) - \boldsymbol{k}_{\mathrm{on}}^{\mathrm{B}} c_{\mathrm{F}}(x)\Delta\rho_{m-1,n}(x) = \sum_{p=m}^{m_{\max}}\left[-\frac{dJ_{p,n}(x)}{dx} + \omega_{p,n}(x)\right],$$

$$\vdots$$

$$\boldsymbol{k}_{\mathrm{off}}^{\mathrm{B}}\Delta\rho_{1,n}(x) - \boldsymbol{k}_{\mathrm{on}}^{\mathrm{B}} c_{\mathrm{F}}(x)\Delta\rho_{0,n}(x) = \sum_{p=1}^{m_{\max}}\left[-\frac{dJ_{p,n}(x)}{dx} + \omega_{p,n}(x)\right]. \tag{S32}$$

Now, let us compute the RHS of Eq. (1) to leading order in $\lambda$. The second and third term on the RHS read

$$-k_{\mathrm{on}}^{\mathrm{B}} c_{\mathrm{F}}(x)\sum_{m=0}^{m_{\max}-1}\sum_{n=0}^{n_{\max}}\rho_{m,n}(x) + k_{\mathrm{off}}^{\mathrm{B}}\sum_{m=1}^{m_{\max}}\sum_{n=0}^{n_{\max}} m\, \rho_{m,n}(x) = \sum_{n=0}^{n_{\max}}\sum_{m=1}^{m_{\max}}[-\boldsymbol{k}_{\mathrm{on}}^{\mathrm{B}} c_{\mathrm{F}}(x)\Delta\rho_{m-1,n}(x) + \boldsymbol{k}_{\mathrm{off}}^{\mathrm{B}} m\, \Delta\rho_{m,n}(x)]$$
$$= \sum_{n=0}^{n_{\max}}\sum_{m=1}^{m_{\max}}\sum_{p=m}^{m_{\max}}\left[-\frac{dJ_{p,n}(x)}{dx} + \omega_{p,n}(x)\right]$$
$$= \sum_{n=0}^{n_{\max}}\sum_{m=1}^{m_{\max}} m\left[-\frac{dJ_{m,n}(x)}{dx} + \omega_{m,n}(x)\right], \tag{S33}$$

where in the first line we used Eq. (S29), in the second line we used Eqs. (S31), ..., (S32), and in the last line we rewrote the double sum over $m$, $p$ as a single sum over $m$ by using the identity

$$\sum_{m=1}^{m_{\max}}\sum_{p=m}^{m_{\max}} a_p = \sum_{m=1}^{m_{\max}} m\, a_m, \tag{S34}$$

which is valid for any sequence $a_m$. The first term in the last line of Eq. (S33) can be computed for large $m_{\max}$ along

the lines of Eq. (S26):

$$\sum_{n=0}^{n_{\max}} \sum_{m=1}^{m_{\max}} m \, \frac{dJ_{m,n}(x)}{dx} =$$

$$-D\frac{d}{dx}\sum_{n=0}^{n_{\max}}\sum_{m=1}^{\infty}\left\{\left[\frac{1}{m!}m^2\left(\frac{k_{\mathrm{on}}^{\mathrm{B}}c_{\mathrm{F}}(x)}{k_{\mathrm{off}}^{\mathrm{B}}}\right)^{m-1}\frac{k_{\mathrm{on}}^{\mathrm{B}}}{k_{\mathrm{off}}^{\mathrm{B}}}\frac{dc_{\mathrm{F}}(x)}{dx}\rho_n(x)+\frac{1}{m!}m\left(\frac{k_{\mathrm{on}}^{\mathrm{B}}c_{\mathrm{F}}(x)}{k_{\mathrm{off}}^{\mathrm{B}}}\right)^{m}\frac{d\rho_n(x)}{dx}\right]v_{\mathrm{R}}(x)[v_{\mathrm{F}}(x)]^{m+n}\right.$$

$$\left.-\frac{1}{m!}m\left(\frac{k_{\mathrm{on}}^{\mathrm{B}}c_{\mathrm{F}}(x)}{k_{\mathrm{off}}^{\mathrm{B}}}\right)^{m}\rho_n(x)\left(\frac{dv_{\mathrm{R}}(x)}{dx}[v_{\mathrm{F}}(x)]^{m+n}+v_{\mathrm{R}}(x)(m+n)[v_{\mathrm{F}}(x)]^{m+n-1}\frac{dv_{\mathrm{F}}(x)}{dx}\right)\right\}=$$

$$-D\frac{d}{dx}\sum_{n=0}^{n_{\max}}\left\{\exp\left(\frac{k_{\mathrm{on}}^{\mathrm{B}}c_{\mathrm{F}}(x)}{k_{\mathrm{off}}^{\mathrm{B}}}v_{\mathrm{F}}(x)\right)[v_{\mathrm{F}}(x)]^{n}\left[\left(1+\frac{k_{\mathrm{on}}^{\mathrm{B}}c_{\mathrm{F}}(x)}{k_{\mathrm{off}}^{\mathrm{B}}}v_{\mathrm{F}}(x)\right)\frac{k_{\mathrm{on}}^{\mathrm{B}}}{k_{\mathrm{off}}^{\mathrm{B}}}\frac{dc_{\mathrm{F}}(x)}{dx}\rho_n(x)v_{\mathrm{R}}(x)v_{\mathrm{F}}(x)\right.\right.$$

$$+\frac{k_{\mathrm{on}}^{\mathrm{B}}c_{\mathrm{F}}(x)}{k_{\mathrm{off}}^{\mathrm{B}}}\frac{d\rho_n(x)}{dx}v_{\mathrm{R}}(x)v_{\mathrm{F}}(x)-\frac{k_{\mathrm{on}}^{\mathrm{B}}c_{\mathrm{F}}(x)}{k_{\mathrm{off}}^{\mathrm{B}}}\rho_n(x)\frac{dv_{\mathrm{R}}(x)}{dx}v_{\mathrm{F}}(x)$$

$$\left.\left.-\frac{k_{\mathrm{on}}^{\mathrm{B}}c_{\mathrm{F}}(x)}{k_{\mathrm{off}}^{\mathrm{B}}}\left(1+\frac{k_{\mathrm{on}}^{\mathrm{B}}c_{\mathrm{F}}(x)}{k_{\mathrm{off}}^{\mathrm{B}}}v_{\mathrm{F}}(x)\right)\rho_n(x)v_{\mathrm{R}}(x)\frac{dv_{\mathrm{F}}(x)}{dx}-\frac{k_{\mathrm{on}}^{\mathrm{B}}c_{\mathrm{F}}(x)}{k_{\mathrm{off}}^{\mathrm{B}}}\rho_n(x)v_{\mathrm{R}}(x)n\,\frac{dv_{\mathrm{F}}(x)}{dx}\right]\right\}. \quad (S35)$$

The second term in the last line of Eq. (S33) reads

$$\sum_{n=0}^{n_{\max}} \sum_{m=1}^{m_{\max}} m \, \omega_{m,n}(x) = -\beta \sum_{n=0}^{n_{\max}} \sum_{m=1}^{m_{\max}} m \, \boldsymbol{\rho}_{m,n}(x), \quad (S36)$$

where we exchanged the sums with respect to $n$, $m$, and we observed that the only term in $\omega_{m,n}$ that is nonzero when summed with respect to $m$ and $n$ is the term proportional to $\beta$, see Eqs. (S24). The remaining terms in the RHS of Eq. (1) can be computed by using Eqs. (S20) and (S25), by which we calculate the exponential sums with respect to $m$. Using Eqs. (1), (S33), (S35), and (S36), we obtain the final equation for the concentration of F ribosomes in the rapid-equilibrium limit

$$D_{\mathrm{F}}\left[\frac{d^2c_{\mathrm{F}}(x)}{dx^2}v_{\mathrm{F}}(x)-c_{\mathrm{F}}(x)\frac{d^2v_{\mathrm{F}}(x)}{dx^2}\right]$$

$$+D\frac{d}{dx}\sum_{n=0}^{n_{\max}}\left\{\exp\left(\frac{k_{\mathrm{on}}^{\mathrm{B}}c_{\mathrm{F}}(x)}{k_{\mathrm{off}}^{\mathrm{B}}}v_{\mathrm{F}}(x)\right)[v_{\mathrm{F}}(x)]^{n}\left[\left(1+\frac{k_{\mathrm{on}}^{\mathrm{B}}c_{\mathrm{F}}(x)}{k_{\mathrm{off}}^{\mathrm{B}}}v_{\mathrm{F}}(x)\right)\left(\frac{dc_{\mathrm{F}}(x)}{dx}v_{\mathrm{F}}(x)-c_{\mathrm{F}}(x)\frac{dv_{\mathrm{F}}(x)}{dx}\right)\right.\right.$$

$$\times\frac{k_{\mathrm{on}}^{\mathrm{B}}}{k_{\mathrm{off}}^{\mathrm{B}}}\rho_n(x)v_{\mathrm{R}}(x)+\frac{k_{\mathrm{on}}^{\mathrm{B}}c_{\mathrm{F}}(x)}{k_{\mathrm{off}}^{\mathrm{B}}}\left(\frac{d\rho_n(x)}{dx}v_{\mathrm{R}}(x)-\rho_n(x)\frac{dv_{\mathrm{R}}(x)}{dx}\right)v_{\mathrm{F}}(x)-\left.\left.\frac{k_{\mathrm{on}}^{\mathrm{B}}c_{\mathrm{F}}(x)}{k_{\mathrm{off}}^{\mathrm{B}}}\rho_n(x)v_{\mathrm{R}}(x)n\,\frac{dv_{\mathrm{F}}(x)}{dx}\right]\right\}$$

$$+\exp\left(\frac{k_{\mathrm{on}}^{\mathrm{B}}c_{\mathrm{F}}(x)}{k_{\mathrm{off}}^{\mathrm{B}}}\right)\left(-k_{\mathrm{on}}^{\mathrm{T}}c_{\mathrm{F}}(x)\sum_{n=0}^{n_{\max}-1}\rho_n(x)+k_{\mathrm{off}}^{\mathrm{T}}\sum_{n=1}^{n_{\max}}n\,\rho_n(x)+\beta\sum_{n=0}^{n_{\max}}n\,\rho_n(x)\right)=0, \quad (S37)$$

where the terms in the second and third line represent the diffusive flux of B ribosomes carried by diffusing mRNAs.

Let us now discuss the boundary conditions for the rapid-equilibrium equations: The boundary conditions for Eq. (S27) are

$$\left[\frac{d\rho_n(x)}{dx}v_{\mathrm{R}}(x)[v_{\mathrm{F}}(x)]^{n}-\rho_n(x)\frac{d[v_{\mathrm{R}}(x)[v_{\mathrm{F}}(x)]^{n}]}{dx}\right]\bigg|_{x=0,\ell}=0. \quad (S38)$$

In Eq. (S37), the no-flux condition boundary condition at the right pole reads

$$\left[\frac{dc_{\mathrm{F}}(x)}{dx}v_{\mathrm{F}}(x)-c_{\mathrm{F}}(x)\frac{dv_{\mathrm{F}}(x)}{dx}\right]\bigg|_{x=\ell}=0. \quad (S39)$$

while the constraint on total ribosome number can be written as

$$N_{\mathrm{tot}} = 2\int_0^\ell dx\left[c_{\mathrm{F}}(x)+\sum_{m=0}^{\infty}\sum_{n=0}^{n_{\max}}(m+n)\boldsymbol{\rho}_{m,n}(x)\right]$$

$$= 2\int_0^\ell dx\left[c_{\mathrm{F}}(x)+\exp\left(\frac{k_{\mathrm{on}}^{\mathrm{B}}c_{\mathrm{F}}(x)}{k_{\mathrm{off}}^{\mathrm{B}}}\right)\sum_{n=0}^{n_{\max}}\left(\frac{k_{\mathrm{on}}^{\mathrm{B}}c_{\mathrm{F}}(x)}{k_{\mathrm{off}}^{\mathrm{B}}}+n\right)\rho_n(x)\right], \quad (S40)$$

where in the second line we summed with respect to $m$ by using Eqs. (S20) and (S25).

Overall, the $n_{\max} + 2$ reaction-diffusion equations (S27), (S37) and their boundary conditions (S38), (S39), (S40) completely characterize the solution $\rho_n(x)$, $c_{\mathrm{F}}(x)$. Using Eqs. (S20) and (S25), we obtain from $\rho_n(x)$ and $c_{\mathrm{F}}(x)$ the full set of mRNA concentrations $\rho_{m,n}(x)$ for any $m$ and $0 \leq n \leq n_{\max}$ as well as all other physical quantities of interest, such as the total mRNA concentration

$$
\begin{aligned}
\rho_{\mathrm{tot}}(x) &= \sum_{m=0}^{\infty} \sum_{n=0}^{n_{\max}} \boldsymbol{\rho}_{m,n}(x) \\
&= \exp\left(\frac{k_{\mathrm{on}}^{\mathrm{B}} c_{\mathrm{F}}(x)}{k_{\mathrm{off}}^{\mathrm{B}}}\right) \sum_{n=0}^{n_{\max}} \rho_n(x),
\end{aligned}
\tag{S41}
$$

the concentrations of T and B ribosomes

$$
\begin{aligned}
c_{\mathrm{T}}(x) &= \sum_{m=0}^{\infty} \sum_{n=0}^{n_{\max}} n\, \boldsymbol{\rho}_{m,n}(x) \\
&= \exp\left(\frac{k_{\mathrm{on}}^{\mathrm{B}} c_{\mathrm{F}}(x)}{k_{\mathrm{off}}^{\mathrm{B}}}\right) \sum_{n=0}^{n_{\max}} n\, \rho_n(x),
\end{aligned}
\tag{S42}
$$

$$
\begin{aligned}
c_{\mathrm{B}}(x) &= \sum_{m=0}^{\infty} \sum_{n=0}^{n_{\max}} m\, \boldsymbol{\rho}_{m,n}(x) \\
&= \frac{k_{\mathrm{on}}^{\mathrm{B}} c_{\mathrm{F}}(x)}{k_{\mathrm{off}}^{\mathrm{B}}} \exp\left(\frac{k_{\mathrm{on}}^{\mathrm{B}} c_{\mathrm{F}}(x)}{k_{\mathrm{off}}^{\mathrm{B}}}\right) \sum_{n=0}^{n_{\max}} \rho_n(x),
\end{aligned}
\tag{S43}
$$

the total mRNA flux

$$
\begin{aligned}
J_{\mathrm{mRNA}}(x) &= \sum_{m=0}^{\infty} \sum_{n=0}^{n_{\max}} J_{m,n}(x) \\
&= -D \exp\left(\frac{k_{\mathrm{on}}^{\mathrm{B}} c_{\mathrm{F}}(x)}{k_{\mathrm{off}}^{\mathrm{B}}} v_{\mathrm{F}}(x)\right) \sum_{n=0}^{n_{\max}} [v_{\mathrm{F}}(x)]^n \left[\frac{k_{\mathrm{on}}^{\mathrm{B}}}{k_{\mathrm{off}}^{\mathrm{B}}} \rho_n(x) v_{\mathrm{R}}(x) \left(\frac{dc_{\mathrm{F}}(x)}{dx} v_{\mathrm{F}}(x) - c_{\mathrm{F}}(x) \frac{dv_{\mathrm{F}}(x)}{dx}\right) \right. \\
&\quad \left. + \frac{d\rho_n(x)}{dx} v_{\mathrm{R}}(x) - \rho_n(x) \frac{dv_{\mathrm{R}}(x)}{dx} - n\, \rho_n(x) v_{\mathrm{R}}(x) \frac{1}{v_{\mathrm{F}}(x)} \frac{dv_{\mathrm{F}}(x)}{dx}\right],
\end{aligned}
\tag{S44}
$$

and the fluxes of T and B ribosomes

$$
\begin{aligned}
J_{\mathrm{T}}(x) &= \sum_{m=0}^{\infty} \sum_{n=0}^{n_{\max}} n\, J_{m,n}(x) \\
&= -D \exp\left(\frac{k_{\mathrm{on}}^{\mathrm{B}} c_{\mathrm{F}}(x)}{k_{\mathrm{off}}^{\mathrm{B}}} v_{\mathrm{F}}(x)\right) \sum_{n=0}^{n_{\max}} n[v_{\mathrm{F}}(x)]^n \left[\frac{k_{\mathrm{on}}^{\mathrm{B}}}{k_{\mathrm{off}}^{\mathrm{B}}} \rho_n(x) v_{\mathrm{R}}(x) \left(\frac{dc_{\mathrm{F}}(x)}{dx} v_{\mathrm{F}}(x) - c_{\mathrm{F}}(x) \frac{dv_{\mathrm{F}}(x)}{dx}\right) \right. \\
&\quad \left. + \frac{d\rho_n(x)}{dx} v_{\mathrm{R}}(x) - \rho_n(x) \frac{dv_{\mathrm{R}}(x)}{dx} - n\, \rho_n(x) v_{\mathrm{R}}(x) \frac{1}{v_{\mathrm{F}}(x)} \frac{dv_{\mathrm{F}}(x)}{dx}\right],
\end{aligned}
\tag{S45}
$$

$$
\begin{aligned}
J_{\mathrm{B}}(x) &= \sum_{m=0}^{\infty} \sum_{n=0}^{n_{\max}} m\, J_{m,n}(x) \\
&= -D \exp\left(\frac{k_{\mathrm{on}}^{\mathrm{B}} c_{\mathrm{F}}(x)}{k_{\mathrm{off}}^{\mathrm{B}}} v_{\mathrm{F}}(x)\right) \sum_{n=0}^{n_{\max}} [v_{\mathrm{F}}(x)]^n \left[ \left(1 + \frac{k_{\mathrm{on}}^{\mathrm{B}} c_{\mathrm{F}}(x)}{k_{\mathrm{off}}^{\mathrm{B}}} v_{\mathrm{F}}(x)\right) \left(\frac{dc_{\mathrm{F}}(x)}{dx} v_{\mathrm{F}}(x) - c_{\mathrm{F}}(x) \frac{dv_{\mathrm{F}}(x)}{dx}\right) \right. \\
&\quad \left. \times \frac{k_{\mathrm{on}}^{\mathrm{B}}}{k_{\mathrm{off}}^{\mathrm{B}}} \rho_n(x) v_{\mathrm{R}}(x) + \frac{k_{\mathrm{on}}^{\mathrm{B}} c_{\mathrm{F}}(x)}{k_{\mathrm{off}}^{\mathrm{B}}} \left(\frac{d\rho_n(x)}{dx} v_{\mathrm{R}}(x) - \rho_n(x) \frac{dv_{\mathrm{R}}(x)}{dx}\right) v_{\mathrm{F}}(x) - \frac{k_{\mathrm{on}}^{\mathrm{B}} c_{\mathrm{F}}(x)}{k_{\mathrm{off}}^{\mathrm{B}}} \rho_n(x) v_{\mathrm{R}}(x) n\, \frac{dv_{\mathrm{F}}(x)}{dx}\right].
\end{aligned}
\tag{S46}
$$

The solution of the rapid-equilibrium equations allows for computing another important quantity, the ribosome efficiency, which we define as the average number of proteins translated per unit time per ribosome. We thus estimate

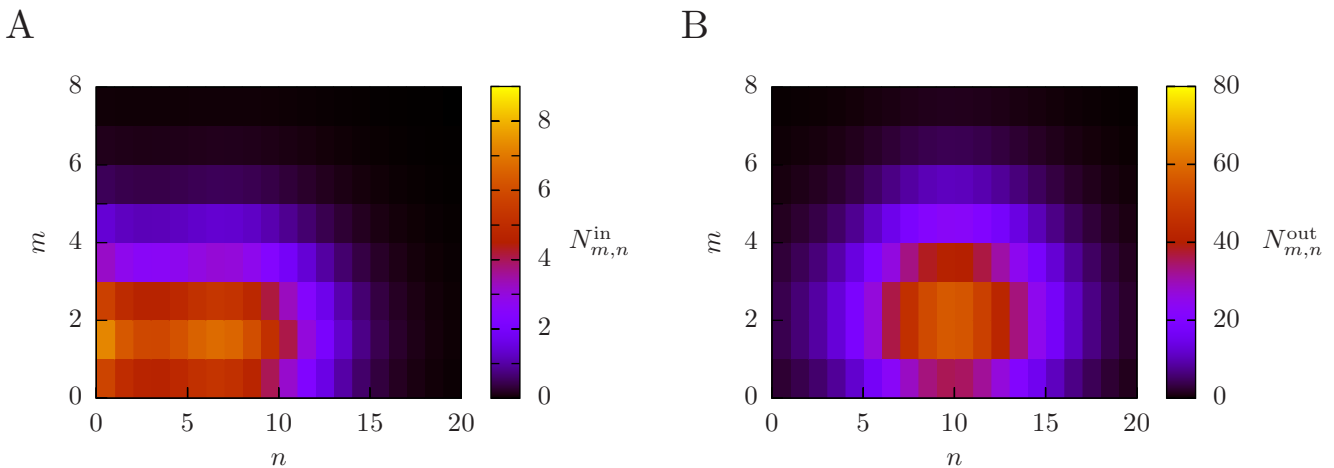

FIG. S2: Comparison of mRNA species inside versus outside the nucleoid. (A) Heat map of the number of mRNAs of species $m$, $n$ inside the nucleoid $N_{m,n}^{\mathrm{in}} = \int_0^{\ell/2} \rho_{m,n}(x)$ in the right half of the cell as a function of the numbers $m$, $n$ of transiently bound and translating ribosomes, respectively. (B) Same plot as in the left panel for the number of mRNAs outside the nucleoid $N_{m,n}^{\mathrm{out}} = \int_{\ell/2}^{\ell} \rho_{m,n}(x)$.

the total rate of protein production per cell in our model as the total rate at which T ribosomes fall of mRNAs after completing translation, i.e.

$$
\begin{aligned}
\Sigma &= 2 \, k_{\mathrm{off}}^{\mathrm{T}} \sum_{m=0}^{\infty} \sum_{n=1}^{n_{\max}} n \int_0^{\ell} dx \, \boldsymbol{\rho}_{m,n}(x) \\
&= 2 \, k_{\mathrm{off}}^{\mathrm{T}} \sum_{n=1}^{n_{\max}} n \int_0^{\ell} dx \exp\left( \frac{k_{\mathrm{on}}^{\mathrm{B}} c_{\mathrm{F}}(x)}{k_{\mathrm{off}}^{\mathrm{B}}} \right) \rho_n(x),
\end{aligned}
\tag{S47}
$$

and we define the ribosome efficiency as $\varepsilon = \Sigma / N_{\mathrm{tot}}$.

## S7  mRNA species inside versus outside the nucleoid

In Fig. S2 we show the distribution of mRNA species in the nucleoid region and in the polar region as heat maps for the numbers $N_{m,n}^{\mathrm{in}}$, $N_{m,n}^{\mathrm{out}}$ of mRNAs of species $m$, $n$ for $0 \leq x \leq \ell/2$ and $l/2 \leq x \leq \ell$, respectively.

## S8  Estimate of ribosome flux

In this section we present an analytical estimate of the flux of mRNA-bound ribosomes flowing from the nucleoid region toward one cell pole. The ribosome flux can be estimated by assuming that every newly synthesized mRNA migrates from the nucleoid to the pole, and carries an average load of ribosomes, yielding an estimated ribosome flux

$$
J_{\mathrm{T+B}}^{\mathrm{est}} = \frac{\alpha_{\mathrm{tot}}}{2} \frac{N_{\mathrm{T}} + N_{\mathrm{B}}}{N_{\mathrm{mRNA}}},
\tag{S48}
$$

where $\alpha_{\mathrm{tot}} = 2 \int_0^{\ell} dx \, \alpha(x) \sim 17/\mathrm{s}$ is the total mRNA transcription rate in the nucleoid region, and the second factor is the average number of ribosomes carried by each mRNA, with $N_{\mathrm{T}} = 4.8 \times 10^4$, $N_{\mathrm{B}} = 1.1 \times 10^4$, $N_{\mathrm{mRNA}} = 5 \times 10^3$. Equation (S48) predicts a flow of $J_{\mathrm{T+B}}^{\mathrm{est}} \sim 100/\mathrm{s}$, in rough agreement with our numerical result $J_{\mathrm{T+B}} = J_{\mathrm{T}} + J_{\mathrm{B}} \sim 30/\mathrm{s}$ for the flux of T and B ribosomes at the nucleoid edge shown in Fig. 4. The factor of $\sim 3$ discrepancy is not due to our estimate of the mRNA flux $J_{\mathrm{mRNA}}^{\mathrm{est}} \simeq \alpha_{\mathrm{tot}}/2 \simeq 8.5/\mathrm{s}$, which agrees reasonably well with the numerical result $J_{\mathrm{mRNA}} \simeq 7/\mathrm{s}$ at the edge of the nucleoid. Rather, the difference is due to the fact that mRNAs with fewer ribosomes diffuse faster due to the strong dependence of the available volume factor $v_{m,n}$ on $m + n$, which means that most of

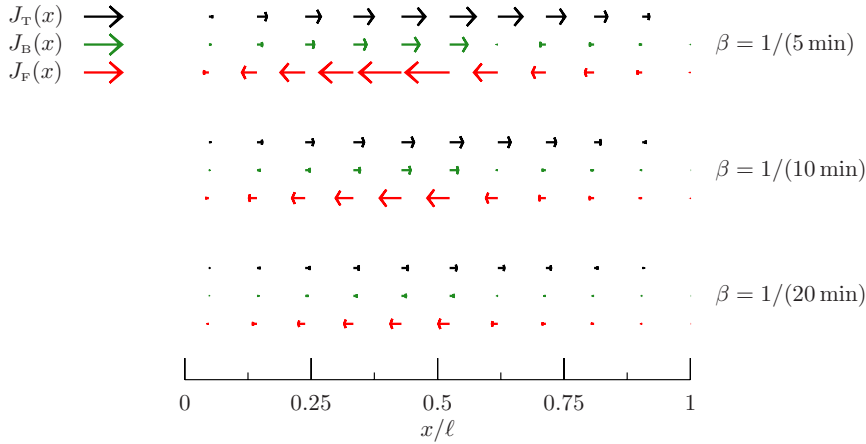

FIG. S3: Fluxes of translating (T), transiently bound (B) and free (F) ribosomes along the cell's long axis: The arrow length is proportional to local ribosome flux, and the arrows in the legend correspond to a flux of 20/s. The fluxes are shown for different values of mRNA degradation rate $\beta$, where $\beta$ and the total mRNA production rate $\alpha_{\mathrm{tot}}$ are varied together, keeping the total mRNA number constant.

the diffusive flux of mRNAs is due to polysomes with a number of ribosomes substantially smaller than average. Thus our use of the average ribosome loading $(N_{\mathrm{T}} + N_{\mathrm{B}})/N_{\mathrm{mRNA}}$ in Eq. (S48) substantially overestimates the ribosome flux carried by mRNAs. However, the estimate $J_{\mathrm{T+B}}^{\mathrm{est}}$ in Eq. (S48) does correctly capture the full numerical result that the ribosome flux is proportional to the total mRNA production rate $\alpha_{\mathrm{tot}}$, see Fig. S3. This proportionality leads us to conclude that ribosome circulation is driven by the flux of new mRNAs from the nucleoid to the poles, and not by binding of mature mRNAs and F ribosomes in the nucleoid with subsequent expulsion from the nucleoid due to excluded-volume effects as discussed in the main text.

## S9  Model including nucleoid-bound mRNAs

In this section, we extend our model to include the fact that mRNAs remain bound to the nucleoid while being transcribed. We introduce an additional set of nucleoid-bound mRNA species with $m$ B ribosomes and $n$ T ribosomes, whose densities we denote by $\rho_{m,n}^*(x,t)$. The reaction-diffusion equations for the nucleoid-bound mRNA densities are

$$
\begin{aligned}
\frac{\partial \rho_{m,n}^*(x,t)}{\partial t} =\ & -k_{\mathrm{on}}^{\mathrm{B}} c_{\mathrm{F}}(x,t) \rho_{m,n}^*(x,t) - k_{\mathrm{off}}^{\mathrm{B}}\, m\, \rho_{m,n}^*(x,t) - k_{\mathrm{on}}^{\mathrm{T}} c_{\mathrm{F}}(x,t) \rho_{m,n}^*(x,t) - k_{\mathrm{off}}^{\mathrm{T}}\, n\, \rho_{m,n}^*(x,t) \\
& + k_{\mathrm{on}}^{\mathrm{B}} c_{\mathrm{F}}(x,t) \rho_{m-1,n}^*(x,t) + k_{\mathrm{off}}^{\mathrm{B}}\, (m+1)\, \rho_{m+1,n}^*(x,t) + k_{\mathrm{on}}^{\mathrm{T}} c_{\mathrm{F}}(x,t) \rho_{m,n-1}^*(x,t) + k_{\mathrm{off}}^{\mathrm{T}}\, (n+1)\, \rho_{m,n+1}^*(x,t) \\
& + \delta_{m,0}\, \delta_{n,0}\, \alpha(x) - \beta\, \rho_{m,n}^*(x,t) - \gamma\, \rho_{m,n}^*(x,t).
\end{aligned}
\tag{S49}
$$

Since the mRNA species $\rho_{m,n}^*$ are bound to the nucleoid, there is no diffusion term. The terms in the first and second lines on the RHS correspond to ribosome binding and unbinding as in Eq. (2). The first term in the last line represents mRNA transcription, the second term co-transcriptional degradation [13], while the last term represents unbinding of an mRNA from the nucleoid at rate $\gamma$. The dynamics of free mRNA species, i.e. mRNAs that are not bound to the nucleoid, is still described by Eq. (2), where the mRNA-transcription term in the last line is replaced by the source term $\gamma\, \rho_{m,n}^*(x)$, which represents a nucleoid-bound mRNA becoming a free mRNA:

$$
\begin{aligned}
\frac{\partial \rho_{m,n}(x,t)}{\partial t} =\ & D\left[\frac{\partial^2 \rho_{m,n}(x,t)}{\partial x^2} v_{m+n}(x) - \rho_{m,n}(x,t) \frac{d^2 v_{m+n}(x)}{dx^2}\right] \\
& - k_{\mathrm{on}}^{\mathrm{B}} c_{\mathrm{F}}(x,t) \rho_{m,n}(x,t) - k_{\mathrm{off}}^{\mathrm{B}}\, m\, \rho_{m,n}(x,t) - k_{\mathrm{on}}^{\mathrm{T}} c_{\mathrm{F}}(x,t) \rho_{m,n}(x,t) - k_{\mathrm{off}}^{\mathrm{T}}\, n\, \rho_{m,n}(x,t) \\
& + k_{\mathrm{on}}^{\mathrm{B}} c_{\mathrm{F}}(x,t) \rho_{m-1,n}(x,t) + k_{\mathrm{off}}^{\mathrm{B}}\, (m+1)\, \rho_{m+1,n}(x,t) + k_{\mathrm{on}}^{\mathrm{T}} c_{\mathrm{F}}(x,t) \rho_{m,n-1}(x,t) + k_{\mathrm{off}}^{\mathrm{T}}\, (n+1)\, \rho_{m,n+1}(x,t) \\
& + \gamma\, \rho_{m,n}^*(x,t) - \beta\, \rho_{m,n}(x,t).
\end{aligned}
\tag{S50}
$$

We take the unbinding rate of nucleoid-bound mRNAs to be $\gamma = 1/\mathrm{min}$, i.e. the inverse of the average time for mRNA transcription, compare section II. Finally, the dynamics of F ribosomes is described by the analog of Eq. (1):

$$\frac{\partial c_{\mathrm{F}}(x,t)}{\partial t} = D_{\mathrm{F}} \left[ \frac{\partial^2 c_{\mathrm{F}}(x,t)}{\partial x^2} v_{\mathrm{F}}(x) - c_{\mathrm{F}}(x,t) \frac{d^2 v_{\mathrm{F}}(x)}{dx^2} \right]$$

$$-k_{\mathrm{on}}^{\mathrm{B}} c_{\mathrm{F}}(x,t) \sum_{m=0}^{m_{\max}-1} \sum_{n=0}^{n_{\max}} [\rho_{m,n}(x,t) + \rho_{m,n}^*(x,t)] + k_{\mathrm{off}}^{\mathrm{B}} \sum_{m=1}^{m_{\max}} \sum_{n=0}^{n_{\max}} m \, [\rho_{m,n}(x,t) + \rho_{m,n}^*(x,t)]$$

$$-k_{\mathrm{on}}^{\mathrm{T}} c_{\mathrm{F}}(x,t) \sum_{m=0}^{m_{\max}} \sum_{n=0}^{n_{\max}-1} [\rho_{m,n}(x,t) + \rho_{m,n}^*(x,t)] + k_{\mathrm{off}}^{\mathrm{T}} \sum_{m=0}^{m_{\max}} \sum_{n=1}^{n_{\max}} n \, [\rho_{m,n}(x,t) + \rho_{m,n}^*(x,t)]$$

$$+\beta \sum_{m=0}^{m_{\max}} \sum_{n=0}^{n_{\max}} (m+n)[\rho_{m,n}(x,t) + \rho_{m,n}^*(x,t)]. \tag{S51}$$

We now consider the steady state of the equations above for $\rho_{m,n}$, $\rho_{m,n}^*$ and $c_{\mathrm{F}}$, and we assume that B ribosomes are in rapid equilibrium with F ribosomes. Proceeding along the lines of section S6, in the rapid-equilibrium limit we have

$$\rho_{m,n}(x) = \frac{1}{m!} \left( \frac{k_{\mathrm{on}}^{\mathrm{B}} c_{\mathrm{F}}(x)}{k_{\mathrm{off}}^{\mathrm{B}}} \right)^m \rho_n(x), \tag{S52}$$

$$\rho_{m,n}^*(x) = \frac{1}{m!} \left( \frac{k_{\mathrm{on}}^{\mathrm{B}} c_{\mathrm{F}}(x)}{k_{\mathrm{off}}^{\mathrm{B}}} \right)^m \rho_n^*(x), \tag{S53}$$

where $\rho_n(x) = \rho_{0,n}(x)$, and $\rho_n^*(x) = \rho_{0,n}^*(x)$. In the rapid-equilibrium limit, the densities $\rho_n(x)$ satisfy the following reaction-diffusion equation

$$D \frac{d}{dx} \left\{ \exp\left( \frac{k_{\mathrm{on}}^{\mathrm{B}} c_{\mathrm{F}}(x)}{k_{\mathrm{off}}^{\mathrm{B}}} v_{\mathrm{F}}(x) \right) [v_{\mathrm{F}}(x)]^n \left[ \frac{k_{\mathrm{on}}^{\mathrm{B}}}{k_{\mathrm{off}}^{\mathrm{B}}} \rho_n(x) v_{\mathrm{R}}(x) \left( \frac{dc_{\mathrm{F}}(x)}{dx} v_{\mathrm{F}}(x) - c_{\mathrm{F}}(x) \frac{dv_{\mathrm{F}}(x)}{dx} \right) \right. \right. \tag{S54}$$

$$\left. \left. + \frac{d\rho_n(x)}{dx} v_{\mathrm{R}}(x) - \rho_n(x) \frac{dv_{\mathrm{R}}(x)}{dx} - n \, \rho_n(x) v_{\mathrm{R}}(x) \frac{1}{v_{\mathrm{F}}(x)} \frac{dv_{\mathrm{F}}(x)}{dx} \right] \right\}$$

$$+ \exp\left( \frac{k_{\mathrm{on}}^{\mathrm{B}} c_{\mathrm{F}}(x)}{k_{\mathrm{off}}^{\mathrm{B}}} \right) [-k_{\mathrm{on}}^{\mathrm{T}} c_{\mathrm{F}}(x) \rho_n(x) - k_{\mathrm{off}}^{\mathrm{T}} n \, \rho_n(x) + k_{\mathrm{on}}^{\mathrm{T}} c_{\mathrm{F}}(x) \rho_{n-1}(x) + k_{\mathrm{off}}^{\mathrm{T}} (n+1) \, \rho_{n+1}(x) - \beta \rho_n(x) + \gamma \rho_n^*(x)] = 0,$$

while $\rho_n^*(x)$ satisfies the following equation involving only reaction terms

$$\exp\left( \frac{k_{\mathrm{on}}^{\mathrm{B}} c_{\mathrm{F}}(x)}{k_{\mathrm{off}}^{\mathrm{B}}} \right) [-k_{\mathrm{on}}^{\mathrm{T}} c_{\mathrm{F}}(x) \rho_n^*(x) - k_{\mathrm{off}}^{\mathrm{T}} n \, \rho_n^*(x) + k_{\mathrm{on}}^{\mathrm{T}} c_{\mathrm{F}}(x) \rho_{n-1}^*(x) + k_{\mathrm{off}}^{\mathrm{T}} (n+1) \, \rho_{n+1}^*(x) - (\beta + \gamma) \rho_n^*(x)] \tag{S55}$$

$$+ \alpha(x) = 0.$$

Finally, the rapid-equilibrium reaction-diffusion equation for $c_{\mathrm{F}}(x)$ reads

$$D_{\mathrm{F}} \left[ \frac{d^2 c_{\mathrm{F}}(x)}{dx^2} v_{\mathrm{F}}(x) - c_{\mathrm{F}}(x) \frac{d^2 v_{\mathrm{F}}(x)}{dx^2} \right] \tag{S56}$$

$$+ D \frac{d}{dx} \sum_{n=0}^{n_{\max}} \left\{ \exp\left( \frac{k_{\mathrm{on}}^{\mathrm{B}} c_{\mathrm{F}}(x)}{k_{\mathrm{off}}^{\mathrm{B}}} v_{\mathrm{F}}(x) \right) [v_{\mathrm{F}}(x)]^n \left[ \left( 1 + \frac{k_{\mathrm{on}}^{\mathrm{B}} c_{\mathrm{F}}(x)}{k_{\mathrm{off}}^{\mathrm{B}}} v_{\mathrm{F}}(x) \right) \left( \frac{dc_{\mathrm{F}}(x)}{dx} v_{\mathrm{F}}(x) - c_{\mathrm{F}}(x) \frac{dv_{\mathrm{F}}(x)}{dx} \right) \right. \right.$$

$$\left. \times \frac{k_{\mathrm{on}}^{\mathrm{B}}}{k_{\mathrm{off}}^{\mathrm{B}}} \rho_n(x) v_{\mathrm{R}}(x) + \frac{k_{\mathrm{on}}^{\mathrm{B}} c_{\mathrm{F}}(x)}{k_{\mathrm{off}}^{\mathrm{B}}} \left( \frac{d\rho_n(x)}{dx} v_{\mathrm{R}}(x) - \rho_n(x) \frac{dv_{\mathrm{R}}(x)}{dx} \right) v_{\mathrm{F}}(x) - \frac{k_{\mathrm{on}}^{\mathrm{B}} c_{\mathrm{F}}(x)}{k_{\mathrm{off}}^{\mathrm{B}}} \rho_n(x) v_{\mathrm{R}}(x) n \frac{dv_{\mathrm{F}}(x)}{dx} \right] \right\}$$

$$\exp\left( \frac{k_{\mathrm{on}}^{\mathrm{B}} c_{\mathrm{F}}(x)}{k_{\mathrm{off}}^{\mathrm{B}}} \right) \left\{ -k_{\mathrm{on}}^{\mathrm{T}} c_{\mathrm{F}}(x) \sum_{n=0}^{n_{\max}-1} [\rho_n(x) + \rho_n^*(x)] + k_{\mathrm{off}}^{\mathrm{T}} \sum_{n=1}^{n_{\max}} n \, [\rho_n(x) + \rho_n^*(x)] \right.$$

$$\left. + \beta \sum_{n=0}^{n_{\max}} n [\rho_n(x) + \rho_n^*(x)] \right\} = 0.$$

For both free and nucleoid-bound mRNAs, we enforce no-flux boundary conditions at the cell pole and at the cell

center

$$\left[\frac{d\rho_n(x)}{dx}v_{\mathrm{R}}(x)[v_{\mathrm{F}}(x)]^n - \rho_n(x)\frac{d[v_{\mathrm{R}}(x)[v_{\mathrm{F}}(x)]^n]}{dx}\right]\Big|_{x=0,\ell} = 0, \tag{S57}$$

$$\left[\frac{d\rho_n^*(x)}{dx}v_{\mathrm{R}}(x)[v_{\mathrm{F}}(x)]^n - \rho_n^*(x)\frac{d[v_{\mathrm{R}}(x)[v_{\mathrm{F}}(x)]^n]}{dx}\right]\Big|_{x=0,\ell} = 0. \tag{S58}$$

For ribosome concentrations, we impose the no-flux boundary condition at the cell pole, Eq. (S39), and a constraint on the total number of ribosomes

$$\begin{aligned}
N_{\mathrm{tot}} &= 2\int_0^\ell dx \left\{ c_{\mathrm{F}}(x) + \sum_{m=0}^{\infty}\sum_{n=0}^{n_{\max}}(m+n)[\rho_{m,n}(x) + \rho_{m,n}^*(x)] \right\} \\
&= 2\int_0^\ell dx \left\{ c_{\mathrm{F}}(x) + \exp\left(\frac{k_{\mathrm{on}}^{\mathrm{B}}c_{\mathrm{F}}(x)}{k_{\mathrm{off}}^{\mathrm{B}}}\right)\sum_{n=0}^{n_{\max}}\left(\frac{k_{\mathrm{on}}^{\mathrm{B}}c_{\mathrm{F}}(x)}{k_{\mathrm{off}}^{\mathrm{B}}} + n\right)[\rho_n(x) + \rho_n^*(x)]\right\}. 
\end{aligned} \tag{S59}$$

Finally, the ribosome efficiency reads

$$\begin{aligned}
\Sigma &= 2\,k_{\mathrm{off}}^{\mathrm{T}}\sum_{m=0}^{\infty}\sum_{n=1}^{n_{\max}}n\int_0^\ell dx[\rho_{m,n}(x) + \rho_{m,n}^*(x)] \\
&= 2\,k_{\mathrm{off}}^{\mathrm{T}}\sum_{n=1}^{n_{\max}}n\int_0^\ell dx\exp\left(\frac{k_{\mathrm{on}}^{\mathrm{B}}c_{\mathrm{F}}(x)}{k_{\mathrm{off}}^{\mathrm{B}}}\right)[\rho_n(x) + \rho_n^*(x)].
\end{aligned} \tag{S60}$$

We solved numerically Eqs. (S54), (S55), and (S56) for $\rho_n$, $\rho_n^*$ and $c_{\mathrm{F}}$, by fixing the maximal number of allowed T ribosomes per mRNA at $n_{\max} = 24$. The resulting mRNA profiles and ribosome concentrations are shown in Figs. S4 and S5, respectively. The results confirm the qualitative behavior of the model with only free mRNAs, see Figs. 3 and 4. The plots of $\rho_{\mathrm{tot}}(x)$ and $\rho_{\mathrm{tot}}^*(x) = \sum_{m=0}^{\infty}\sum_{n=0}^{n_{\max}}\rho_{m,n}^*(x)$ in Fig. S4 show that the large majority, i.e. $\sim 83\%$, of mRNAs are free, while the remaining $\sim 17\%$ of mRNAs are bound to the nucleoid, where the latter fraction constitutes also an estimate for the percentage of mRNAs that are degraded co-transcriptionally [13]. Free mRNAs are strongly segregated from the nucleoid, while nucleoid-bound mRNAs, are by construction localized in the nucleoid region. For free mRNAs, the larger $m$, $n$, the stronger the segregation away from the nucleoid, and the overall free polysome distribution is still peaked around $\sim$10 T ribosomes and $\sim$2 B ribosomes per mRNA as in Fig. 3. For nucleoid-bound mRNAs, the vast majority of mRNAs have low $m, n$, with the distribution peaked at $m \sim 1$ and $n \sim 3 - 4$, reflecting the somewhat lower concentration of F ribosomes in the nucleoid (Fig. S5) and, more importantly, the limited time for T ribosomes to bind before the mRNAs become free. The localization pattern of these typical bound mRNAs matches the profile of the nucleoid, i.e. the profile of mRNA production. By contrast, the spatial distributions of the few bound mRNA species with large $m, n$ are strongly peaked at the edge of the nucleoid, reflecting the higher density of free ribosomes outside versus inside the nucleoid.

Figure S5 shows the steady-state concentrations, $c_{\mathrm{T}}(x)$ and $c_{\mathrm{B}}(x)$, of T and B ribosomes loaded on free mRNAs, given by Eqs. (S42) and (S43), the concentration of F ribosomes $c_{\mathrm{F}}(x)$, as well as the fluxes $J_{\mathrm{T}}(x)$, $J_{\mathrm{B}}(x)$ given by Eqs. (S45), (S46), and $J_{\mathrm{F}}(x)$. The figure also shows the concentrations of T and B ribosomes loaded on nucleoid-bound mRNAs, $c_{\mathrm{T}}^*(x) = \sum_{m=0}^{\infty}\sum_{n=1}^{n_{\max}}n\,\rho_{m,n}^*(x)$ and $c_{\mathrm{B}}^*(x) = \sum_{m=1}^{\infty}\sum_{n=0}^{n_{\max}}m\,\rho_{m,n}^*(x)$, respectively. While T and B ribosomes loaded on free mRNAs are strongly excluded from the nucleoid, F ribosomes can penetrate the nucleoid region. In addition, the concentration profiles above provide an estimate of the fractions of ribosomes in the nucleoid region that translate mRNAs co-transcriptionally and post-transcriptionally. The number of ribosomes that translate co- and post-transcriptionally is $N_{\mathrm{co}} = 2\sum_{m=0}^{\infty}\sum_{n=1}^{n_{\max}}n\int_0^{\ell/2}dx\rho_{m,n}^*(x) \sim 3.6\times 10^3$ and $N_{\mathrm{post}} = 2\sum_{m=0}^{\infty}\sum_{n=1}^{n_{\max}}n\int_0^{\ell/2}dx\rho_{m,n}(x) \sim 3.9\times 10^3$, respectively, while the total number of ribosomes in the nucleoid region $N_{\mathrm{in}} \sim 1.1\times 10^4$ can be obtained by replacing $\ell$ with $\ell/2$ in the expression (S59) for the total number of ribosomes in the cell. Thus, $N_{\mathrm{co}}/N_{\mathrm{in}} \sim 34\%$ of ribosomes carry out co-transcriptional translation in the DNA-rich region, while a fraction of $N_{\mathrm{post}}/N_{\mathrm{in}} \sim 37\%$ translates post-transcriptionally. As shown in Fig. S4A, the larger the loading number of a free mRNA, the stronger its nucleoid segregation: it follows that such post-transcriptional translation in the nucleoid occurs mostly on free mRNAs loaded with a relatively small number of ribosomes, which could be either newly transcribed mRNAs that did not have time to diffuse out of the nucleoid, or old mRNAs with small loading number which penetrated into the nucleoid. Figure S5 also shows that there is a net poleward flux of T and B ribosomes loaded on free mRNAs and a net flux of F ribosomes toward the cell center, in agreement with the model with only free mRNAs. Finally, T and B ribosomes loaded on nucleoid-bound mRNAs are localized approximately uniformly throughout the nucleoid region.

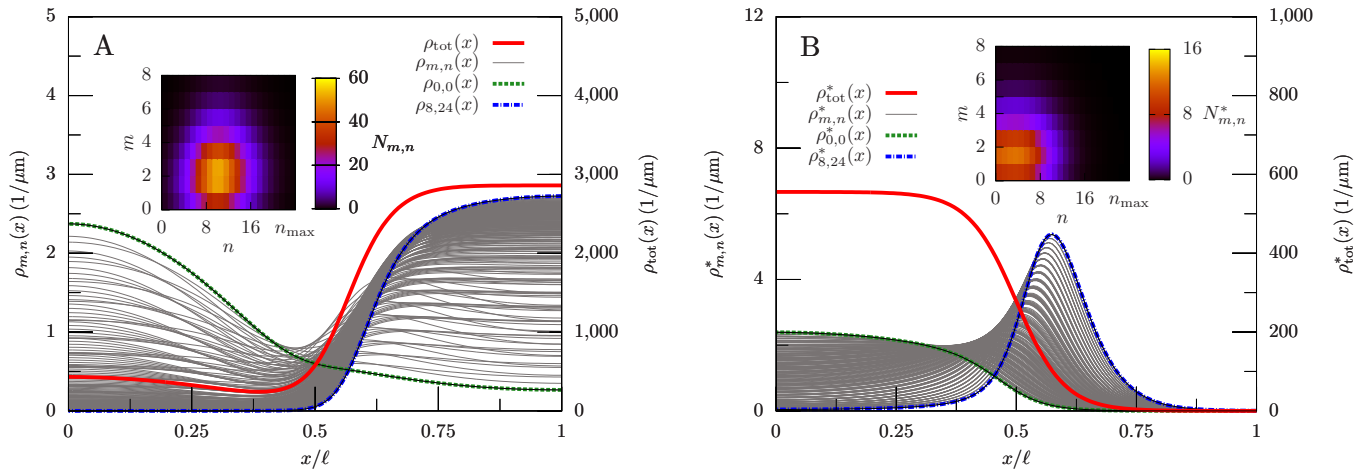

FIG. S4: Steady-state mRNA and polysome distributions for the extended model including both free and nucleoid-bound mRNAs, where we fixed the maximal number of allowed translating (T) ribosomes per mRNA at $n \leq n_{\max} = 24$, and we show mRNA species with $m \leq 8$ transiently bound (B) ribosomes. (A) mRNA and polysome distribution for free mRNAs. Total mRNA density $\rho_{\rm tot}(x)$ (red) and density $\rho_{m,n}(x)$ of mRNAs with $m$ B ribosomes and $n$ T ribosomes (gray). The density $\rho_{0,0}(x)$ of ribosome-free mRNAs (green) and the density $\rho_{8,24}(x)$ of mRNAs with the largest number of B, T ribosomes considered (blue) are also shown. The profiles $\rho_{m,n}(x)$, $\rho_{0,0}(x)$, and $\rho_{8,24}(x)$ are normalized to unit area. Inset: distribution of mRNA species, shown as a heat map of the number $N_{m,n} = \int_0^{\ell} dx \rho_{m,n}(x)$ of mRNAs with $m$ B ribosomes and $n$ T ribosomes in the right half of the cell. The maximal number of allowed T ribosomes per mRNA, $n_{\max} = 24$, is also marked. (B) Same as panel A for nucleoid-bound mRNAs, with $N_{m,n}^* = \int_0^{\ell} dx \rho_{m,n}^*(x)$ .

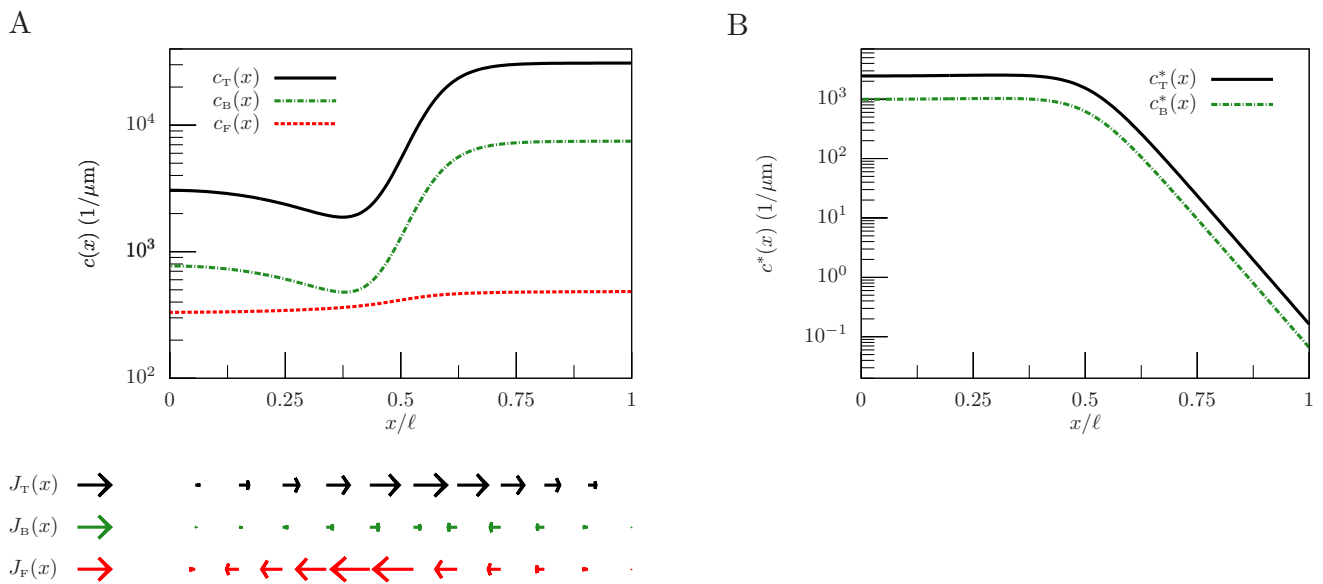

FIG. S5: Steady-state ribosome concentrations and ribosome fluxes for the extended model including both free and nucleoid-bound mRNAs. (A) Concentrations and fluxes for translating (T) and transiently bound (B) ribosomes loaded on free mRNAs, and for free (F) ribosomes. On the top, we show the concentrations $c_{\rm T}(x)$, $c_{\rm B}(x)$, and $c_{\rm F}(x)$ of T, B, and F ribosomes, for the right half of the cell. Bottom: Fluxes of T, B, and F ribosomes along the cell's long axis depicted in the top panel, where the arrow length is proportional to local ribosome flux, and the arrows in the legends correspond to a flux of 30/s. (B) Concentrations $c_{\rm T}^*(x)$, $c_{\rm B}^*(x)$ of T and B ribosomes loaded on nucleoid-bound mRNAs.

We next make use of the model with nucleoid-bound mRNAs to study the contribution of co-transcriptional translation to ribosome efficiency. To do so, we remove co-transcriptional translation by fiat in our model by eliminating binding of T ribosomes to nucleoid-bound mRNAs, but still allowing rapid-equilibrium binding and unbinding of B ribosomes to these mRNAs. First, without co-transcriptional translation the first four terms on the left-hand side

(LHS) of Eq. (S55) vanish: as a result, one has

$$\exp\left(\frac{k_{\mathrm{on}}^{\mathrm{B}} c_{\mathrm{F}}(x)}{k_{\mathrm{off}}^{\mathrm{B}}}\right) \rho_n^*(x) = \delta_{n,0}\alpha(x)/(\beta+\gamma). \tag{S61}$$

Substituting this in the reaction-diffusion equation (S54) for $\rho_n(x)$, we recover the intuitive result that the reaction-diffusion equations and boundary conditions for the two mRNA species with no co-transcriptional translation are equivalent to those for free mRNAs only, Eq. (S27), with a reduced rate of mRNA production, $\alpha(x) \to \alpha(x)\gamma/(\beta+\gamma)$, accounting for the decay of some mRNAs before transcription is complete. Second, in the absence of co-transcriptional translation the terms proportional to $\rho^*$ in the reaction-diffusion equation (S56) vanish, and the reaction-diffusion equation for $c_{\mathrm{F}}$ also reduces to Eq. (S37) for the model with free mRNAs only. Finally, let us consider the boundary conditions of Eq. (S56) in the absence of co-transcriptional translation: While the no-flux condition at the cell pole, Eq. (S39), still holds, the constraint (S59) on the total number of ribosomes should be modified to take account of the fact that there are no T ribosomes on nucleoid-bound mRNAs:

$$
\begin{aligned}
N_{\mathrm{tot}} &= 2\int_0^\ell dx \left\{ c_{\mathrm{F}}(x) + \sum_{m=0}^\infty \sum_{n=0}^{n_{\max}} [(m+n)\rho_{m,n}(x) + m\,\rho_{m,n}^*(x)] \right\} \\
&= 2\int_0^\ell dx \left\{ c_{\mathrm{F}}(x) + \exp\left(\frac{k_{\mathrm{on}}^{\mathrm{B}} c_{\mathrm{F}}(x)}{k_{\mathrm{off}}^{\mathrm{B}}}\right) \sum_{n=0}^{n_{\max}} \left(\frac{k_{\mathrm{on}}^{\mathrm{B}} c_{\mathrm{F}}(x)}{k_{\mathrm{off}}^{\mathrm{B}}} + n\right) \rho_n(x) + \frac{k_{\mathrm{on}}^{\mathrm{B}} c_{\mathrm{F}}(x)}{k_{\mathrm{off}}^{\mathrm{B}}} \exp\left(\frac{k_{\mathrm{on}}^{\mathrm{B}} c_{\mathrm{F}}(x)}{k_{\mathrm{off}}^{\mathrm{B}}}\right) \sum_{n=0}^{n_{\max}} \rho_n^*(x) \right\} \\
&= 2\int_0^\ell dx \left[ c_{\mathrm{F}}(x)\left(1 + \frac{k_{\mathrm{on}}^{\mathrm{B}}}{k_{\mathrm{off}}^{\mathrm{B}}} \frac{\alpha(x)}{\beta+\gamma}\right) + \exp\left(\frac{k_{\mathrm{on}}^{\mathrm{B}} c_{\mathrm{F}}(x)}{k_{\mathrm{off}}^{\mathrm{B}}}\right) \sum_{n=0}^{n_{\max}} \left(\frac{k_{\mathrm{on}}^{\mathrm{B}} c_{\mathrm{F}}(x)}{k_{\mathrm{off}}^{\mathrm{B}}} + n\right) \rho_n(x) \right],
\end{aligned}
\tag{S62}
$$

where in the second line we used Eqs. (S52) and (S53), and in the third line we substituted Eq. (S61). The constraint (S62) for the total ribosome number and its analog (S40) for the model with no nucleoid-bound mRNAs differ by the term proportional to $\alpha(x)$: intuitively, this term represents the total number of B ribosomes carried by mRNAs that are being transcribed in the nucleoid.

We will now estimate the ribosome efficiency, thus quantifying the change in efficiency due to the loss of co-transcriptional translation. Without co-transcriptional translation, ribosomes are not allowed to translate nucleoid-bound mRNAs, thus the expression for the total rate of protein production $\Sigma$ is given by Eq. (S47), and the ribosome efficiency is given by $\varepsilon = \Sigma/N_{\mathrm{tot}} = 1.88 \times 10^{-2}/\mathrm{s}$, which is $\sim 3\%$ smaller than the efficiency $\varepsilon = 1.94 \times 10^{-2}/\mathrm{s}$ in the case with co-transcriptional translation. This loss of efficiency is mostly due to the fact that nucleoid-bound mRNAs carry B ribosomes: indeed, compared to the case with co-transcriptional translation, this additional fraction of B ribosomes implies a reduced pool of F ribosomes, thus decreasing the F $\to$ T rate, and consequently the ribosome efficiency.

## S10    Model including 30S and 50S ribosomal subunits

In this section, we extend our model to include the fact that each bacterial ribosome is composed of a 30S and a 50S subunit. In section S10 A we introduce the reaction-diffusion equations for the two-subunit model, whose parameters will be discussed in section S10 B. In section S10 C we discuss the rapid-equilibrium approximation for B 30S and 50S subunits, and in section S10 D we present the numerical solution of the resulting reaction-diffusion equations.

### A.    Reaction-diffusion equations

The reaction-diffusion equations for the mRNA densities are

$$\frac{\partial \rho_{l,m,n}(x,t)}{\partial t} = D\left[\frac{\partial^2 \rho_{l,m,n}(x,t)}{\partial x^2}v_{l+m+n}(x) - \rho_{l,m,n}(x,t)\frac{d^2 v_{l+m+n}(x)}{dx^2}\right]$$

$$-k_{\mathrm{on}}^{30S}c_{30S\,\mathrm{F}}(x,t)\rho_{l,m,n}(x,t) - k_{\mathrm{off}}^{30S}\,l\,\rho_{l,m,n}(x,t)$$

$$-k_{\mathrm{on}}^{50S}c_{50S\,\mathrm{F}}(x,t)\rho_{l,m,n}(x,t) - k_{\mathrm{off}}^{50S}\,m\,\rho_{l,m,n}(x,t)$$

$$+k_{\mathrm{on}}^{30S}c_{30S\,\mathrm{F}}(x,t)\rho_{l-1,m,n}(x,t) + k_{\mathrm{off}}^{30S}\,(l+1)\,\rho_{l+1,m,n}(x,t)$$

$$+k_{\mathrm{on}}^{50S}c_{50S\,\mathrm{F}}(x,t)\rho_{l,m-1,n}(x,t) + k_{\mathrm{off}}^{50S}\,(m+1)\,\rho_{l,m+1,n}(x,t)$$

$$-k_{\mathrm{off}}^{70S}\,n\,\rho_{l,m,n}(x,t) + k_{\mathrm{off}}^{70S}\,(n+1)\,\rho_{l,m,n+1}(x,t)$$

$$-k_{\mathrm{on}}'^{30S}c_{30S\,\mathrm{F}}(x,t)\rho_{l,m,n}(x,t) + k_{\mathrm{off}}'^{30S}\rho_{l,m,n}'(x,t) + k_{\mathrm{on}}'^{50S}c_{50S\,\mathrm{F}}(x,t)\rho_{l,m,n-1}'(x,t)$$

$$+\delta_{l,0}\,\delta_{m,0}\,\delta_{n,0}\,\alpha(x) - \beta\,\rho_{l,m,n}(x,t), \tag{S63}$$

$$\frac{\partial \rho_{l,m,n}'(x,t)}{\partial t} = D\left[\frac{\partial^2 \rho_{l,m,n}'(x,t)}{\partial x^2}v_{l+m+n}'(x) - \rho_{l,m,n}'(x,t)\frac{d^2 v_{l+m+n}'(x)}{dx^2}\right]$$

$$-k_{\mathrm{on}}^{30S}c_{30S\,\mathrm{F}}(x,t)\rho_{l,m,n}'(x,t) - k_{\mathrm{off}}^{30S}\,l\,\rho_{l,m,n}'(x,t)$$

$$-k_{\mathrm{on}}^{50S}c_{50S\,\mathrm{F}}(x,t)\rho_{l,m,n}'(x,t) - k_{\mathrm{off}}^{50S}\,m\,\rho_{l,m,n}'(x,t)$$

$$+k_{\mathrm{on}}^{30S}c_{30S\,\mathrm{F}}(x,t)\rho_{l-1,m,n}'(x,t) + k_{\mathrm{off}}^{30S}\,(l+1)\,\rho_{l+1,m,n}'(x,t)$$

$$+k_{\mathrm{on}}^{50S}c_{50S\,\mathrm{F}}(x,t)\rho_{l,m-1,n}'(x,t) + k_{\mathrm{off}}^{50S}\,(m+1)\,\rho_{l,m+1,n}'(x,t)$$

$$-k_{\mathrm{off}}^{70S}\,n\,\rho_{l,m,n}'(x,t) + k_{\mathrm{off}}^{70S}\,(n+1)\,\rho_{l,m,n+1}'(x,t)$$

$$+k_{\mathrm{on}}'^{30S}c_{30S\,\mathrm{F}}(x,t)\rho_{l,m,n}(x,t) - k_{\mathrm{off}}'^{30S}\rho_{l,m,n}'(x,t) - k_{\mathrm{on}}'^{50S}c_{50S\,\mathrm{F}}(x,t)\rho_{l,m,n}'(x,t)$$

$$-\beta\,\rho_{l,m,n}'(x,t). \tag{S64}$$

In both Eqs. (S63) and (S64), the terms in the first line represent mRNA diffusion, where $v_{l+m+n}$, $v_{l+m+n}'$ are the available-volume profiles for mRNAs without or with a 30S subunit occupying the translation-initiation site, respectively. The second, third, fourth, and fifth line represent binding and unbinding of 30S and 50S subunits at rates $k_{\mathrm{on}}^{30S}$, $k_{\mathrm{off}}^{30S}$ and $k_{\mathrm{on}}^{50S}$, $k_{\mathrm{off}}^{50S}$ respectively, compare Eq. (2). The terms in the sixth line represent unbinding of 70S pairs [14] at rate $k_{\mathrm{off}}^{70S}$. The first two terms in the seventh line represent a 30S subunit binding and unbinding the mRNA translation-initiation site at rates $k_{\mathrm{on}}'^{30S}$, $k_{\mathrm{off}}'^{30S}$ respectively, while the third term represents a 50S subunit binding to the 30S subunit on the initiation site at rate $k_{\mathrm{on}}'^{50S}$ and forming a T 70S pair. Finally, the terms in the last line of Eq. (S63) represent mRNA transcription and degradation, and the term in the last line of Eq. (S64) represents mRNA degradation. The reaction-diffusion equation for the concentration of free 30S subunits reads

$$\frac{\partial c_{30S\,\mathrm{F}}(x,t)}{\partial t} = D_{\mathrm{F}}\left[\frac{\partial^2 c_{30S\,\mathrm{F}}(x,t)}{\partial x^2}v_{\mathrm{F}}(x) - c_{30S\,\mathrm{F}}(x,t)\frac{d^2 v_{\mathrm{F}}(x)}{dx^2}\right]$$

$$-k_{\mathrm{on}}^{30S}c_{30S\,\mathrm{F}}(x,t)\sum_{l=0}^{l_{\max}-1}\sum_{m=0}^{m_{\max}}\sum_{n=0}^{n_{\max}}[\rho_{l,m,n}(x,t)+\rho_{l,m,n}'(x,t)] + k_{\mathrm{off}}^{30S}\sum_{l=1}^{l_{\max}}\sum_{m=0}^{m_{\max}}\sum_{n=0}^{n_{\max}}l\,[\rho_{l,m,n}(x,t)+\rho_{l,m,n}'(x,t)]$$

$$+k_{\mathrm{off}}^{70S}\sum_{l=0}^{l_{\max}}\sum_{m=0}^{m_{\max}}\sum_{n=1}^{n_{\max}}n\,[\rho_{l,m,n}(x,t)+\rho_{l,m,n}'(x,t)]$$

$$-k_{\mathrm{on}}'^{30S}c_{30S\,\mathrm{F}}(x,t)\sum_{l=0}^{l_{\max}}\sum_{m=0}^{m_{\max}}\sum_{n=0}^{n_{\max}}\rho_{l,m,n}(x,t) + k_{\mathrm{off}}'^{30S}\sum_{l=0}^{l_{\max}}\sum_{m=0}^{m_{\max}}\sum_{n=0}^{n_{\max}}\rho_{l,m,n}'(x,t)$$

$$+\beta\sum_{l=0}^{l_{\max}}\sum_{m=0}^{m_{\max}}\sum_{n=0}^{n_{\max}}[(l+n)\,\rho_{l,m,n}(x,t)+(l+n+1)\rho_{l,m,n}'(x,t)]. \tag{S65}$$

In Eq. (S65), only mRNA species with allowed values of $0 \leq l \leq l_{\max}$, $0 \leq m \leq m_{\max}$, and $0 \leq n \leq n_{\max}$ are considered. The terms in the second line represent 30S subunits transiently binding and unbinding mRNAs, compare Eq. (1). The third line represents unbinding of a 30S subunit coming from a 70S pair, while the two terms in the fourth line represent a 30S subunit binding and unbinding the initiation site, respectively. Finally, the last line represent 30S subunits being freed from mRNA molecules as these are degraded. The reaction-diffusion equation for

the concentration of free 50S subunits reads

$$
\frac{\partial c_{50S\,F}(x,t)}{\partial t} =
$$

$$
D_F \left[ \frac{\partial^2 c_{50S\,F}(x,t)}{\partial x^2} v_F(x) - c_{50S\,F}(x,t) \frac{d^2 v_F(x)}{dx^2} \right]
$$

$$
-k_{on}^{50S} c_{50S\,F}(x,t) \sum_{l=0}^{l_{max}} \sum_{m=0}^{m_{max}-1} \sum_{n=0}^{n_{max}} [\rho_{l,m,n}(x,t) + \rho'_{l,m,n}(x,t)] + k_{off}^{50S} \sum_{l=0}^{l_{max}} \sum_{m=1}^{m_{max}} \sum_{n=0}^{n_{max}} m \, [\rho_{l,m,n}(x,t) + \rho'_{l,m,n}(x,t)]
$$

$$
+k_{off}^{70S} \sum_{l=0}^{l_{max}} \sum_{m=0}^{m_{max}} \sum_{n=1}^{n_{max}} n[\rho_{l,m,n}(x,t) + \rho'_{l,m,n}(x,t)] - k'^{50S}_{on} c_{50S\,F}(x,t) \sum_{l=0}^{l_{max}} \sum_{m=0}^{m_{max}} \sum_{n=0}^{n_{max}-1} \rho'_{l,m,n}(x,t)
$$

$$
+\beta \sum_{l=0}^{l_{max}} \sum_{m=0}^{m_{max}} \sum_{n=0}^{n_{max}} (m+n)[\rho_{l,m,n}(x,t) + \rho'_{l,m,n}(x,t)]. \tag{S66}
$$

The terms in the second line represent 50S subunits transiently binding and unbinding mRNAs. In the third line, the first term represents unbinding of a 50S subunit coming from a 70S pair, and the second term represents a 50S subunit binding the 30S subunit on the initiation site. Finally, the term in the fourth line represents 50S subunits being freed from degraded mRNAs.

We consider Eqs. (S63)-(S66) at steady state, and in Eqs. (S63), (S64) we set a no-flux boundary condition at the cell pole and at the cell center:

$$
\left[ \frac{d\rho_{l,m,n}(x)}{dx} v_{l+m+n}(x) - \rho_{l,m,n}(x) \frac{dv_{l+m+n}(x)}{dx} \right] \bigg|_{x=0,\ell} = 0, \tag{S67}
$$

$$
\left[ \frac{d\rho'_{l,m,n}(x)}{dx} v'_{l+m+n}(x) - \rho'_{l,m,n}(x) \frac{dv'_{l+m+n}(x)}{dx} \right] \bigg|_{x=0,\ell} = 0. \tag{S68}
$$

In Eqs. (S65), (S66), we set no-flux boundary conditions at the cell pole

$$
\left[ \frac{dc_{30S\,F}(x)}{dx} v_F(x) - c_{30S\,F}(x) \frac{dv_F(x)}{dx} \right] \bigg|_{x=\ell} = 0, \tag{S69}
$$

$$
\left[ \frac{dc_{50S\,F}(x)}{dx} v_F(x) - c_{50S\,F}(x) \frac{dv_F(x)}{dx} \right] \bigg|_{x=\ell} = 0, \tag{S70}
$$

and we enforce a constraint on the total number of 30S and 50S ribosomal subunits

$$
2 \int_0^\ell dx \left\{ c_{30S\,F}(x) + \sum_{l=0}^{l_{max}} \sum_{m=0}^{m_{max}} \sum_{n=0}^{n_{max}} \left[ (l+n)\rho_{l,m,n}(x) + (l+n+1)\rho'_{l,m,n}(x) \right] \right\} = N_{30S\,tot}, \tag{S71}
$$

$$
2 \int_0^\ell dx \left\{ c_{50S\,F}(x) + \sum_{l=0}^{l_{max}} \sum_{m=0}^{m_{max}} \sum_{n=0}^{n_{max}} (m+n) \left[ \rho_{l,m,n}(x) + \rho'_{l,m,n}(x) \right] \right\} = N_{50S\,tot}. \tag{S72}
$$

## B. Model parameters

The model parameters in Eqs. (S63)-(S72) are fixed as follows: The diffusion coefficients $D$, $D_F$, the degradation rate $\beta$, and the ribosomal available volume $v_F(x)$ are taken to be the same as in the model with one ribosomal unit, see section II. Assuming for simplicity a radius of $10\,\mathrm{nm}$ for both the 30S and 50S subunits, the available volume $v_{l+m+n}(x)$ is given by Eq. (S12). It follows that the available volume for mRNAs with an additional 30S subunit at the initiation site is $v'_{l+m+n}(x) = v_{l+m+n+1}(x)$. We fixed $N_{30S\,tot}$ and $N_{50S\,tot}$ from the total number of ribosomes and from the 30S:50S stoichiometric ratio, which can be estimated as follows: First, every 30S or 50S subunit includes one 16S or one 23S ribosomal RNA (rRNA), respectively. As a result, the 30S:50S ratio can be estimated from the stoichiometry of 16S- and 23S-rRNA transcript abundances, which is found to be close to 1:1 [15]. Second, the 1:1 ratio above is generally supported by studies of the abundance of ribosomal proteins in the two subunits. Third, the experimental observation that $\sim 80\%$ of 30S ribosomal subunits are engaged in translation [5], with similar results for 50S subunits [16], implies that the great majority of ribosomes must come as pairs of 30S and 50S subunits, thus

confirming a 30S:50S ratio close to 1:1. We thus assume a 1:1 ratio of 30S:50S subunits. Consequently, for a total ribosome number of $N_{\text{tot}} = 6 \times 10^4$ [5], we obtain $N_{\text{30S tot}} = N_{\text{50S tot}} = N_{\text{tot}}$. The transient-binding rates $k_{\text{on}}^{\text{30S}}$, $k_{\text{off}}^{\text{30S}}$, $k_{\text{on}}^{\text{50S}}$, $k_{\text{off}}^{\text{50S}}$ will be assumed to be significantly faster than any other relevant time scale, and they will be treated in the rapid-equilibrium limit as in section S6. In section S10 C, we will show that in this limit the reaction-diffusion equations depend only on the ratio between the binding and the unbinding rates, which can be computed along the lines of section II, providing the estimate $k_{\text{on}}^{\text{30S}}/k_{\text{off}}^{\text{30S}} \simeq k_{\text{on}}^{\text{50S}}/k_{\text{off}}^{\text{50S}} \simeq 5.4 \times 10^{-3}\,\mu\text{m}$. The binding constant $k_{\text{on}}^{\prime\,\text{30S}}$ of 30S subunits on the initiation site is estimated as follows: An in vitro kinetic analysis of the assembly of the 30S initiation complex quantified the binding rate of 30S subunits to a single mRNA as $k_{\text{3D}}\,c_{\text{3D}}$, i.e. the rate is proportional to the three-dimensional 30S concentration $c_{\text{3D}}$, and to an associated rate constant $k_{\text{3D}}$. The value of $k_{\text{3D}}$ was estimated from the slope of the 30S-subunit binding rate as a function of the three-dimensional 30S concentration, and was found to be $k_{\text{3D}} \sim 0.1\,\mu\text{m}^3/\text{s}$ [17]. The 30S binding rate in our one-dimensional model is then obtained by dividing $k_{\text{3D}}$ by the area of the two-dimensional cell slice perpendicular to the cell axis: $k_{\text{on}}^{\prime\,\text{30S}} = k_{\text{3D}}/(\pi R^2) = 0.13\,\mu\text{m/s}$, where $R = 0.5\,\mu\text{m}$ is the radius of a circular cell slice. The unbinding rate of 30S subunits from the initiation site $k_{\text{off}}^{\prime\,\text{30S}} = 2/\text{s}$ is taken from the same in vitro kinetic analysis [17]. The rate of unbinding of 70S pairs can be estimated along the same lines as $k_{\text{off}}^{\text{T}}$ in the model for one ribosomal unit in section II, yielding $k_{\text{off}}^{\text{70S}} = k_{\text{off}}^{\text{T}}$. Finally, the binding rate of 50S subunits to 30S subunits on the initiation site can be obtained from the other parameters by imposing the conservation of the total number of 30S and 50S subunits:

$$k_{\text{off}}^{\text{70S}} N_{\text{70S}} - k_{\text{on}}^{\prime\,\text{30S}} N_{\text{30S F}} \overline{\rho} + k_{\text{off}}^{\prime\,\text{30S}} \ell \,\overline{\rho'} = 0, \tag{S73}$$

$$k_{\text{off}}^{\text{70S}} N_{\text{70S}} - k_{\text{on}}^{\prime\,\text{50S}} N_{\text{50S F}} \overline{\rho'} = 0, \tag{S74}$$

where $N_{\text{30S F}}$, $N_{\text{50S F}}$, and $N_{\text{70S}}$ are the total number of F 30S and 50S subunits and 70S pairs, and $\overline{\rho}$, $\overline{\rho'}$ are the average total axial densities of mRNAs without or with a 30S subunit on the initiation site, respectively. Equation (S73) imposes the balance between the overall binding and unbinding rates of 30S subunits in the cell: The first term in the LHS represents the total rate at which 30S subunits in a 70S pair unbind from mRNAs, while the second and the third term are the total rates at which 30S subunits bind and unbind from the translation-initiation site, respectively. Note that the rates of binding and unbinding of transiently bound 30S subunits do not appear in Eq. (S73) because the sum of these rates is zero. Denoting by $N_{\text{30S}'}$ the number of 30S subunits bound to the translation-initiation site, in Eq. (S73) we omitted the rate $\beta(N_{\text{70S}} + N_{\text{30S}'})$ at which 70S subunits and 30S subunits bound to the translation-initiation site are freed by mRNAs that are degraded, because this rate is significantly smaller than $k_{\text{off}}^{\text{70S}} N_{\text{70S}}$. Equation (S74) represents the balance between the total binding and unbinding rates of 50S subunits. Similarly to Eq. (S73), we omitted the term $\beta N_{\text{70S}} \ll k_{\text{off}}^{\text{70S}} N_{\text{70S}}$ associated with mRNA decay. The rate $k_{\text{on}}^{\prime\,\text{50S}}$ at which 50S subunits bind to 30S subunits on the translation-initiation site is then determined as follows: We set $N_{\text{30S F}} = N_{\text{50S F}} = 2\,\%\,N_{\text{tot}}$ and $N_{\text{70S}} = 80\,\%\,N_{\text{tot}}$ [5, 16], we use the fact that the mRNA densities $\overline{\rho}$ and $\overline{\rho'}$ are related by the constraint on the total mRNA number $\overline{\rho} + \overline{\rho'} = N_{\text{mRNA}}/(2\,\ell)$, and we solve Eqs. (S73) and (S74) for $\overline{\rho}$, $\overline{\rho'}$, and $k_{\text{on}}^{\prime\,\text{50S}}$.

### C. Rapid equilibrium for transiently bound 30S and 50S subunits

In this section we will derive the reaction-diffusion equations assuming that B 30S and 50S subunits are in rapid equilibrium with their F counterparts. We present only the main steps of the derivation, which follows along the lines of the rapid-equilibrium approximation for the model with one ribosomal subunit discussed in section S6.

In the rapid-equilibrium limit, we set

$$k_{\text{on}}^{\text{30S}} = \lambda \,\boldsymbol{k}_{\text{on}}^{\text{30S}}, \; k_{\text{off}}^{\text{30S}} = \lambda \,\boldsymbol{k}_{\text{off}}^{\text{30S}}, \tag{S75}$$

$$k_{\text{on}}^{\text{50S}} = \lambda \,\boldsymbol{k}_{\text{on}}^{\text{50S}}, \; k_{\text{off}}^{\text{50S}} = \lambda \,\boldsymbol{k}_{\text{off}}^{\text{50S}}, \tag{S76}$$

where $\lambda \gg 1$ and the boldface binding-unbinding rates are of order unity.

To leading order in $\lambda$, we obtain

$$[k_{\text{off}}^{\text{30S}}(l+1)\rho_{l+1,m,n}(x) - k_{\text{on}}^{\text{30S}} c_{\text{30S F}}(x)\rho_{l,m,n}(x)] - [k_{\text{off}}^{\text{30S}} l\,\rho_{l,m,n}(x) - k_{\text{on}}^{\text{30S}} c_{\text{30S F}}(x)\rho_{l-1,m,n}(x)]$$
$$+ [k_{\text{off}}^{\text{50S}}(m+1)\rho_{l,m+1,n}(x) - k_{\text{on}}^{\text{50S}} c_{\text{50S F}}(x)\rho_{l,m,n}(x)] - [k_{\text{off}}^{\text{50S}} m\,\rho_{l,m,n}(x) - k_{\text{on}}^{\text{50S}} c_{\text{50S F}}(x)\rho_{l,m-1,n}(x)] = 0. \tag{S77}$$

The four terms in the LHS of Eq. (S77) represent the mRNA current between species $(l,m)$ and the four species $(l+1,m)$, $(l-1,m)$, $(l,m+1)$ and $(l,m-1)$, thus implying that the sum of the four currents is zero. Note that this condition does not necessarily imply that each current vanishes. Indeed, there could be solutions $\rho_{l,m,n}(x)$ involving net nonequilibrium currents between pairs of mRNA species, that would still satisfy the current-balance condition (S77). However, if we assume that the 30S and 50S transient binding and unbinding is a purely spontaneous, equilibrium

process that is not coupled to any external drive or energy source, all the individual currents between mRNA species must vanish

$$k_{\text{off}}^{30S}(l+1)\rho_{l+1,m,n}(x) - k_{\text{on}}^{30S}c_{30S\,F}(x)\rho_{l,m,n}(x) = 0, \qquad l \le l_{\max} - 1, \tag{S78}$$

$$k_{\text{off}}^{50S}(m+1)\rho_{l,m+1,n}(x) - k_{\text{on}}^{50S}c_{50S\,F}(x)\rho_{l,m,n}(x) = 0, \qquad m \le m_{\max} - 1. \tag{S79}$$

Equations (S78) and (S79) can now be solved along the lines of Eq. (S19) for one ribosomal unit, and their solution is analogous to Eq. (S20), providing the following leading-order expression for $\rho_{l,m,n}(x)$:

$$\rho_{l,m,n}(x) = \frac{1}{l!\,m!}\left(\frac{k_{\text{on}}^{30S}c_{30S\,F}(x)}{k_{\text{off}}^{30S}}\right)^l \left(\frac{k_{\text{on}}^{50S}c_{50S\,F}(x)}{k_{\text{off}}^{50S}}\right)^m \rho_{0,0,n}(x) \equiv \boldsymbol{\rho}_{l,m,n}(x). \tag{S80}$$

Proceeding along the same lines, we obtain the leading-order expression for $\rho'_{l,m,n}(x)$:

$$\rho'_{l,m,n}(x) = \frac{1}{l!\,m!}\left(\frac{k_{\text{on}}^{30S}c_{30S\,F}(x)}{k_{\text{off}}^{30S}}\right)^l \left(\frac{k_{\text{on}}^{50S}c_{50S\,F}(x)}{k_{\text{off}}^{50S}}\right)^m \rho'_{0,0,n}(x) \equiv \boldsymbol{\rho}'_{l,m,n}(x), \tag{S81}$$

Next, we derive the reaction-diffusion equations for

$$\rho_n(x) \equiv \rho_{0,0,n}(x) \tag{S82}$$

and for

$$\rho'_n(x) \equiv \rho'_{0,0,n}(x). \tag{S83}$$

Proceeding along the lines of Eqs. (S21)-(S27), we introduce the local fluxes of mRNAs

$$J_{l,m,n}(x) = -D\left(\frac{d\boldsymbol{\rho}_{l,m,n}(x)}{dx}v_{l+m+n}(x) - \boldsymbol{\rho}_{l,m,n}(x)\frac{dv_{l+m+n}(x)}{dx}\right), \tag{S84}$$

$$J'_{l,m,n}(x) = -D\left(\frac{d\boldsymbol{\rho}'_{l,m,n}(x)}{dx}v'_{l+m+n}(x) - \boldsymbol{\rho}'_{l,m,n}(x)\frac{dv'_{l+m+n}(x)}{dx}\right), \tag{S85}$$

and we define

$$\begin{aligned}
\omega_{m,n}(x) = & -k_{\text{off}}^{70S}\,n\,\boldsymbol{\rho}_{l,m,n}(x) + k_{\text{off}}^{70S}(n+1)\boldsymbol{\rho}_{l,m,n+1}(x) - k_{\text{on}}'^{30S}c_{30S\,F}(x)\boldsymbol{\rho}_{l,m,m}(x) + k_{\text{off}}'^{30S}\boldsymbol{\rho}'_{l,m,n}(x) \\
& +k_{\text{on}}'^{50S}c_{50S\,F}(x)\boldsymbol{\rho}'_{l,m,n-1}(x) - \beta\boldsymbol{\rho}_{l,m,n}(x),
\end{aligned} \tag{S86}$$

$$\begin{aligned}
\omega'_{m,n}(x) = & -k_{\text{off}}^{70S}\,n\,\boldsymbol{\rho}'_{l,m,n}(x) + k_{\text{off}}^{50S}(n+1)\boldsymbol{\rho}'_{l,m,n+1}(x) + k_{\text{on}}'^{30S}c_{30S\,F}(x)\boldsymbol{\rho}_{l,m,m}(x) - k_{\text{off}}'^{30S}\boldsymbol{\rho}'_{l,m,n}(x) \\
& -k_{\text{on}}'^{50S}c_{50S\,F}(x)\boldsymbol{\rho}'_{l,m,n}(x) - \beta\boldsymbol{\rho}'_{l,m,n}(x).
\end{aligned} \tag{S87}$$

We sum Eqs. (S63) and (S64) with respect to $l = 0, \ldots, l_{\max}$ and $m = 0, \ldots, m_{\max}$, and we obtain the following reaction-diffusion equations to leading order in $\lambda$

$$\sum_{l=0}^{l_{\max}}\sum_{m=0}^{m_{\max}}\left[-\frac{dJ_{l,m,n}(x)}{dx} + \omega_{l,m,n}(x)\right] + \delta_{n,0}\alpha(x) = 0, \tag{S88}$$

$$\sum_{l=0}^{l_{\max}}\sum_{m=0}^{m_{\max}}\left[-\frac{dJ'_{l,m,n}(x)}{dx} + \omega'_{l,m,n}(x)\right] = 0. \tag{S89}$$

Proceeding along the lines of Eqs. (S26) and (S27), we rewrite Eq. (S88) as a reaction-diffusion equations for $\rho_n(x)$:

$$\begin{aligned}
& D\frac{d}{dx}\left\{\exp\left[\left(\frac{k_{\text{on}}^{30S}c_{30S\,F}(x)}{k_{\text{off}}^{30S}} + \frac{k_{\text{on}}^{50S}c_{50S\,F}(x)}{k_{\text{off}}^{50S}}\right)v_{\text{F}}(x)\right][v_{\text{F}}(x)]^n\left[\rho_n(x)v_{\text{R}}(x)\left[v_{\text{F}}(x)\frac{d}{dx}\left(\frac{k_{\text{on}}^{30S}c_{30S\,F}(x)}{k_{\text{off}}^{30S}}\right.\right.\right.\right. \tag{S90} \\
& \left.+\frac{k_{\text{on}}^{50S}c_{50S\,F}(x)}{k_{\text{off}}^{50S}}\right) - \frac{dv_{\text{F}}(x)}{dx}\left(\frac{k_{\text{on}}^{30S}c_{30S\,F}(x)}{k_{\text{off}}^{30S}} + \frac{k_{\text{on}}^{50S}c_{50S\,F}(x)}{k_{\text{off}}^{50S}}\right)\bigg] + \frac{d\rho_n(x)}{dx}v_{\text{R}}(x) - \rho_n(x)\frac{dv_{\text{R}}(x)}{dx} \\
& \left.\left.-n\,\rho_n(x)v_{\text{R}}(x)\frac{1}{v_{\text{F}}(x)}\frac{dv_{\text{F}}(x)}{dx}\right]\right\} + \exp\left(\frac{k_{\text{on}}^{30S}c_{30S\,F}(x)}{k_{\text{off}}^{30S}} + \frac{k_{\text{on}}^{50S}c_{50S\,F}(x)}{k_{\text{off}}^{50S}}\right)\left[-n\,k_{\text{off}}^{70S}\rho_n(x)\right. \\
& \left.+k_{\text{off}}^{70S}(n+1)\rho_{n+1}(x) - k_{\text{on}}'^{30S}c_{30S\,F}(x)\rho_n(x) + k_{\text{off}}'^{30S}\rho'_n(x) + k_{\text{on}}'^{50S}c_{50S\,F}(x)\rho'_{n-1}(x) - \beta\rho_n(x)\right] + \delta_{n,0}\,\alpha(x) = 0.
\end{aligned}$$

Similarly, Eq. (S89) implies the following reaction-diffusion equations for $\rho'_n(x)$:

$$D\frac{d}{dx}\left\{\exp\left[\left(\frac{k^{30S}_{on}c_{30S\,F}(x)}{k^{30S}_{off}}+\frac{k^{50S}_{on}c_{50S\,F}(x)}{k^{50S}_{off}}\right)v_F(x)\right][v_F(x)]^{n+1}\left[\rho'_n(x)v_R(x)\left[v_F(x)\frac{d}{dx}\left(\frac{k^{30S}_{on}c_{30S\,F}(x)}{k^{30S}_{off}}\right.\right.\right.\right.\tag{S91}$$

$$\left.\left.+\frac{k^{50S}_{on}c_{50S\,F}(x)}{k^{50S}_{off}}\right)-\frac{dv_F(x)}{dx}\left(\frac{k^{30S}_{on}c_{30S\,F}(x)}{k^{30S}_{off}}+\frac{k^{50S}_{on}c_{50S\,F}(x)}{k^{50S}_{off}}\right)\right]+\frac{d\rho'_n(x)}{dx}v_R(x)-\rho'_n(x)\frac{dv_R(x)}{dx}$$

$$\left.-(n+1)\rho'_n(x)v_R(x)\frac{1}{v_F(x)}\frac{dv_F(x)}{dx}\right]\right\}+\exp\left(\frac{k^{30S}_{on}c_{30S\,F}(x)}{k^{30S}_{off}}+\frac{k^{50S}_{on}c_{50S\,F}(x)}{k^{50S}_{off}}\right)\left[-k^{70S}_{off}n\,\rho'_n(x)\right.$$

$$\left.+k^{70S}_{off}(n+1)\rho'_{n+1}(x)+k^{30S}_{on}c_{30S\,F}(x)\rho_n(x)-k'^{30S}_{off}\rho'_n(x)-k'^{50S}_{on}c_{50S\,F}(x)\rho'_n(x)-\beta\rho'_n(x)\right]\ =\ 0.$$

We now derive the rapid-equilibrium reaction-diffusion equation for the ribosome concentration $c_{30S\,F}(x)$: We introduce the subleading corrections to $\rho_{l,m,n}$ and $\rho'_{l,m,n}$, setting

$$\rho_{l,m,n}(x)\ =\ \boldsymbol{\rho}_{l,m,n}(x)+\frac{1}{\lambda}\Delta\rho_{l,m,n}(x),\tag{S92}$$

$$\rho'_{l,m,n}(x)\ =\ \boldsymbol{\rho}'_{l,m,n}(x)+\frac{1}{\lambda}\Delta\rho'_{l,m,n}(x).\tag{S93}$$

We rewrite the steady-state reaction-diffusion equations (S63) for $\rho_{l,m,n}(x)$ as follows:

$$[\boldsymbol{k}^{30S}_{off}(l+1)\,\Delta\rho_{l+1,m,n}(x)-\boldsymbol{k}^{30S}_{on}c_{30S\,F}(x)\Delta\rho_{l,m,n}(x)]-[\boldsymbol{k}^{30S}_{off}\,l\,\Delta\rho_{l,m,n}(x)-\boldsymbol{k}^{30S}_{on}c_{30S\,F}(x)\Delta\rho_{l-1,m,n}(x)]$$

$$+[k^{50S}_{off}(m+1)\,\rho_{l,m+1,n}(x)-k^{50S}_{on}c_{50S\,F}(x)\rho_{l,m,n}(x)]-[k^{50S}_{off}\,m\,\rho_{l,m,n}(x)-k^{50S}_{on}c_{50S\,F}(x)\rho_{l,m-1,n}(x)]$$

$$+D\frac{d}{dx}\left(\frac{d\rho_{l,m,n}(x)}{dx}v_{l+m+n}-\rho_{l,m,n}(x)\frac{dv_{l+m+n}(x)}{dx}\right)-k^{70S}_{off}n\,\rho_{l,m,n}(x)+k^{70S}_{off}(n+1)\,\rho_{l,m,n+1}(x)$$

$$-k'^{30S}_{on}c_{30S\,F}(x)\rho_{l,m,n}(x)+k'^{30S}_{off}\rho'_{l,m,n}(x)+k'^{50S}_{on}c_{50S\,F}(x)\rho'_{l,m,n-1}(x)+\delta_{l,0}\,\delta_{m,0}\,\delta_{n,0}\,\alpha(x)-\beta\,\rho_{l,m,n}(x)\ =\ 0,\tag{S94}$$

where in the first two lines we collected all the terms involving fast processes related to B 30S and 50S subunits, and in the last two lines we collected all the remaining terms describing slow processes, such as diffusion and unbinding of 70S pairs. Also, in the first line we used Eq. (S92), and we observed that the terms to leading order in $\lambda$ cancel out because of Eq. (S80). To derive the reaction-diffusion for $c_{30S\,F}(x)$, we sum Eq. (S94) with respect to $m=0,\ldots,m_{max}$, and we obtain to leading order in $\lambda$:

$$[\boldsymbol{k}^{30S}_{off}\,l\,\Delta\rho^{30S}_{l,n}(x)-\boldsymbol{k}^{30S}_{on}c_{30S\,F}(x)\Delta\rho^{30S}_{l-1,n}(x)]-[\boldsymbol{k}^{30S}_{off}(l+1)\Delta\rho^{30S}_{l+1,n}(x)-\boldsymbol{k}^{30S}_{on}c_{30S\,F}(x)\Delta\rho^{30S}_{l,n}(x)]\ =$$

$$\sum_{m=0}^{m_{max}}\left[-\frac{dJ_{l,m,n}(x)}{dx}+\omega_{l,m,n}(x)\right]+\delta_{l,0}\delta_{n,0}\alpha(x),\tag{S95}$$

where

$$\Delta\rho^{30S}_{l,n}(x)=\sum_{m=0}^{m_{max}}\Delta\rho_{l,m,n}(x),\tag{S96}$$

and the superscript '30S' refers to the fact that $\Delta\rho^{30S}_{l,n}$ is the total subleading density of mRNAs with $l$ 30S subunits, i.e. it results from a sum over all contributions of 50S subunits in the RHS of Eq. (S96). Intuitively, by summing Eq. (S94) with respect to $m$, we averaged out the contribution of 50S subunits, thus obtaining the overall current between mRNA species with $l$ 30S subunits and species with $l-1$ and $l+1$ 30S subunits, compare the two terms in the LHS of Eq. (S95), respectively. Given that the contribution of 50S subunits has been averaged out, Eq. (S95) now has the same form as Eq. (S30) for a single ribosomal unit, and it can thus be solved iteratively along the lines of Eqs. (S31)-(S32): the result is

$$\boldsymbol{k}^{30S}_{off}\,l\,\Delta\rho^{30S}_{l,n}(x)-\boldsymbol{k}^{30S}_{on}c_{30S\,F}(x)\Delta\rho^{30S}_{l-1,n}(x)=\sum_{p=l}^{l_{max}}\sum_{m=0}^{m_{max}}\left[-\frac{dJ_{p,m,n}(x)}{dx}+\omega_{p,m,n}(x)\right],\qquad 1\le l\le l_{max}.\tag{S97}$$

Importantly, to derive the reaction-diffusion equation for $c_{30S\,F}(x)$, it is not necessary to determine all the currents between mRNA species $(l,m)$ and species $(l+1,m)$, $(l-1,m)$, $(l,m+1)$ and $(l,m-1)$ that appear in Eq. (S77): the

average current in Eq. (S97) is all we need. Indeed, let us consider the second line of Eq. (S65): the terms involving $\rho_{l,m,n}$ can be rewritten in terms of the current in Eq. (S97) as follows:

$$
\begin{aligned}
-k_{\mathrm{on}}^{30\mathrm{S}} c_{30\mathrm{S\,F}}(x) \sum_{l=0}^{l_{\max}-1} \sum_{m=0}^{m_{\max}} \sum_{n=0}^{n_{\max}} \rho_{l,m,n}(x) + k_{\mathrm{off}}^{30\mathrm{S}} \sum_{l=1}^{l_{\max}} \sum_{m=0}^{m_{\max}} \sum_{n=0}^{n_{\max}} l\, \rho_{l,m,n}(x) &= \\
-\boldsymbol{k}_{\mathrm{on}}^{30\mathrm{S}} c_{30\mathrm{S\,F}}(x) \sum_{l=0}^{l_{\max}-1} \sum_{m=0}^{m_{\max}} \sum_{n=0}^{n_{\max}} \Delta\rho_{l,m,n}(x) + \boldsymbol{k}_{\mathrm{off}}^{30\mathrm{S}} \sum_{l=1}^{l_{\max}} \sum_{m=0}^{m_{\max}} \sum_{n=0}^{n_{\max}} l\, \Delta\rho_{l,m,n}(x) &= \\
\sum_{l=1}^{l_{\max}} \sum_{n=0}^{n_{\max}} [\boldsymbol{k}_{\mathrm{off}}^{30\mathrm{S}} l\, \Delta\rho_{l,n}^{30\mathrm{S}}(x) - \boldsymbol{k}_{\mathrm{on}}^{30\mathrm{S}} c_{30\mathrm{S\,F}}(x) \Delta\rho_{l-1,n}^{30\mathrm{S}}(x)] &= \\
\sum_{l=1}^{l_{\max}} \sum_{p=l}^{l_{\max}} \sum_{m=0}^{m_{\max}} \sum_{n=0}^{n_{\max}} \left[ -\frac{dJ_{p,m,n}(x)}{dx} + \omega_{p,m,n}(x) \right] &= \\
\sum_{l=1}^{l_{\max}} l \sum_{m=0}^{m_{\max}} \sum_{n=0}^{n_{\max}} \left[ -\frac{dJ_{l,m,n}(x)}{dx} + \omega_{l,m,n}(x) \right],
\end{aligned}
\tag{S98}
$$

where in the second line we used Eqs. (S80) and (S92) and we observed that the terms of order $\lambda$ cancel out, in the third line we used Eq. (S96), in the fourth line we used Eq. (S97), and in the fifth line we used the identity (S34). Similarly, the terms in the second line of Eq. (S65) involving $\rho'_{l,m,n}$ read:

$$
-k_{\mathrm{on}}^{30\mathrm{S}} c_{30\mathrm{S\,F}}(x) \sum_{l=0}^{l_{\max}-1} \sum_{m=0}^{m_{\max}} \sum_{n=0}^{n_{\max}} \rho'_{l,m,n}(x) + k_{\mathrm{off}}^{30\mathrm{S}} \sum_{l=1}^{l_{\max}} \sum_{m=0}^{m_{\max}} \sum_{n=0}^{n_{\max}} l\, \rho'_{l,m,n}(x) = \sum_{l=1}^{l_{\max}} l \sum_{m=0}^{m_{\max}} \sum_{n=0}^{n_{\max}} \left[ -\frac{dJ'_{l,m,n}(x)}{dx} + \omega'_{l,m,n}(x) \right].
\tag{S99}
$$

We now compute explicitly the RHS of Eqs. (S98) and (S99) for large $l_{\max}$, $m_{\max}$ along the lines of Eqs. (S35) and (S36), we substitute the result in Eq. (S65), and we obtain the final reaction-diffusion equation for the concentration of F 30S subunits

$$
D_{\mathrm{F}} \left[ \frac{d^2 c_{30\mathrm{S\,F}}(x)}{dx^2} v_{\mathrm{F}}(x) - c_{30\mathrm{S\,F}}(x) \frac{d^2 v_{\mathrm{F}}(x)}{dx^2} \right]
\tag{S100}
$$

$$
\begin{aligned}
&+D \frac{d}{dx} \sum_{n=0}^{n_{\max}} \Bigg\{ \exp\left[ \left( \frac{k_{\mathrm{on}}^{30\mathrm{S}} c_{30\mathrm{S\,F}}(x)}{k_{\mathrm{off}}^{30\mathrm{S}}} + \frac{k_{\mathrm{on}}^{50\mathrm{S}} c_{50\mathrm{S\,F}}(x)}{k_{\mathrm{off}}^{50\mathrm{S}}} \right) v_{\mathrm{F}}(x) \right] [v_{\mathrm{F}}(x)]^n \left[ \left( 1 + \frac{k_{\mathrm{on}}^{30\mathrm{S}} c_{30\mathrm{S\,F}}(x)}{k_{\mathrm{off}}^{30\mathrm{S}}} v_{\mathrm{F}} \right) \frac{k_{\mathrm{on}}^{30\mathrm{S}}}{k_{\mathrm{off}}^{30\mathrm{S}}} \rho_n(x) v_{\mathrm{R}}(x) \right. \\
&\times \left( \frac{dc_{30\mathrm{S\,F}}(x)}{dx} v_{\mathrm{F}}(x) - c_{30\mathrm{S\,F}}(x) \frac{dv_{\mathrm{F}}(x)}{dx} \right) + \frac{k_{\mathrm{on}}^{50\mathrm{S}}}{k_{\mathrm{off}}^{50\mathrm{S}}} \rho_n(x) v_{\mathrm{R}}(x) v_{\mathrm{F}}(x) \frac{k_{\mathrm{on}}^{30\mathrm{S}} c_{30\mathrm{S\,F}}(x)}{k_{\mathrm{off}}^{30\mathrm{S}}} \left( \frac{dc_{50\mathrm{S\,F}}(x)}{dx} v_{\mathrm{F}}(x) - c_{50\mathrm{S\,F}}(x) \frac{dv_{\mathrm{F}}(x)}{dx} \right) \\
&+ \frac{k_{\mathrm{on}}^{30\mathrm{S}} c_{30\mathrm{S\,F}}(x)}{k_{\mathrm{off}}^{30\mathrm{S}}} \left( \frac{d\rho_n(x)}{dx} v_{\mathrm{R}}(x) - \rho_n(x) \frac{dv_{\mathrm{R}}(x)}{dx} \right) v_{\mathrm{F}}(x) - \left. \frac{k_{\mathrm{on}}^{30\mathrm{S}} c_{30\mathrm{S\,F}}(x)}{k_{\mathrm{off}}^{30\mathrm{S}}} \rho_n(x) v_{\mathrm{R}}(x) n \frac{dv_{\mathrm{F}}(x)}{dx} \right] \Bigg\} \\
&+D \frac{d}{dx} \sum_{n=0}^{n_{\max}} \Bigg\{ \exp\left[ \left( \frac{k_{\mathrm{on}}^{30\mathrm{S}} c_{30\mathrm{S\,F}}(x)}{k_{\mathrm{off}}^{30\mathrm{S}}} + \frac{k_{\mathrm{on}}^{50\mathrm{S}} c_{50\mathrm{S\,F}}(x)}{k_{\mathrm{off}}^{50\mathrm{S}}} \right) v_{\mathrm{F}}(x) \right] [v_{\mathrm{F}}(x)]^{n+1} \left[ \left( 1 + \frac{k_{\mathrm{on}}^{30\mathrm{S}} c_{30\mathrm{S\,F}}(x)}{k_{\mathrm{off}}^{30\mathrm{S}}} v_{\mathrm{F}} \right) \frac{k_{\mathrm{on}}^{30\mathrm{S}}}{k_{\mathrm{off}}^{30\mathrm{S}}} \rho'_n(x) v_{\mathrm{R}}(x) \right. \\
&\times \left( \frac{dc_{30\mathrm{S\,F}}(x)}{dx} v_{\mathrm{F}}(x) - c_{30\mathrm{S\,F}}(x) \frac{dv_{\mathrm{F}}(x)}{dx} \right) + \frac{k_{\mathrm{on}}^{50\mathrm{S}}}{k_{\mathrm{off}}^{50\mathrm{S}}} \rho'_n(x) v_{\mathrm{R}}(x) v_{\mathrm{F}}(x) \frac{k_{\mathrm{on}}^{30\mathrm{S}} c_{30\mathrm{S\,F}}(x)}{k_{\mathrm{off}}^{30\mathrm{S}}} \left( \frac{dc_{50\mathrm{S\,F}}(x)}{dx} v_{\mathrm{F}}(x) - c_{50\mathrm{S\,F}}(x) \frac{dv_{\mathrm{F}}(x)}{dx} \right) \\
&+ \frac{k_{\mathrm{on}}^{30\mathrm{S}} c_{30\mathrm{S\,F}}(x)}{k_{\mathrm{off}}^{30\mathrm{S}}} \left( \frac{d\rho'_n(x)}{dx} v_{\mathrm{R}}(x) - \rho'_n(x) \frac{dv_{\mathrm{R}}(x)}{dx} \right) v_{\mathrm{F}}(x) - \left. \frac{k_{\mathrm{on}}^{30\mathrm{S}} c_{30\mathrm{S\,F}}(x)}{k_{\mathrm{off}}^{30\mathrm{S}}} \rho'_n(x) v_{\mathrm{R}}(x)(n+1) \frac{dv_{\mathrm{F}}(x)}{dx} \right] \Bigg\} \\
&+ \exp\left( \frac{k_{\mathrm{on}}^{30\mathrm{S}} c_{30\mathrm{S\,F}}(x)}{k_{\mathrm{off}}^{30\mathrm{S}}} + \frac{k_{\mathrm{on}}^{50\mathrm{S}} c_{50\mathrm{S\,F}}(x)}{k_{\mathrm{off}}^{50\mathrm{S}}} \right) \Bigg\{ k_{\mathrm{off}}^{70\mathrm{S}} \sum_{n=1}^{n_{\max}} n[\rho_n(x) + \rho'_n(x)] + \beta \sum_{n=1}^{n_{\max}} [n\, \rho_n(x) + (n+1)\rho'_n(x)] \\
&\qquad\qquad\qquad\qquad -k_{\mathrm{on}}'^{30\mathrm{S}} c_{30\mathrm{S\,F}}(x) \sum_{n=0}^{n_{\max}} \rho_n(x) + k_{\mathrm{off}}'^{30\mathrm{S}} \sum_{n=0}^{n_{\max}} \rho'_n(x) \Bigg\} = 0.
\end{aligned}
$$

In Eq. (S100), the terms in the second, third, and fourth lines represent the diffusive flux of B 30S subunits carried by diffusing mRNAs with no 30S subunits on the initiation site, and the terms in the fifth, sixth and seventh lines represent the diffusive flux of B 30S subunits carried by mRNAs with a 30S unit at the initiation site.

The reaction-diffusion equation for the concentration of F 50S subunits can be obtained along the same lines as Eq. (S100), and it reads

$$D_{\mathrm{F}}\left[\frac{d^2 c_{\mathrm{50S\,F}}(x)}{dx^2}v_{\mathrm{F}}(x)-c_{\mathrm{50S\,F}}(x)\frac{d^2 v_{\mathrm{F}}(x)}{dx^2}\right] \tag{S101}$$

$$+D\frac{d}{dx}\sum_{n=0}^{n_{\max}}\left\{\exp\left[\left(\frac{k_{\mathrm{on}}^{\mathrm{30S}}c_{\mathrm{30S\,F}}(x)}{k_{\mathrm{off}}^{\mathrm{30S}}}+\frac{k_{\mathrm{on}}^{\mathrm{50S}}c_{\mathrm{50S\,F}}(x)}{k_{\mathrm{off}}^{\mathrm{50S}}}\right)v_{\mathrm{F}}(x)\right][v_{\mathrm{F}}(x)]^n\left[\left(1+\frac{k_{\mathrm{on}}^{\mathrm{50S}}c_{\mathrm{50S\,F}}(x)}{k_{\mathrm{off}}^{\mathrm{50S}}}v_{\mathrm{F}}(x)\right)\frac{k_{\mathrm{on}}^{\mathrm{50S}}}{k_{\mathrm{off}}^{\mathrm{50S}}}\rho_n(x)v_{\mathrm{R}}(x)\right.\right.$$

$$\times\left(\frac{dc_{\mathrm{50S\,F}}(x)}{dx}v_{\mathrm{F}}(x)-c_{\mathrm{50S\,F}}(x)\frac{dv_{\mathrm{F}}(x)}{dx}\right)+\frac{k_{\mathrm{on}}^{\mathrm{30S}}}{k_{\mathrm{off}}^{\mathrm{30S}}}\rho_n(x)v_{\mathrm{R}}(x)v_{\mathrm{F}}(x)\frac{k_{\mathrm{on}}^{\mathrm{50S}}c_{\mathrm{50S\,F}}(x)}{k_{\mathrm{off}}^{\mathrm{50S}}}\left(\frac{dc_{\mathrm{30S\,F}}(x)}{dx}v_{\mathrm{F}}(x)-c_{\mathrm{30S\,F}}(x)\frac{dv_{\mathrm{F}}(x)}{dx}\right)$$

$$\left.\left.+\frac{k_{\mathrm{on}}^{\mathrm{50S}}c_{\mathrm{50S\,F}}(x)}{k_{\mathrm{off}}^{\mathrm{50S}}}\left(\frac{d\rho_n(x)}{dx}v_{\mathrm{R}}(x)-\rho_n(x)\frac{dv_{\mathrm{R}}(x)}{dx}\right)v_{\mathrm{F}}(x)-\frac{k_{\mathrm{on}}^{\mathrm{50S}}c_{\mathrm{50S\,F}}(x)}{k_{\mathrm{off}}^{\mathrm{50S}}}\rho_n(x)v_{\mathrm{R}}(x)n\frac{dv_{\mathrm{R}}(x)}{dx}\right]\right\}$$

$$+D\frac{d}{dx}\sum_{n=0}^{n_{\max}}\left\{\exp\left[\left(\frac{k_{\mathrm{on}}^{\mathrm{30S}}c_{\mathrm{30S\,F}}(x)}{k_{\mathrm{off}}^{\mathrm{30S}}}+\frac{k_{\mathrm{on}}^{\mathrm{50S}}c_{\mathrm{50S\,F}}(x)}{k_{\mathrm{off}}^{\mathrm{50S}}}\right)v_{\mathrm{F}}(x)\right][v_{\mathrm{F}}(x)]^{n+1}\left[\left(1+\frac{k_{\mathrm{on}}^{\mathrm{50S}}c_{\mathrm{50S\,F}}(x)}{k_{\mathrm{off}}^{\mathrm{50S}}}v_{\mathrm{F}}(x)\right)\frac{k_{\mathrm{on}}^{\mathrm{50S}}}{k_{\mathrm{off}}^{\mathrm{50S}}}\rho_n'(x)v_{\mathrm{R}}(x)\right.\right.$$

$$\times\left(\frac{dc_{\mathrm{50S\,F}}(x)}{dx}v_{\mathrm{F}}(x)-c_{\mathrm{50S\,F}}(x)\frac{dv_{\mathrm{F}}(x)}{dx}\right)+\frac{k_{\mathrm{on}}^{\mathrm{30S}}}{k_{\mathrm{off}}^{\mathrm{30S}}}\rho_n'(x)v_{\mathrm{R}}(x)v_{\mathrm{F}}(x)\frac{k_{\mathrm{on}}^{\mathrm{50S}}c_{\mathrm{50S\,F}}(x)}{k_{\mathrm{off}}^{\mathrm{50S}}}\left(\frac{dc_{\mathrm{30S\,F}}(x)}{dx}v_{\mathrm{F}}(x)-c_{\mathrm{30S\,F}}(x)\frac{dv_{\mathrm{F}}(x)}{dx}\right)$$

$$\left.\left.+\frac{k_{\mathrm{on}}^{\mathrm{50S}}c_{\mathrm{50S\,F}}(x)}{k_{\mathrm{off}}^{\mathrm{50S}}}\left(\frac{d\rho_n'(x)}{dx}v_{\mathrm{R}}(x)-\rho_n'(x)\frac{dv_{\mathrm{R}}(x)}{dx}\right)v_{\mathrm{F}}(x)-\frac{k_{\mathrm{on}}^{\mathrm{50S}}c_{\mathrm{50S\,F}}(x)}{k_{\mathrm{off}}^{\mathrm{50S}}}\rho_n'(x)v_{\mathrm{R}}(x)(n+1)\frac{dv_{\mathrm{R}}(x)}{dx}\right]\right\}$$

$$+\exp\left(\frac{k_{\mathrm{on}}^{\mathrm{30S}}c_{\mathrm{30S\,F}}(x)}{k_{\mathrm{off}}^{\mathrm{30S}}}+\frac{k_{\mathrm{on}}^{\mathrm{50S}}c_{\mathrm{50S\,F}}(x)}{k_{\mathrm{off}}^{\mathrm{50S}}}\right)\left\{k_{\mathrm{off}}^{\mathrm{70S}}\sum_{n=1}^{n_{\max}}n[\rho_n(x)+\rho_n'(x)]+\beta\sum_{n=1}^{n_{\max}}n\left[\rho_n(x)+\rho_n'(x)\right]\right.$$

$$\left.-k_{\mathrm{on}}'^{\,\mathrm{50S}}c_{\mathrm{50S\,F}}(x)\sum_{n=0}^{n_{\max}-1}\rho_n'(x)\right\}\;=\;0.$$

The boundary conditions for Eqs. (S90), (S91), (S100) and (S101) are obtained from Eqs. (S67)-(S72) and from the leading-order expressions (S80) and (S81) for the mRNA densities. The boundary conditions for Eqs. (S90) and (S91) read

$$\left[\frac{d\rho_n(x)}{dx}v_{\mathrm{R}}(x)[v_{\mathrm{F}}(x)]^n-\rho_n(x)\frac{d[v_{\mathrm{R}}(x)[v_{\mathrm{F}}(x)]^n]}{dx}\right]\Bigg|_{x=0,\ell}\;=\;0, \tag{S102}$$

$$\left[\frac{d\rho_n'(x)}{dx}v_{\mathrm{R}}(x)[v_{\mathrm{F}}(x)]^{n+1}-\rho_n'(x)\frac{d[v_{\mathrm{R}}(x)[v_{\mathrm{F}}(x)]^{n+1}]}{dx}\right]\Bigg|_{x=0,\ell}\;=\;0. \tag{S103}$$

The no-flux boundary conditions for Eqs. (S100) and (S101) are given by Eqs. (S69) and (S70), while the constraints (S71), (S72) on the total number of 30S and 50S ribosomal subunits read

$$2\int_0^\ell dx\left\{c_{\mathrm{30S\,F}}(x)+\exp\left(\frac{k_{\mathrm{on}}^{\mathrm{30S}}c_{\mathrm{30S\,F}}(x)}{k_{\mathrm{off}}^{\mathrm{30S}}}+\frac{k_{\mathrm{on}}^{\mathrm{50S}}c_{\mathrm{50S\,F}}(x)}{k_{\mathrm{off}}^{\mathrm{50S}}}\right)\sum_{n=0}^{n_{\max}}\left[\left(\frac{k_{\mathrm{on}}^{\mathrm{30S}}c_{\mathrm{30S\,F}}(x)}{k_{\mathrm{off}}^{\mathrm{30S}}}+n\right)\rho_n(x)\right.\right.$$

$$\left.\left.+\left(\frac{k_{\mathrm{on}}^{\mathrm{30S}}c_{\mathrm{30S\,F}}(x)}{k_{\mathrm{off}}^{\mathrm{30S}}}+n+1\right)\rho_n'(x)\right]\right\}\;=\;N_{\mathrm{30S\,tot}}, \tag{S104}$$

$$2\int_0^\ell dx\left\{c_{\mathrm{50S\,F}}(x)+\exp\left(\frac{k_{\mathrm{on}}^{\mathrm{30S}}c_{\mathrm{30S\,F}}(x)}{k_{\mathrm{off}}^{\mathrm{30S}}}+\frac{k_{\mathrm{on}}^{\mathrm{50S}}c_{\mathrm{50S\,F}}(x)}{k_{\mathrm{off}}^{\mathrm{50S}}}\right)\sum_{n=0}^{n_{\max}}\left(\frac{k_{\mathrm{on}}^{\mathrm{50S}}c_{\mathrm{50S\,F}}(x)}{k_{\mathrm{off}}^{\mathrm{50S}}}+n\right)[\rho_n(x)+\rho_n'(x)]\right\}\;=\;N_{\mathrm{50S\,tot}}. \tag{S105}$$

Overall, Eqs. (S69), (S70), (S90), (S91), (S100), (S101) and (S102)–(S105) characterize the steady-state behavior of the reaction-diffusion model with two ribosomal subunits in the rapid-equilibrium limit. Compared to the model with a single ribosomal subunit discussed in section II, this two-submit model provides a more realistic characterization of the transcriptional-translational process, describing explicitly the assembly and disassembly of 70S pairs on mRNA molecules. More realistic models describing the internal degrees of freedom of ribosomes and mRNAs could be obtained with full molecular-dynamics simulations, rather than with reaction-diffusion equations for local concentrations. Even if they are likely to be limited to a single ribosome and mRNA molecule, such molecular-dynamics simulations may be useful for describing several interesting features of the transcriptional-translational machinery on a microscopic level,

such as recycling of ribosomes which translate mRNA molecules forming loops, where initiation and termination sites are in close proximity [18].

Proceeding along the lines of section S6, all physical quantities of interest can be expressed in terms of $\rho_n$, $\rho'_n$, $c_{30S\,F}$, $c_{50S\,F}$ by using Eqs. (S80)-(S83). Some of these quantities are the total mRNA densities

$$\rho_{\text{tot}}(x) = \sum_{l=0}^{\infty} \sum_{m=0}^{\infty} \sum_{n=0}^{n_{\max}} \boldsymbol{\rho}_{l,m,n}(x), \tag{S106}$$

$$\rho'_{\text{tot}}(x) = \sum_{l=0}^{\infty} \sum_{m=0}^{\infty} \sum_{n=0}^{n_{\max}} \boldsymbol{\rho}'_{l,m,n}(x), \tag{S107}$$

the concentration of 70S subunits

$$c_{70S}(x) = \sum_{l=0}^{\infty} \sum_{m=0}^{\infty} \sum_{n=0}^{n_{\max}} n[\boldsymbol{\rho}_{l,m,n}(x) + \boldsymbol{\rho}'_{l,m,n}(x)], \tag{S108}$$

the concentrations of B 30S and 50S subunits

$$c_{30S\,B}(x) = \sum_{l=0}^{\infty} \sum_{m=0}^{\infty} \sum_{n=0}^{n_{\max}} l[\boldsymbol{\rho}_{l,m,n}(x) + \boldsymbol{\rho}'_{l,m,n}(x)], \tag{S109}$$

$$c_{50S\,B}(x) = \sum_{l=0}^{\infty} \sum_{m=0}^{\infty} \sum_{n=0}^{n_{\max}} m[\boldsymbol{\rho}_{l,m,n}(x) + \boldsymbol{\rho}'_{l,m,n}(x)], \tag{S110}$$

and the concentration of 30S subunits bound to the translation-initiation site $c'_{30S}(x) = \rho'_{\text{tot}}(x)$. Other quantities are the fluxes of F 30S and 50S subunits, that we denote by $J_{30S\,F}(x)$ and $J_{50S\,F}(x)$, and that are obtained from Eq. (S3) by replacing $c_F$ with $c_{30S\,F}$ and $c_{50S\,F}$, respectively, the flux of 70S subunits

$$J_{70S}(x) = \sum_{l=0}^{\infty} \sum_{m=0}^{\infty} \sum_{n=0}^{n_{\max}} n\left[J_{l,m,n}(x) + J'_{l,m,n}(x)\right], \tag{S111}$$

the fluxes of B 30S and 50S subunits

$$J_{30S\,B}(x) = \sum_{l=0}^{\infty} \sum_{m=0}^{\infty} \sum_{n=0}^{n_{\max}} l\left[J_{l,m,n}(x) + J'_{l,m,n}(x)\right], \tag{S112}$$

$$J_{50S\,B}(x) = \sum_{l=0}^{\infty} \sum_{m=0}^{\infty} \sum_{n=0}^{n_{\max}} m\left[J_{l,m,n}(x) + J'_{l,m,n}(x)\right], \tag{S113}$$

and the flux of 30S subunits bound to the translation-initiation site

$$J'_{30S}(x) = \sum_{l=0}^{\infty} \sum_{m=0}^{\infty} \sum_{n=0}^{n_{\max}} J'_{l,m,n}(x). \tag{S114}$$

### D. Results

We numerically solved Eqs. (S90), (S91), (S100), and (S101) by fixing the maximal number of allowed translating 70S pairs per mRNA at $n_{\max} = 24$, and the results are shown in Figs. 5, S18, and S7. In addition to the features of these results discussed in section II, here we observe that the total mRNA distributions in Fig. S18 have a dip at $x/\ell \sim 0.4$. This minimum is due to the fact that the total mRNA profile is the superposition of two peaks: One peak in the nucleoid region due to mRNAs that are being transcribed and are nearly free of ribosomal subunits, and one peak at the cell poles due to mRNAs loaded with multiple subunits, which are strongly excluded from the nucleoid. In Fig. S18, there is an average number of $\sim 14$ subunits, i.e. 30S, 50S and 70S pairs, per mRNA, implying excluded-volume

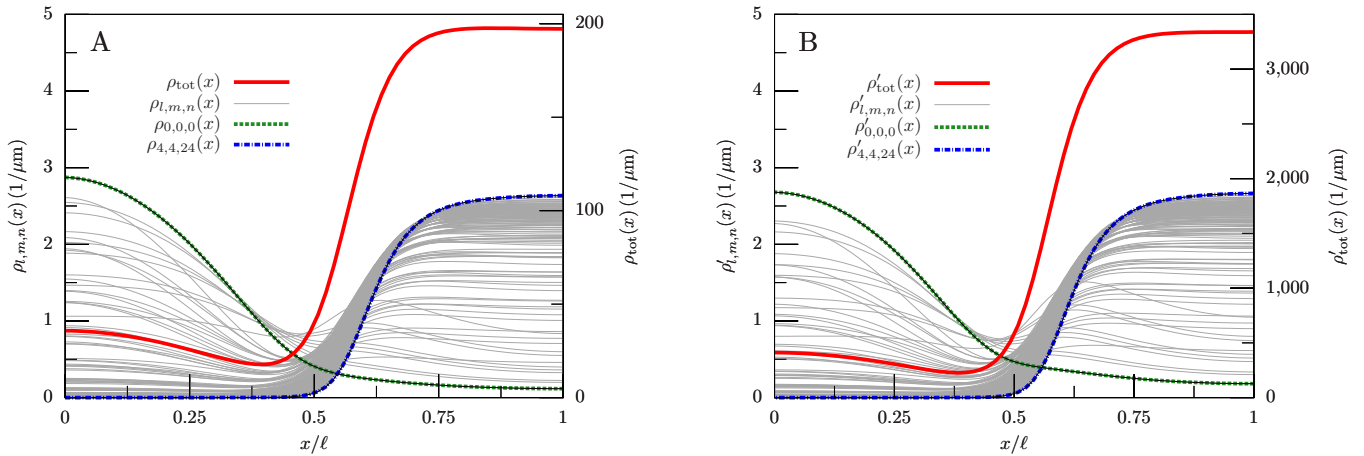

FIG. S6: Steady-state mRNA distributions for the model including 30S and 50S ribosomal subunits, where we fixed the maximal number of allowed translating (T) 30S-50S (70S) pairs per mRNA at $n \leq n_{\max} = 24$, and we show mRNA species with $l \leq 4$ and $m \leq 4$ transiently bound (B) 30S and 50S subunits, respectively. (A) mRNA profiles for mRNAs with no 30S subunit bound to the translation-initiation site. Total mRNA density $\rho_{\text{tot}}(x)$ (red) and density $\rho_{l,m,n}(x)$ of mRNAs with $l$ B 30S subunits, $m$ B 50S subunits, and $n$ 70S pairs (gray), where we show only the curves with $l$, $m$ and $n$ even for greater clarity. The density $\rho_{0,0,0}(x)$ of ribosome-free mRNAs (green) and the density $\rho_{4,4,24}(x)$ of mRNAs with the largest number of B, T subunits considered (blue) are also shown. The profiles $\rho_{l,m,n}(x)$, $\rho_{0,0,0}(x)$, and $\rho_{4,4,24}(x)$ are normalized to unit area. (B) Same as panel A for mRNAs with a 30S subunit bound to the initiation site.

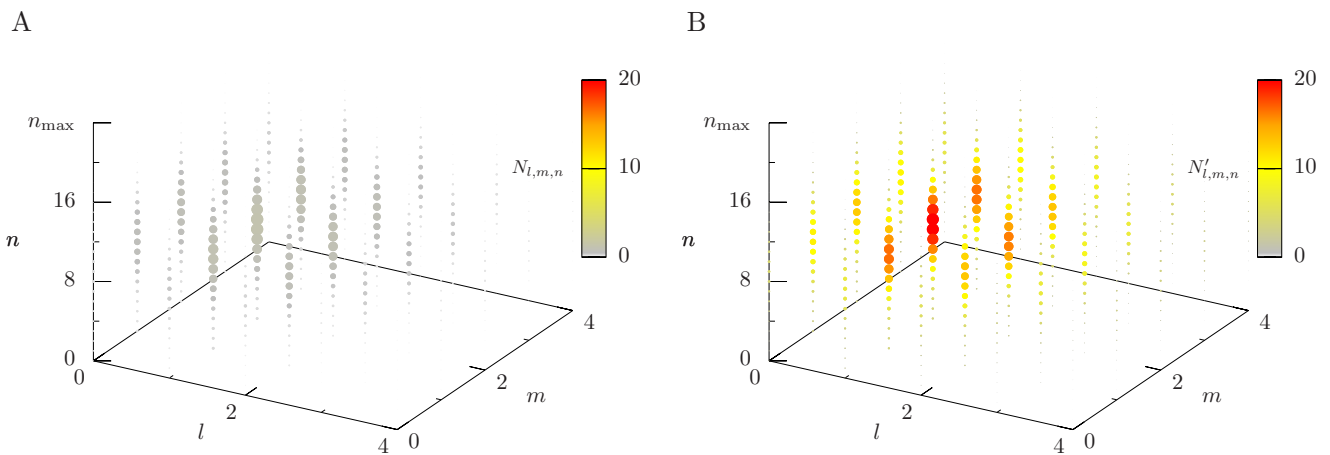

FIG. S7: Polysome distributions for the model including 30S and 50S ribosomal subunits, where we fixed the maximal number of allowed translating (T) 30S-50S (70S) pairs per mRNA at $n \leq n_{\max} = 24$, and we show mRNA species with $l \leq 4$ and $m \leq 4$ transiently bound (B) 30S and 50S subunits, respectively. (A) distribution of mRNA species with no 30S subunit bound to the translation-initiation site, shown as a heat map of the number $N_{l,m,n} = \int_0^\ell dx \rho_{l,m,n}(x)$ of mRNAs with $l$ B 30S subunits, $m$ B 50S subunits, and $n$ 70S pairs in the right half of the cell. The diameter of each dot is proportional to $N_{l,m,n}$, where the largest dot in the plot corresponds to $N_{l,m,n} \sim 1$. (B) Same as panel A for the number $N'_{l,m,n} = \int_0^\ell dx \rho'_{l,m,n}(x)$ of mRNAs with a 30S subunit bound to the initiation site, where the largest dot in the plot corresponds to $N'_{l,m,n} \sim 20$.

effects strong enough that the second peak shrinks toward the cell poles, thus leaving a dip between the cell center and the poles—see also section S14.

We estimate the total translation rate in our model with two subunits as

$$\Sigma = 2 \, k_{\text{off}}^{70S} \sum_{l=0}^{\infty} \sum_{m=0}^{\infty} \sum_{n=1}^{n_{\max}} n \int_0^\ell dx [\rho_{l,m,n}(x) + \rho'_{l,m,n}(x)] \tag{S115}$$

and the translation efficiency as $\varepsilon = \Sigma / N_{50S\,\text{tot}} = 1.94 \times 10^{-2}/\text{s}$, which is practically identical to the efficiency $\varepsilon = 1.94 \times 10^{-2}/\text{s}$ for the model with a single ribosomal unit discussed in section II.

We conclude this section by discussing how our results pertain to in vivo conditions. In this regard, we recall that all the model parameters discussed in section S10 B have been estimated from in vivo observations, except the

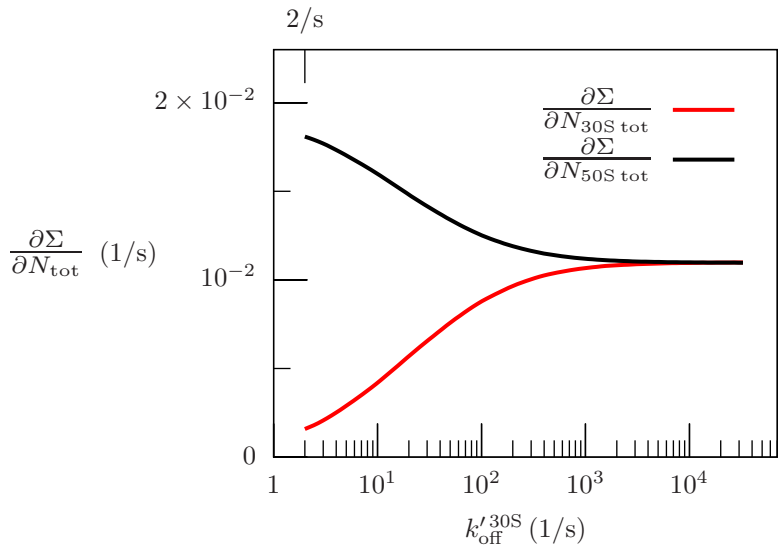

FIG. S8: Translation-rate increases $\partial\Sigma/\partial N_{30\text{S tot}}$ and $\partial\Sigma/\partial N_{50\text{S tot}}$ due to the addition of a 30S and a 50S subunit, respectively, as functions of the unbinding rate $k'^{30\text{S}}_{\text{off}}$ of 30S subunits from the translation-initiation site. The in vitro value of the unbinding rate $k'^{30\text{S}}_{\text{off}} = 2/\text{s}$ is marked on the top horizontal axis.

binding-unbinding rates of 30S subunits to the translation-initiation site, $k'^{30\text{S}}_{\text{on}}$ and $k'^{30\text{S}}_{\text{off}}$, which have been estimated in vitro [17]. First, we note that the higher concentration of solvated molecules present in vivo may significantly affect the binding-unbinding rates: a relevant example is that of DNA-bound proteins, whose dimeric nature allows for partial dissociation and thus enhanced exchange in the presence of competing proteins in solution [19–21]. It is natural to hypothesize that the same mechanism might apply to 30S subunits bound to the translation-initiation site, resulting in a significant increase in the unbinding rate $k'^{30\text{S}}_{\text{off}}$ in vivo compared to the in vitro value $k'^{30\text{S}}_{\text{off}} = 2/\text{s}$ discussed in section S10 B. Second, using the in vitro values of $k'^{30\text{S}}_{\text{on}}$ and $k'^{30\text{S}}_{\text{off}}$ and the solution of Eq. (S73), (S74), we obtain that the fraction of mRNAs with a 30S subunit bound to the translation-initiation site is $\overline{\rho'}/(\overline{\rho}+\overline{\rho'}) \sim 96\,\%$, i.e. the vast majority of mRNAs. It follows that the overall translation rate $\Sigma$ is mostly limited by the number of 50S subunits rather than by the number of 30S subunits. Indeed, let us consider the quantities $\partial\Sigma/\partial N_{30\text{S tot}}$ and $\partial\Sigma/\partial N_{50\text{S tot}}$, which represent the translation-rate increase due to the addition of a 30S or a 50S subunit, respectively. These two translation-rate increases are shown in Fig. S8 as functions of the unbinding rate $k'^{30\text{S}}_{\text{off}}$. For values of the unbinding rate close to the in vitro value, we have $\partial\Sigma/\partial N_{30\text{S tot}} \ll \partial\Sigma/\partial N_{50\text{S tot}}$: in this case, there is a "waste" of 30S subunits, i.e. the number of 30S subunits could be significantly reduced without significantly affecting the translation rate. On the other hand, for larger values of $k'^{30\text{S}}_{\text{off}} \gtrsim 10^3/\text{s}$, the two translation-rate increases are similar, i.e. $\partial\Sigma/\partial N_{30\text{S tot}} \sim \partial\Sigma/\partial N_{50\text{S tot}}$: in this case, the 30S and 50S subunits are co-rate limiting for translation, reflecting an efficient allocation of cellular resources which is generally expected in vivo [22, 23]. Overall, these observations provide an example of how our model can be used to relate microscopic parameters such as the binding and unbinding rates of ribosomal subunits to macroscopic observables such as subunit stoichiometries and ultimately to functional performance, namely the protein-translation rate.

## S11   mRNA degradation by RNAse enzymes

In this section, we incorporate in the two-subunit model above a more realistic description of mRNA degradation based on the mechanism of mRNA decay suggested by in vivo studies in *E. coli*: mRNA degradation is triggered by RNase enzymes, which target the 5′ end of mRNA and compete with ribosomes to bind the transcript and degrade it [13]. In addition, RNase enzymes have been suggested to localize to the cell membrane [24]: In our one-dimensional reaction-diffusion model, such a distribution of RNase on the three-dimensional membrane will result in an effective, one-dimensional distribution of RNase on the long cell axis. To obtain this distribution, we model the cell as a cylinder with two hemispherical endcaps of radius $R$, we assume that RNase is uniformly distributed on the membrane, and we introduce the angle $\theta$ at the hemisphere center between a point on the hemisphere and the long cell axis, see Fig. S9. Given an angle element $d\theta$, we consider corresponding length element $dx$ along the $x$ axis. We consider the amount of RNase along the $\theta$ angle, which is proportional to $R\,d\theta$, and we project it on the length element

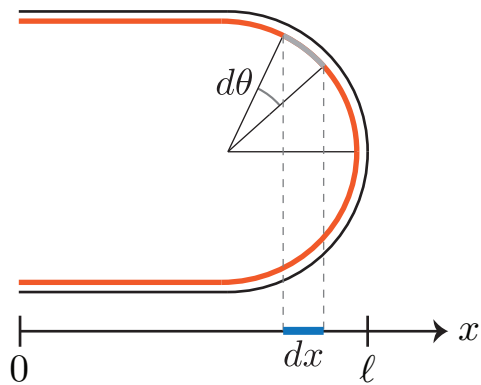

FIG. S9: Modeled distribution of RNase on the cell membrane, and its projection on the long cell axis. RNase is uniformly distributed on the membrane, and it is represented as an orange layer. The angle element $d\theta$ describes an arc on the cell membrane (gray) which is then projected on the $x$ axis, where it corresponds to the length element $dx$ (blue).

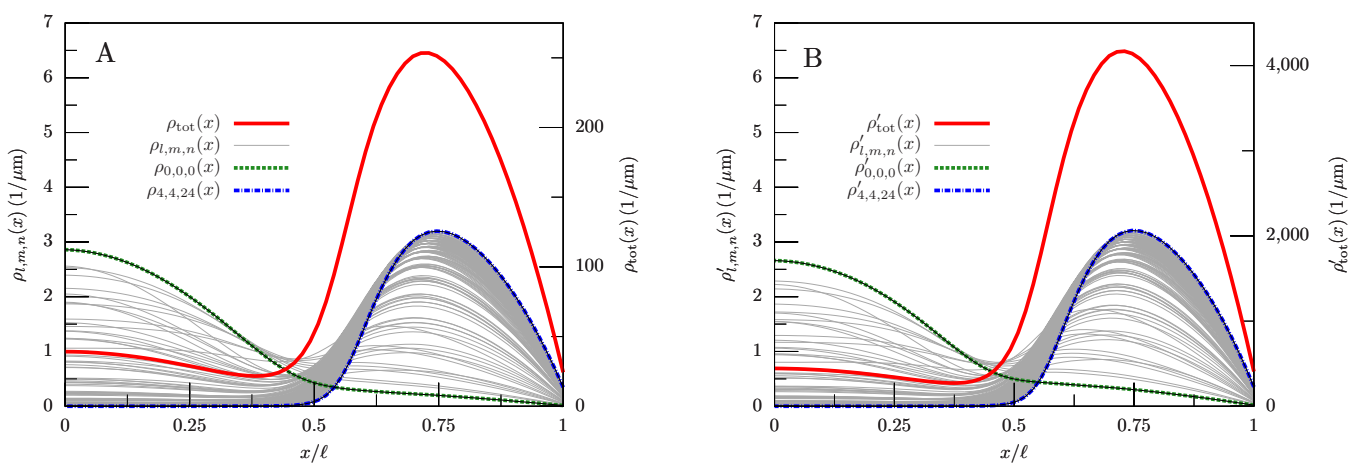

FIG. S10: Steady-state mRNA distributions for the model with 30S and 50S ribosomal subunits and spatially dependent mRNA decay, where we fixed the maximal number of allowed translating (T) 30S-50S (70S) pairs per mRNA at $n \leq n_{\max} = 24$, and we show mRNA species with $l \leq 4$ and $m \leq 4$ transiently bound (B) 30S and 50S subunits, respectively. (A) mRNA profiles for mRNAs with no 30S subunit bound to the translation-initiation site. Total mRNA density $\rho_{\text{tot}}(x)$ (red) and density $\rho_{l,m,n}(x)$ of mRNAs with $l$ B 30S subunits, $m$ B 50S subunits, and $n$ 70S pairs (gray),where we show only the curves with $l$, $m$, and $n$ even for greater clarity. The density $\rho_{0,0,0}(x)$ of ribosome-free mRNAs (green) and the density $\rho_{4,4,24}(x)$ of mRNAs with the largest number of B, T subunits considered (blue) are also shown. The profiles $\rho_{l,m,n}(x)$, $\rho_{0,0,0}(x)$, and $\rho_{4,4,24}(x)$ are normalized to unit area. (B) Same as panel A for mRNAs with a 30S subunit bound to the initiation site.

$dx$. As a result, in the hemispherical region $\ell - R \leq x < \ell$ we obtain an effective, one-dimensional RNase density proportional to $\frac{R\,d\theta}{dx} = 1/\sqrt{\sigma(x)}$, where $\sigma(x) = \frac{\pi(R\sin\theta)^2}{\pi R^2} = \frac{\ell - x}{R}\left(2 - \frac{\ell - x}{R}\right)$ is the ratio between the cell cross section in the hemispherical endcap and the cross section at midcell. In the central region $0 \leq x < \ell - R$, the effective RNase density is constant, reflecting the uniform RNase distribution in the surrounding cylindrical cell membrane. The axial RNase distribution above will result in an effective, space-dependent mRNA degradation rate $\beta(x)$, which we will assume to be proportional to the RNase profile:

$$\beta(x) = \begin{cases} \beta_* & 0 \leq x < \ell - R, \\ \frac{\beta_*}{\sqrt{\sigma(x)}} & \ell - R \leq x < \ell, \end{cases} \tag{S116}$$

where the constant $\beta_*$ has been chosen so as to obtain an average mRNA decay rate $(1/\ell)\int_0^\ell dx\,\beta(x)$ equal to the uniform rate $\beta = 3 \times 10^{-3}/\text{s}$ of the model discussed in section II.

In addition, we observe that the hemispherical shape of the cell endcaps results in a reduced cross section of the cell perpendicular to the long cell axis, and thus in a smaller available volume for both ribosomes and mRNAs at the cell poles. It follows that in the polar region the available volume will be reduced by an amount equal to $\sigma(x)$, and we set $v_{\text{F}}(x) \to v_{\text{F}}(x)\sigma(x)$ and $v_{l+m+n}(x) \to v_{l+m+n}(x)\sigma(x)$.

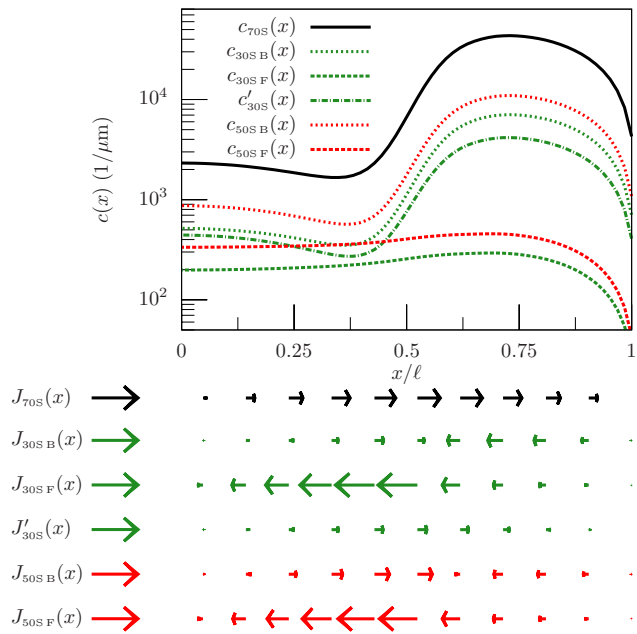

FIG. S11: Steady-state ribosomal-subunit concentrations and fluxes for the model with 30S and 50S ribosomal subunits and spatially dependent mRNA decay. Top: Concentrations $c_{70S}(x)$, $c_{30S\,B}(x)$, $c_{30S\,F}(x)$, $c'_{30S}(x)$ of 70S, transiently bound (B) 30S, free (F) 30S subunits, and of 30S subunits bound to the translation-initiation site, respectively. We also show the concentrations $c_{50S\,B}(x)$ and $c_{50S\,F}(x)$ of B and F 50S subunits, respectively. Bottom: Fluxes $J_{70S}(x)$, $J_{30S\,B}(x)$, $J_{30S\,F}(x)$ of 70S, B and F 30S subunits, and flux $J'_{30S}(x)$ of 30S subunits bound to the translation-initiation site. The fluxes $J_{50S\,B}(x)$, $J_{50S\,F}(x)$ of B and F 50S subunits are also shown. Fluxes are represented along the cell's long axis depicted in the top panel, the arrow length is proportional to local ribosome flux, and the arrows in the legends correspond to a flux of 30/s.

Finally, the hemispherical shape of the cell endcaps modifies the binding rates. Indeed, consider a binding rate $k_{on}$ and the binder concentration $c(x)$ in our one-dimensional model. First, $k_{on}$ and $c(x)$ can be related to the three-dimensional binding rate and concentration $k_{3D}$ and $c_{3D}(\vec{x})$ by equating the number of bindings per unit time: $k_{on}c(x) = k_{3D}c_{3D}(\vec{x})$. Second, the one- and three-dimensional concentrations are related by $c(x) = c_{3D}(\vec{x})\pi(R\sin\theta)^2 = c_{3D}(\vec{x})\sigma(x)\pi R^2$. Putting everything together, we obtain that the binding rates in the model with hemispherical cell endcaps are given by $k_{on} = k^*_{on}/\sigma(x)$, where $k^*_{on} = k_{3D}/(\pi R^2)$ is the binding rate for a cylindrical cell used in section S10 B.

The mRNA densities, ribosomal concentrations and fluxes, and polysome distributions for the model above with spatially dependent mRNA decay are shown in Figs. S10, S11, and S12, respectively, where the maximal number of allowed 70S pairs per mRNA has been fixed at $n_{max} = 24$.

## S12 Results for the late phase of the cell division cycle

In this section we extend the results from the two-subunit model discussed in section S10 to a cell in the late phase of its division cycle. To achieve this, we manually selected 21 dividing cells with a markedly double-lobed nucleoid, rescaled their DNA fluorescence profiles to their median cell length of $2\ell \sim 4.35\,\mu m$, and estimated the nucleoid profile along the long cell axis by averaging over multiple cells, see Fig. S13. We chose a density of DNA length $\varphi(x) \propto 1/\{1 + \exp[\zeta(x/\ell - 2/3)]\}\{\exp[-3\zeta/10(1/2 - x/\ell)^2] + \exp[-3\zeta/10(1/2 + x/\ell)^2]\}$ to reproduce the average DNA fluorescence profile in Fig. S13, and we adjusted the model parameters as follows in order to describe the late phase of the cell cycle. The parameters that are intensive with respect to the cell length, such as the binding and unbinding rates, the diffusion coefficients, and the mRNA decay rate, are taken to be the same as in the model with $2\ell = 3\,\mu m$, see section S10 B. On the other hand, extensive parameters were obtained by considering the corresponding parameter values of the model with $2\ell = 3\,\mu m$, and scaling them linearly with the cell length. As a result, we have $N_{30S\,tot} = N_{50S\,tot} = 8.7 \times 10^4$, $N_{mRNA} = 7.2 \times 10^3$, $N_{30S\,F} = N_{50S\,F} = 2\,\%\,N_{tot}$, and $N_{70S} = 80\,\%\,N_{tot}$. The nucleoid profile $\alpha(x)$ is proportional to $\varphi(x)$, and its normalization is chosen to achieve the total number of mRNAs above, i.e. $\alpha_{tot}/\beta = N_{mRNA}$. Assuming that both the DNA plectonemic length $L$ and the nucleoid volume $V$ scale linearly with

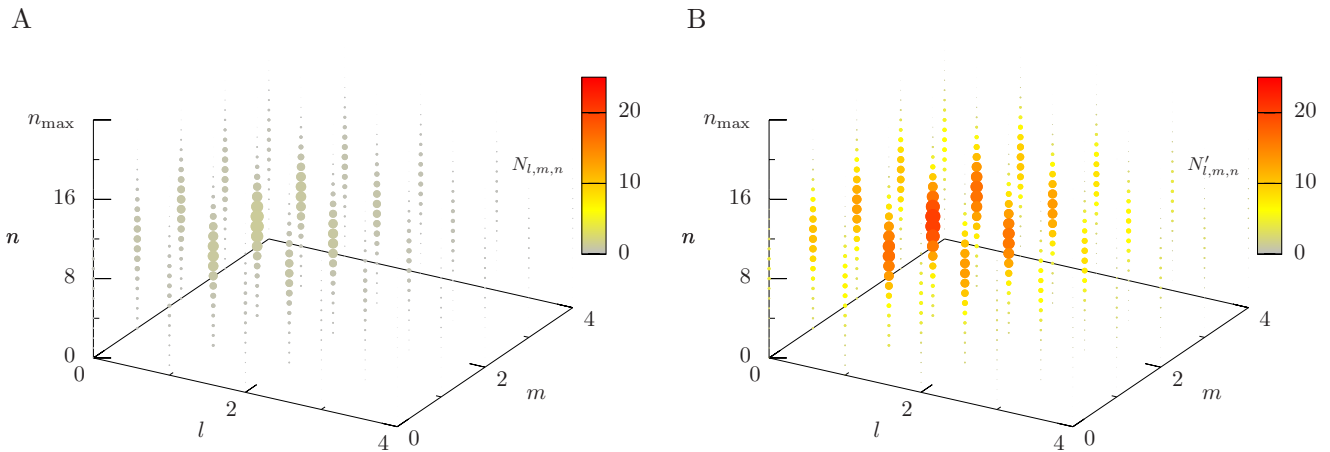

FIG. S12: Polysome distributions for the model with 30S and 50S ribosomal subunits and spatially dependent mRNA decay, where we fixed the maximal number of allowed translating (T) 30S-50S (70S) pairs per mRNA at $n \leq n_{max} = 24$, and we show mRNA species with $l \leq 4$ and $m \leq 4$ transiently bound (B) 30S and 50S subunits, respectively. (A) distribution of mRNA species with no 30S subunit bound to the translation-initiation site, shown as a heat map of the number $N_{l,m,n} = \int_0^\ell dx \rho_{l,m,n}(x)$ of mRNAs with $l$ B 30S subunits, $m$ B 50S subunits, and $n$ T 70S pairs. The diameter of each dot is proportional to $N_{l,m,n}$, where the largest dot in the plot corresponds to $N_{l,m,n} \sim 1$. (B) Same as panel A for the number $N'_{l,m,n} = \int_0^\ell dx \rho'_{l,m,n}(x)$ of mRNAs with a 30S subunit bound to the initiation site, where the largest dot in the plot corresponds to $N'_{l,m,n} \sim 20$.

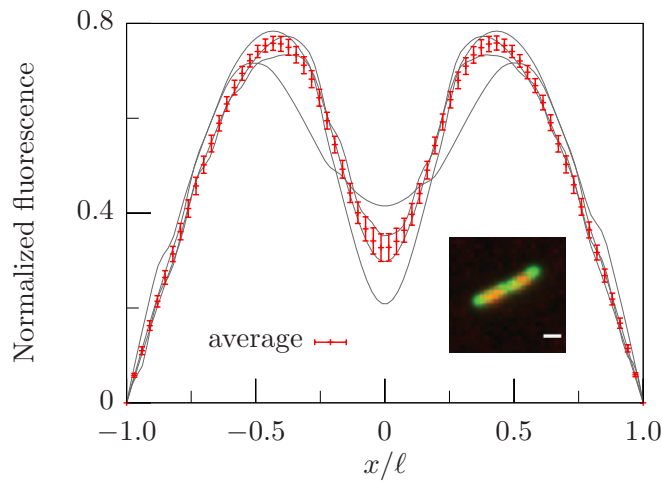

FIG. S13: DNA fluorescence along the long cell axis for *E. coli* cells in the late phase of the division cycle in glucose minimal media. Fluorescence for a few representative cells (gray), and resulting average fluorescence over 21 cells in the same phase of the division cycle with standard error of the mean (red), both normalized to unit area. Inset: Representative cell near late division phase, with ribosomal protein (green) and nucleoid (red). Scale bar: $1\,\mu m$.

the cell length, the normalization of the DNA density $(1/\ell) \int_0^\ell dx\, \varphi(x) = L/V$ is the same as for $2\,\ell = 3\,\mu m$.

With this choice of parameters, we solved the reaction-diffusion equations for the two-subunit model by fixing the maximal number of allowed 70S pairs per mRNA at $n_{max} = 28$: the resulting mRNA densities, ribosomal concentrations and fluxes, and polysome distributions are shown in Figs. S14, S15, and S16, respectively.

## S13   Results for different growth rates

In this section we extend the results from the two-subunit model discussed in section S10 to cells with different growth rates. In particular, we consider cells in glycerol minimal and defined rich media, with growth rates $\sim 0.56$/h and $2.3$/h, respectively, whose DNA fluorescence profiles are shown in Fig. S17. We will thus consider a density of

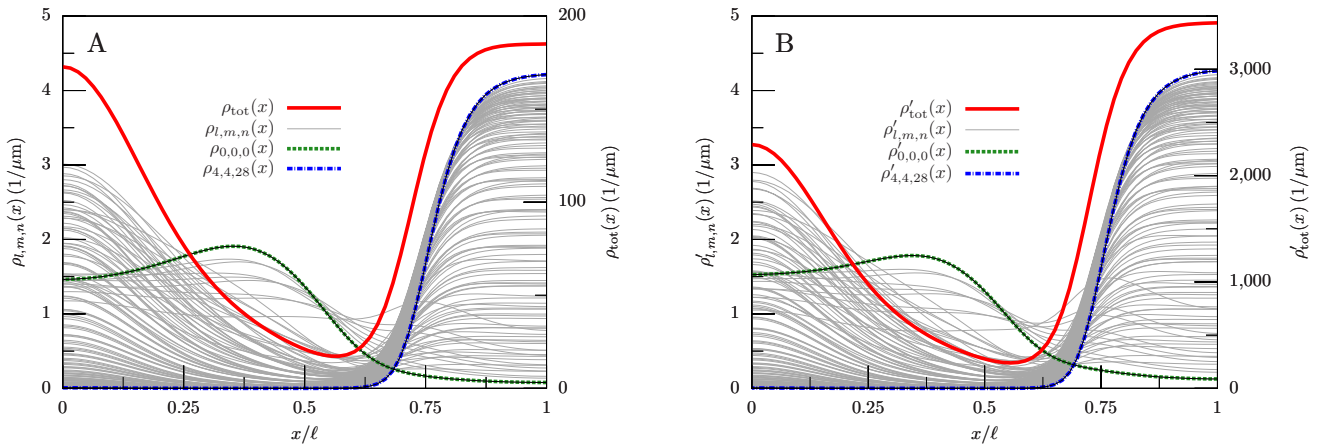

FIG. S14: Steady-state mRNA distributions in the model with 30S and 50S ribosomal subunits for a long cell in the late phase of its division cycle, where we fixed the maximal number of allowed translating (T) 30S-50S (70S) pairs per mRNA at $n \leq n_{\max} = 28$, and we show mRNA species with $l \leq 4$ and $m \leq 4$ transiently bound (B) 30S and 50S subunits, respectively. (A) mRNA profiles for mRNAs with no 30S subunit bound to the translation-initiation site. Total mRNA density $\rho_{\text{tot}}(x)$ (red) and density $\rho_{l,m,n}(x)$ of mRNAs with $l$ B 30S subunits, $m$ B 50S subunits, and $n$ 70S pairs (gray), where we show only the curves with $l$, $m$, and $n$ even for greater clarity. The density $\rho_{0,0,0}(x)$ of ribosome-free mRNAs (green) and the density $\rho_{4,4,28}(x)$ of mRNAs with the largest number of B, T subunits considered (blue) are also shown. The profiles $\rho_{l,m,n}(x)$, $\rho_{0,0,0}(x)$, and $\rho_{4,4,28}(x)$ are normalized to unit area. (B) Same as panel A for mRNAs with a 30S subunit bound to the initiation site.

DNA length $\varphi(x)$ to reproduce the profiles in Fig. S17, and adjust all other parameters to the growth conditions along the lines of section S12.

In glycerol minimal media, we selected cells in the mid-phase of the cell division cycle as follows. We considered a cell length $2\ell$ given by $\ell/\ell_{\text{median glycerol}} = \ell_{\text{glucose}}/\ell_{\text{median glucose}}$, where $2\ell_{\text{glucose}} = 3\,\mu\text{m}$ is the typical, medium length of a cell in the mid-phase of the division cycle in glucose minimal media, see section II. We obtained $2\ell \sim 3\,\mu\text{m}$, selected cells with length within 5% of $2\ell$, rescaled their DNA fluorescence profiles to a cell length of $2\ell$, and estimated the nucleoid profile along the long cell axis by averaging over multiple cells, see Fig. S17A. Given that $\ell$ is close to $\ell_{\text{glucose}}$, and that Figs. S17A and 2 show that the nucleoid profile in glycerol minimal media is close to that in glucose minimal media, all model parameters in glycerol minimal media will be close to those in glucose minimal media, and so will be the model predictions.

Proceeding along the lines of the analysis in glycerol minimal media, for defined rich media we obtained $2\ell \sim 3.5\,\mu\text{m}$, and we chose a density of DNA length $\varphi(x) \propto 1/\{1 + \exp[3/4\,\zeta(x/\ell - 2/3)]\}\{\exp[-\zeta/5(1/2 - x/\ell)^2] + \exp[-\zeta/5(1/2 + x/\ell)^2]\}$ to reproduce the observed DNA fluorescence profile, compare Fig. S17B. The intensive model parameters are the same as for $2\ell = 2\ell_{\text{glucose}} = 3\,\mu\text{m}$, while we estimated the extensive parameters by scaling them with respect to the cell length as in section S12. We obtained $N_{30S\,\text{tot}} = N_{50S\,\text{tot}} = 7 \times 10^4$, $N_{\text{mRNA}} = 5.8 \times 10^3$, $N_{30S\,\text{F}} = N_{50S\,\text{F}} = 2\,\%\,N_{\text{tot}}$, $N_{70S} = 80\,\%\,N_{\text{tot}}$. In addition, the normalization of the mRNA-synthesis profile $\alpha(x) \propto \varphi(x)$ is given by $\alpha_{\text{tot}} = \beta N_{\text{mRNA}}$, while the normalization of the DNA density $(1/\ell)\int_0^\ell dx\,\varphi(x) = L/V$ is the same as for $2\ell = 2\ell_{\text{glucose}}$. With this choice of parameters, we solved the reaction-diffusion equations for the two-subunit model by fixing the maximal number of allowed 70S pairs per mRNA at $n_{\max} = 28$: the resulting mRNA densities, ribosomal concentrations and fluxes, and polysome distributions are shown in Figs. S18, S19, and S20, respectively.

## S14   Model with no transiently bound ribosomes

In this section we disallow both B 30S and B 50S subunits as follows: First, we set $k_{\text{on}}^{30S}/k_{\text{off}}^{30S} = k_{\text{on}}^{50S}/k_{\text{off}}^{50S} = 0$. Second, we solve Eqs. (S73) and (S74) by setting $N_{70S} = 98\,\%\,N_{\text{tot}}$, to take account of the increased number of 70S pairs due to the absence of B subunits.

With this choice of parameters, we solve the reaction-diffusion equations (S90), (S91), (S100), and (S101). Given that there are no B 30S nor B 50S subunits, the only nonzero mRNA densities and mRNA numbers are those with $l = m = 0$, i.e. $\rho_{0,0,n}(x)$, $\rho'_{0,0,n}(x)$, $N_{0,0,n}$ and $N'_{0,0,n}$. In Fig. S21 we show the resulting mRNA profiles, the total mRNA densities, and the mRNA numbers, for both mRNAs with and without a 30S subunit bound to the translation-initiation site, while in Fig. S22 we show the concentrations and fluxes of ribosomal subunits. Overall, these results

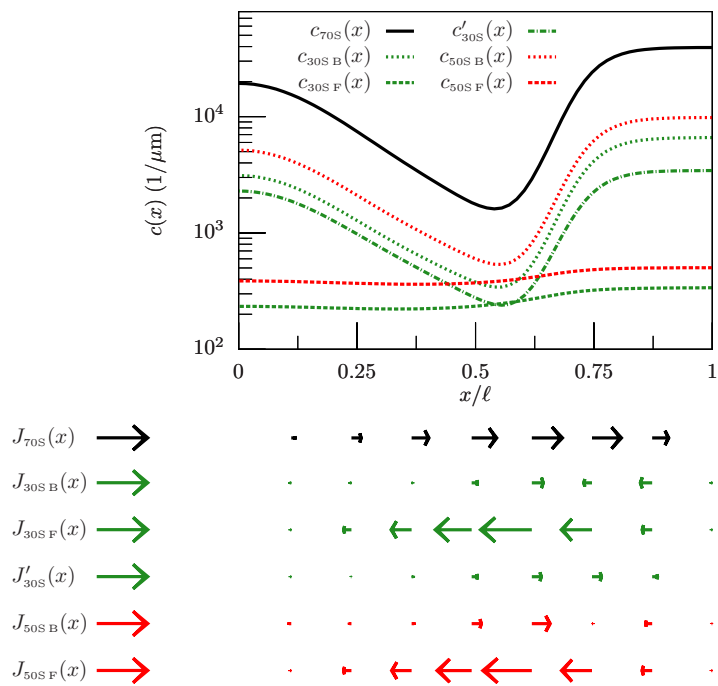

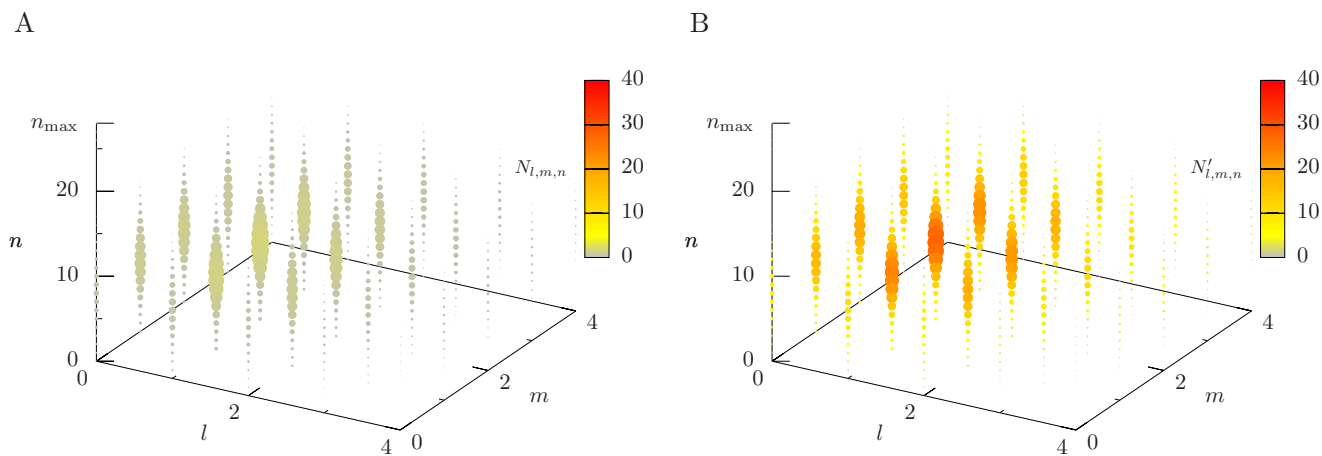

FIG. S15: Steady-state ribosomal-subunit concentrations and fluxes in the model including 30S and 50S ribosomal subunits for a long cell, in the late phase of its division cycle. Top: Concentrations $c_{70S}(x)$, $c_{30S\,B}(x)$, $c_{30S\,F}(x)$, $c'_{30S}(x)$ of 70S, transiently bound (B) 30S, free (F) 30S subunits, and of 30S subunits bound to the translation-initiation site, respectively. We also show the concentrations $c_{50S\,B}(x)$ and $c_{50S\,F}(x)$ of B and F 50S subunits, respectively. Bottom: Fluxes $J_{70S}(x)$, $J_{30S\,B}(x)$, $J_{30S\,F}(x)$ of 70S, B and F 30S subunits, and flux $J'_{30S}(x)$ of 30S subunits bound to the initiation site. The fluxes $J_{50S\,B}(x)$, $J_{50S\,F}(x)$ of B and F 50S subunits are also shown. Fluxes are represented along the cell's long axis depicted in the top panel, the arrow length is proportional to local ribosome flux, and the arrows in the legends correspond to a flux of 30/s.

FIG. S16: Polysome distributions in the model including 30S and 50S ribosomal subunits for a long cell in the late phase of its division cycle, where we fixed the maximal number of allowed translating (T) 30S-50S (70S) pairs per mRNA at $n \leq n_{max} = 28$, and we show mRNA species with $l \leq 4$ and $m \leq 4$ transiently bound (B) 30S and 50S subunits, respectively. (A) distribution of mRNA species with no 30S subunit bound to the translation-initiation site, shown as a heat map of the number $N_{l,m,n} = \int_0^\ell dx \rho_{l,m,n}(x)$ of mRNAs with $l$ B 30S subunits, $m$ B 50S subunits, and $n$ T 70S pairs. The diameter of each dot is proportional to $N_{l,m,n}$, where the largest dot in the plot corresponds to $N_{l,m,n} \sim 2$. (B) Same as panel A for the number $N'_{l,m,n} = \int_0^\ell dx \rho'_{l,m,n}(x)$ of mRNAs with a 30S subunit bound to the initiation site, where the largest dot in the plot corresponds to $N'_{l,m,n} \sim 30$.

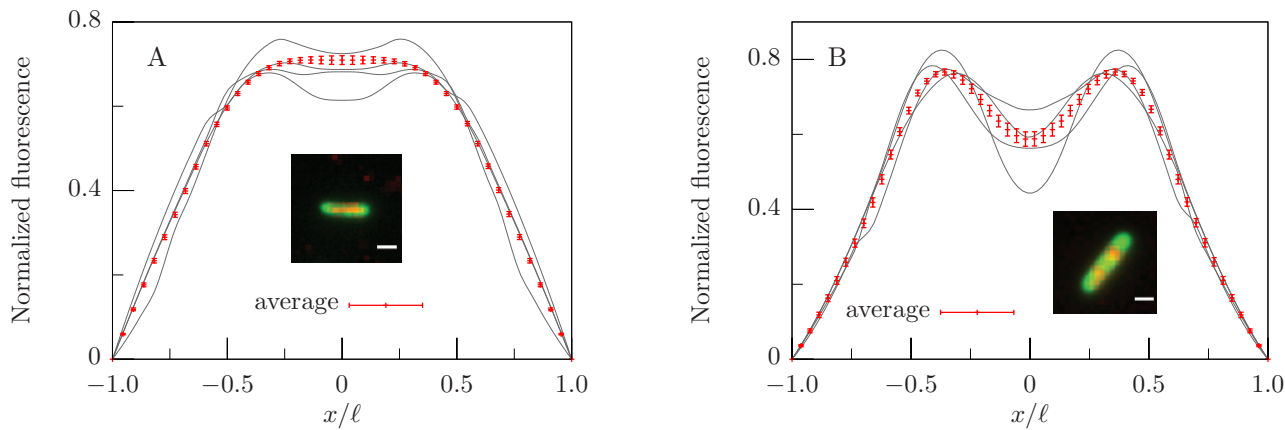

FIG. S17: DNA fluorescence along the long cell axis for *E. coli* in glycerol minimal and defined rich media. (A) Glycerol minimal media: fluorescence for a few representative cells (gray) and resulting average fluorescence over 81 cells with standard error of the mean (red), both symmetrized and normalized to unit area. Inset: representative cell grown in glycerol minimal media, with ribosomal protein (green) and nucleoid (red). Scale bar: $1\,\mu m$. (B) Same as panel A for defined rich media, averaged over 43 cells.

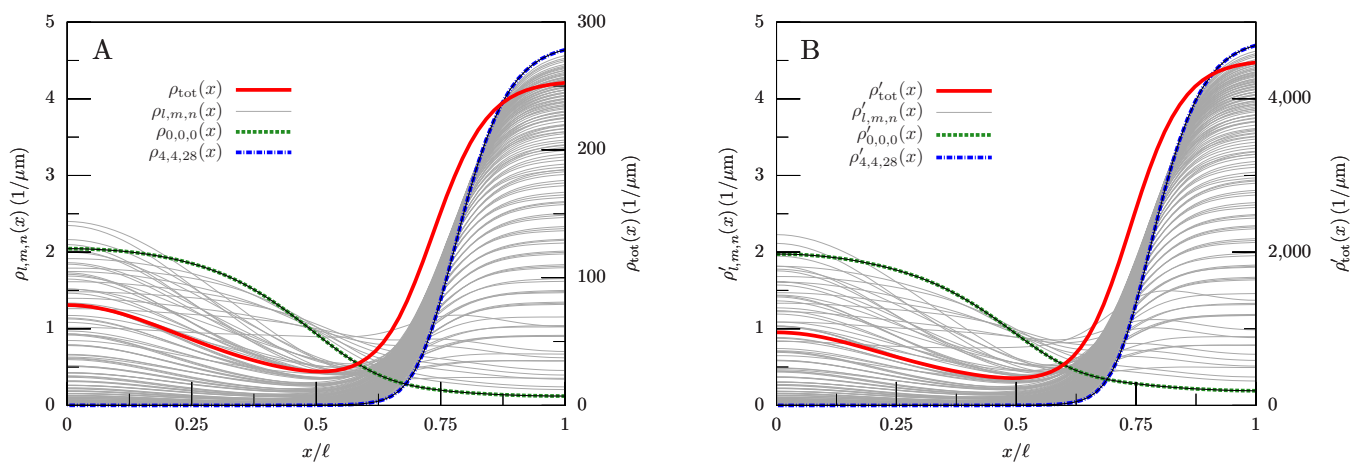

FIG. S18: Steady-state mRNA distributions for the model including 30S and 50S ribosomal subunits for cells in defined rich media with growth rate $\sim 2/\mathrm{h}$, compare Fig. S17B, where we fixed the maximal number of allowed translating (T) 30S-50S (70S) pairs per mRNA at $n \leq n_{\max} = 28$, and we show mRNA species with $l \leq 4$ and $m \leq 4$ transiently bound (B) 30S and 50S subunits, respectively. (A) mRNA profiles for mRNAs with no 30S subunit bound to the translation-initiation site. Total mRNA density $\rho_{\mathrm{tot}}(x)$ (red) and density $\rho_{l,m,n}(x)$ of mRNAs with $l$ B 30S subunits, $m$ B 50S subunits, and $n$ 70S pairs (gray), where we show only the curves with $l$, $m$, and $n$ even for greater clarity. The density $\rho_{0,0,0}(x)$ of ribosome-free mRNAs (green) and the density $\rho_{4,4,28}(x)$ of mRNAs with the largest number of B, T subunits considered (blue) are also shown. The profiles $\rho_{l,m,n}(x)$, $\rho_{0,0,0}(x)$, and $\rho_{4,4,28}(x)$ are normalized to unit area. (B) Same as panel A for mRNAs with a 30S subunit bound to the initiation site.

confirm those including B subunits. An important feature distinguishing the existence from the absence of B ribosomes is the ribosome efficiency: without B ribosomes, this efficiency is $\varepsilon = 2.39 \times 10^{-2}/\mathrm{s}$, which is $\sim 23\%$ higher than the efficiency including B ribosomes, compare Section S10 D. Put simply, as the total number of ribosomal subunits is conserved in this comparison, the inclusion of transient ribosomal binding reduces the number of 30S and 50S subunits that can form translating 70S pairs, thus lowering the ribosome efficiency. In addition, Fig. S21 shows that there is an average number of $\sim 12$ subunits, i.e. 30S and 70S, per mRNA. This average mRNA occupancy is smaller than that obtained including B subunits in section S10 D: The absence of B ribosomes thus implies that excluded-volume effects are slightly weaker, and that the dip in the total mRNA profiles in Fig. S18 is no longer present in Fig. S21.

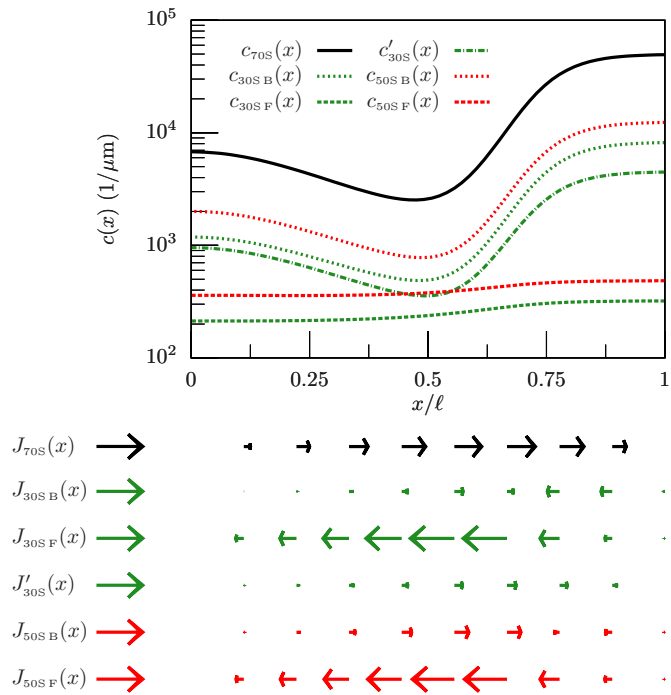

FIG. S19: Steady-state ribosomal-subunit concentrations and fluxes in the model including 30S and 50S ribosomal subunits for cells in defined rich media with growth rate $\sim 2$/hr, compare Fig. S17B. Top: Concentrations $c_{70S}(x)$, $c_{30S\,B}(x)$, $c_{30S\,F}(x)$, $c'_{30S}(x)$ of 70S, transiently bound (B) 30S, free (F) 30S subunits, and of 30S subunits bound to the translation-initiation site, respectively. We also show the concentrations $c_{50S\,B}(x)$ and $c_{50S\,F}(x)$ of B and F 50S subunits, respectively. Bottom: Fluxes $J_{70S}(x)$, $J_{30S\,B}(x)$, $J_{30S\,F}(x)$ of 70S, B and F 30S subunits, and flux $J'_{30S}(x)$ of 30S subunits bound to the initiation site. The fluxes $J_{50S\,B}(x)$, $J_{50S\,F}(x)$ of B and F 50S subunits are also shown. Fluxes are represented along the cell's long axis depicted in the top panel, the arrow length is proportional to local ribosome flux, and the arrows in the legends correspond to a flux of 30/s.

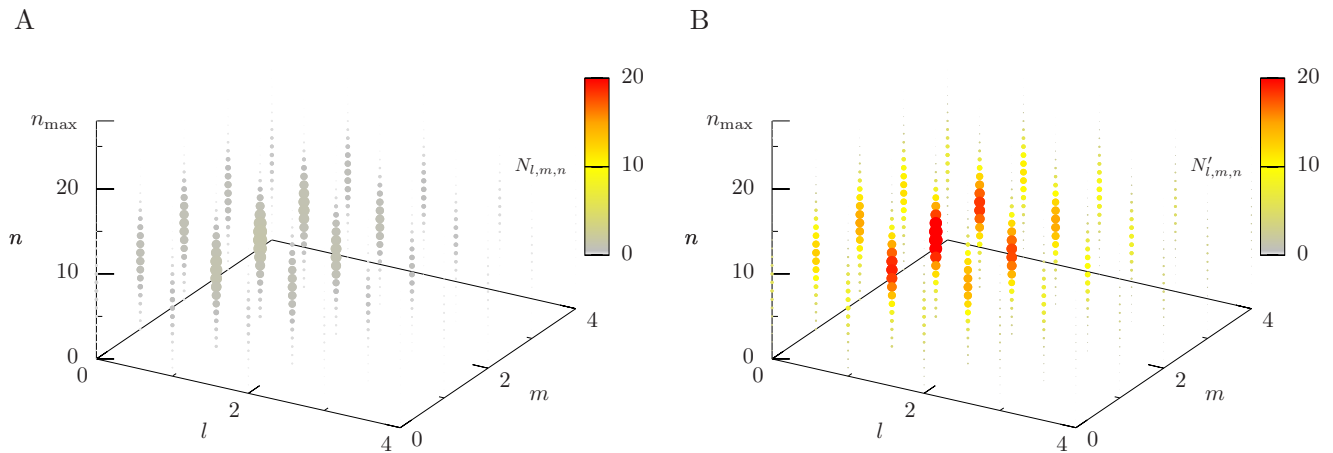

FIG. S20: Polysome distributions for the model with 30S and 50S ribosomal subunits for cells in defined rich media with growth rate $\sim 2$/h, compare Fig. S17B, where we fixed the maximal number of allowed translating (T) 30S-50S (70S) pairs per mRNA at $n \leq n_{\max} = 28$, and we show mRNA species with $l \leq 4$ and $m \leq 4$ transiently bound (B) 30S and 50S subunits, respectively. (A) distribution of mRNA species with no 30S subunit bound to the translation-initiation site, shown as a heat map of the number $N_{l,m,n} = \int_0^\ell dx \rho_{l,m,n}(x)$ of mRNAs with $l$ B 30S subunits, $m$ B 50S subunits, and $n$ T 70S pairs. The diameter of each dot is proportional to $N_{l,m,n}$, where the largest dot in the plot corresponds to $N_{l,m,n} \sim 1$. (B) Same as panel A for the number $N'_{l,m,n} = \int_0^\ell dx \rho'_{l,m,n}(x)$ of mRNAs with a 30S subunit bound to the initiation site, where the largest dot in the plot corresponds to $N'_{l,m,n} \sim 20$.

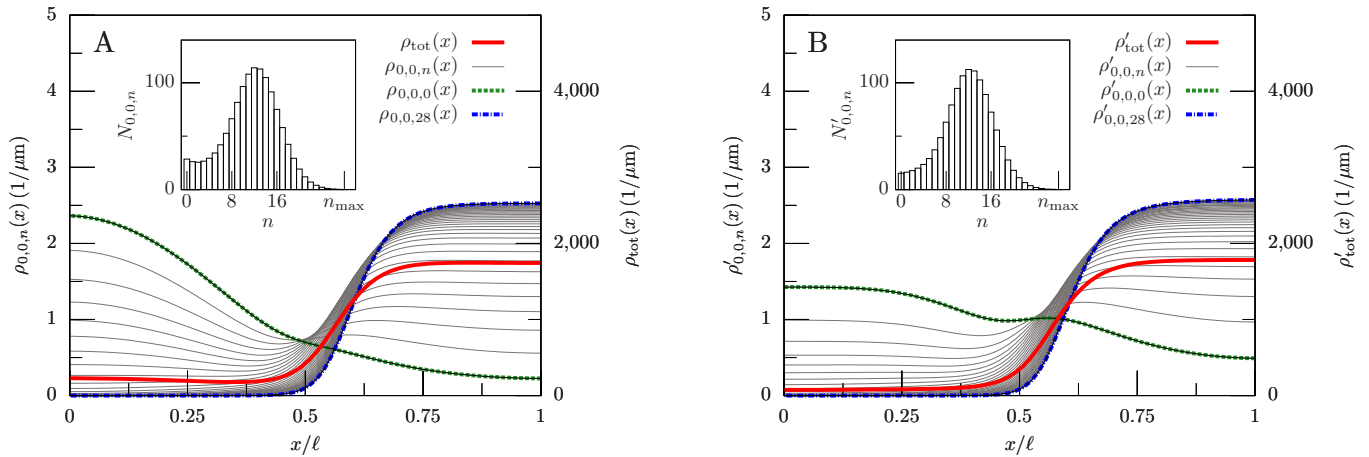

FIG. S21: Steady-state mRNA distributions for the model including 30S and 50S ribosomal subunits, with no transiently bound (B) 30S and 50S subunits. The maximal allowed number of 70S pairs per mRNA is $n_{max} = 28$. (A) mRNA profiles for mRNAs with no 30S subunit bound to the translation-initiation site. Total mRNA density $\rho_{tot}(x)$ (red) and density $\rho_{0,0,n}(x)$ of mRNAs with $n$ 70S pairs (gray). The density $\rho_{0,0,0}(x)$ of ribosome-free mRNAs (green) and the density $\rho_{0,0,28}(x)$ of mRNAs with the largest number of 70S pairs considered (blue) are also shown. The profiles $\rho_{0,0,n}(x)$, $\rho_{0,0,0}(x)$, and $\rho_{0,0,28}(x)$ are normalized to unit area. Inset: distribution $N_{0,0,n} = \int_0^\ell dx \rho_{0,0,n}(x)$ of mRNAs with $n$ 70S pairs in the right half of the cell as a function of $n$. (B) Same as panel A for mRNAs with a 30S subunit bound to the initiation site.

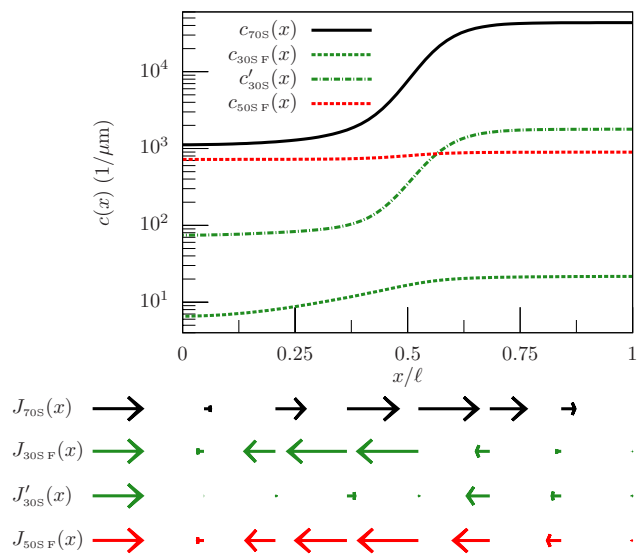

FIG. S22: Steady-state ribosomal-subunit concentrations and fluxes for the model including 30S and 50S ribosomal subunits, with no transiently bound (B) 30S and 50S subunits. Top: Concentrations $c_{70S}(x)$, $c_{30S\,F}(x)$ of 70S and free (F) 30S subunits, concentration $c'_{30S}(x)$ of 30S subunits bound to the translation-initiation site, and concentration $c_{50S\,F}(x)$ of F 50S subunits, for the right half of the cell. Bottom: Fluxes $J_{70S}(x)$, $J_{30S\,F}(x)$ of 70S and F 30S subunits, flux $J'_{30S}(x)$ of 30S subunits bound to the initiation site, and flux $J_{50S\,F}(x)$ of F 50S subunits. Fluxes are represented along the cell's long axis depicted in the top panel, the arrow length is proportional to local ribosome flux, and the arrows in the legends correspond to a flux of 5/s.

## S15 Force balance between nucleoid and polysomes determines nucleoid size

In this section we will calculate the force exerted by polysomes and F ribosomes on the nucleoid as well as the entropy of compaction of the nucleoid, and the resulting force-balance equation will allow us to determine the nucleoid size.

For the sake of clarity, let us first focus on F ribosomes. We compute the total force exerted by the nucleoid on F ribosomes, and the total force by F ribosomes on the nucleoid will necessarily be equal and opposite. To begin, we

observe that the F-ribosome current (S3) is the sum of two terms which can be interpreted as follows: The first term,

$$J_{\mathrm{D}}(x) = -D_{\mathrm{F}} v_{\mathrm{F}}(x) \frac{dc_{\mathrm{F}}(x)}{dx},$$

(S117)

is the current due to the Brownian diffusion of F ribosomes from high- to low-concentration regions, and it is thus proportional to the concentration gradient, with a space-dependent diffusion coefficient $D_{\mathrm{F}} v_{\mathrm{F}}(x)$. The second term,

$$J_{\mathrm{DNA}}(x) = D_{\mathrm{F}} c_{\mathrm{F}}(x) \frac{dv_{\mathrm{F}}(x)}{dx},$$

(S118)

is the current due to the force locally exerted by the nucleoid on F ribosomes; it is proportional to the gradient of $v_{\mathrm{F}}(x)$, thus it represents the tendency of the nucleoid to push ribosomes from regions with lower available volume to regions with higher available volume. The current $J_{\mathrm{DNA}}(x)$ is thus the result of a net force $f(x)$ applied by the nucleoid on each F ribosome at position $x$, with the force and the current related by

$$J_{\mathrm{DNA}}(x) = \mu(x) f(x) c_{\mathrm{F}}(x),$$

(S119)

where the mobility $\mu(x)$ is related to the Brownian diffusion coefficient $D_{\mathrm{F}} v_{\mathrm{F}}(x)$ by the Einstein relation

$$D_{\mathrm{F}} v_{\mathrm{F}}(x) = \mu(x) \, \mathrm{k_B} \, T,$$

(S120)

where $\mathrm{k_B}$ is Boltzmann's constant and $T$ is the temperature [25]. From Eqs. (S118), (S119), and (S120) we obtain $f(x) = \frac{\mathrm{k_B} T}{v_{\mathrm{F}}(x)} \frac{dv_{\mathrm{F}}(x)}{dx}$. Multiplying $f(x)$ by the local F ribosome concentration, integrating with respect to $x$, and reversing the sign of the force, we obtain the total force exerted by F ribosomes in the right half of the cell on the nucleoid, which reads $-\mathrm{k_B} T \int_0^\ell dx \, \frac{c_{\mathrm{F}}(x)}{v_{\mathrm{F}}(x)} \frac{dv_{\mathrm{F}}(x)}{dx}$. Finally, the combined force exerted by both F ribosomes and polysomes on the nucleoid is obtained by incorporating the mRNA concentrations $\rho_{m,n}(x)$ and available volumes $v_{m+n}(x)$ in the expression above, and summing over all mRNA species. We obtain

$$\mathcal{F}_{\mathrm{in}} = -\mathrm{k_B} T \int_0^\ell dx \left[ \frac{c_{\mathrm{F}}(x)}{v_{\mathrm{F}}(x)} \frac{dv_{\mathrm{F}}(x)}{dx} + \sum_{m=0}^{m_{\max}} \sum_{n=0}^{n_{\max}} \frac{\rho_{m,n}(x)}{v_{m+n}(x)} \frac{dv_{m+n}(x)}{dx} \right]$$

$$= -\mathrm{k_B} T \int_0^\ell dx \left\{ \frac{c_{\mathrm{F}}(x)}{v_{\mathrm{F}}(x)} \frac{dv_{\mathrm{F}}(x)}{dx} + \exp \left( \frac{k_{\mathrm{on}}^{\mathrm{B}} c_{\mathrm{F}}(x)}{k_{\mathrm{off}}^{\mathrm{B}}} \right) \sum_{n=0}^{n_{\max}} \left[ \frac{1}{v_{\mathrm{R}}(x)} \frac{dv_{\mathrm{R}}(x)}{dx} + \left( \frac{k_{\mathrm{on}}^{\mathrm{B}} c_{\mathrm{F}}(x)}{k_{\mathrm{off}}^{\mathrm{B}}} + n \right) \frac{1}{v_{\mathrm{F}}(x)} \frac{dv_{\mathrm{F}}(x)}{dx} \right] \rho_n(x) \right\},$$

(S121)

where the subscript 'in' denotes a force directed towards the inner part of the cell, and in the second line we used Eqs. (S20) and (S25), which are valid in the rapid-equilibrium limit, to compute the sum with respect to $m$. Note that $\mathcal{F}_{\mathrm{in}}$ depends on the DNA profile $\varphi(x)$ through the available volumes $v_{\mathrm{F}}(x)$ and $v_{\mathrm{R}}(x)$, compare Eqs. (S9) and (S11).

The force exerted by F ribosomes and polysomes results in the compaction of the nucleoid [8, 26]. At mechanical equilibrium, this force must balance the "spring" force exerted by the compressed nucleoid, which can be estimated in terms of the entropy of a self-avoiding DNA polymer. We approximate this entropy $S$ by treating the plectonemic chromosomal DNA as a sequence of $N$ joined segments, each with length $\xi$, confined in a volume $V$, yielding the confinement dependence of the entropy [27]

$$S = -\mathrm{k_B} \frac{\pi^2}{2} \frac{\xi^2 N^{2\nu}}{V^{2/3}},$$

(S122)

where $\xi = 200 \, \mathrm{nm}$ is twice the estimated persistence length of the plectoneme [8], the total number of monomers $N = L/\xi = 7.5 \times 10^3$ is obtained as the ratio between the total plectoneme length $L$ and the segment length $\xi$, and

$$\nu = 0.592$$

(S123)

is the exponent describing the mean end-to-end distance for a self-avoiding polymer [27]. Given that the exponent (S123) results from a numerical simulation where the polymer is represented as a self-avoiding walk on a cubic lattice [27], the estimate (S122) applies to self-avoiding polymers whose thickness is comparable to the segment length. It follows that for a DNA plectoneme the expression (S122) should be, in principle, modified to account for the difference between the thickness and segment length. However, in what follows we will show that Eq. (S122) quantitatively reproduces the observed nucleoid size, thus showing that the thickness correction above is negligible.

Proceeding along the lines of Eq. (S7), we assume that the DNA is confined in a cylindrical region at midcell: We denote the length of this region by $2x_0$, and the DNA profile is $\varphi(x) \propto 1/\{1 + \exp[\zeta(x - x_0)/\ell]\}$, where $\varphi$ is normalized according to the condition (S8) on the total DNA length. The volume $V$ in Eq. (S122) can thus be estimated as

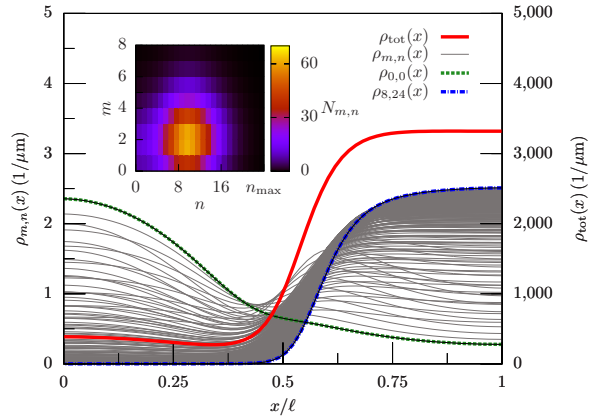

FIG. S23: Steady-state mRNA and polysome distributions for the model including the force balance between nucleoid and polysomes. Total mRNA density $\rho_{\text{tot}}(x)$ (red) and density $\rho_{m,n}(x)$ of mRNAs with $m$ transiently bound (B) ribosomes and $n$ translating (T) ribosomes for $0 \leq m \leq m_{\text{max}} = 8$ and $0 \leq n \leq n_{\text{max}} = 24$ (gray). The density $\rho_{0,0}(x)$ of ribosome-free mRNAs (green) and the density $\rho_{8,24}(x)$ of mRNAs with the largest number of T and B ribosomes considered (blue) are also shown. The profiles $\rho_{m,n}(x)$, $\rho_{0,0}(x)$, and $\rho_{8,24}(x)$ are normalized to unit area. Inset: distribution of mRNA species, shown as a heat map of the number $N_{m,n}$ of mRNAs with $m$ B ribosomes and $n$ T ribosomes in the right half of the cell. The maximal number of T ribosomes per mRNA used in our model, $n_{\text{max}} = 24$, is indicated.

$V = 2\pi R^2 x_0$, and the outward entropic force exerted by the nucleoid on F ribosomes and polysomes in each half of the cell is obtained from the derivative of the entropy as

$$
\begin{aligned}
\mathcal{F}_{\text{out}} &= \frac{1}{2} T \frac{\partial S}{\partial x_0} \\
&= k_{\text{B}} T \frac{\pi^2}{6} \frac{\xi^2 N^{2\nu}}{(2\pi R^2)^{2/3} x_0^{5/3}},
\end{aligned}
\tag{S124}
$$

where the factor $1/2$ in the first line results from the fact that we compute the outward force in one half of the cell.

In what follows we will thus use Eqs. (S121) and (S124), combine the resulting force-balance equation

$$
\mathcal{F}_{\text{in}} + \mathcal{F}_{\text{out}} = 0
\tag{S125}
$$

with the reaction-diffusion Eqs. (S121) and (S124), and solve self-consistently for $\rho_n(x)$, $c_{\text{F}}(x)$, and for the nucleoid size $x_0$. The results are shown in Figs. S23 and S24, where we show mRNA concentrations and polysome distributions, the nucleoid profile $\varphi(x)$, the calculated value of $x_0/\ell \approx 0.5$, as well as the F ribosome concentrations and fluxes.

Finally, we provide a simple analytical estimate for the nucleoid size as a function of the total number of mRNAs, and other relevant parameters. Given that the ribosome-to-mRNA ratio is much larger than unity, compare section II, we assume that mRNAs at the poles are fully excluded from the nucleoid by virtue of having a sufficient number of bound ribosomes: we thus approximate the nucleoid profile in one half of the cell as a step function with the step at $x = x_0$, i.e. $v_{m+n}(x)$ is equal to zero for $0 \leq x < x_0$ and to unity for $x_0 \leq x \leq 1$. Given that the size of an F ribosome is significantly smaller than that of a typical polysome, F ribosomes can easily penetrate the nucleoid, thus the pressure that they exert on the nucleoid is much smaller than the force that polysomes exert on the nucleoid: Indeed, in Eq. (S121) we denote the contribution to $\mathcal{F}_{\text{in}}$ given by F ribosomes and mRNAs by $\mathcal{F}_{\text{in F}}$ and $\mathcal{F}_{\text{in mRNA}}$, respectively, and for the reference conditions of Figs. 2-4, we have $\mathcal{F}_{\text{in F}}/(\mathcal{F}_{\text{in F}} + \mathcal{F}_{\text{in mRNA}}) \sim 2 \times 10^{-2}$. We thus neglect the contribution by F ribosomes in the first line of Eq. (S121), and we obtain

$$
\begin{aligned}
\mathcal{F}_{\text{in}} &= -k_{\text{B}} T \sum_{m=0}^{m_{\text{max}}} \sum_{n=0}^{n_{\text{max}}} \rho_{m,n}^{\text{out}} \int_0^\ell dx \frac{dv_{m+n}(x)}{dx} \\
&= -k_{\text{B}} T \rho_{\text{tot}}^{\text{out}},
\end{aligned}
\tag{S126}
$$

where in the first line we used the fact that at equilibrium $\rho_{m,n}(x)/v_{m+n}(x)$ is independent of $x$, compare section S2, and we denoted by $\rho_{m,n}^{\text{out}}$ the concentration of mRNA species $m$, $n$ outside the nucleoid, and by

$$
\rho_{\text{tot}}^{\text{out}} = \frac{N_{\text{mRNA}}}{2(\ell - x_0)}
\tag{S127}
$$

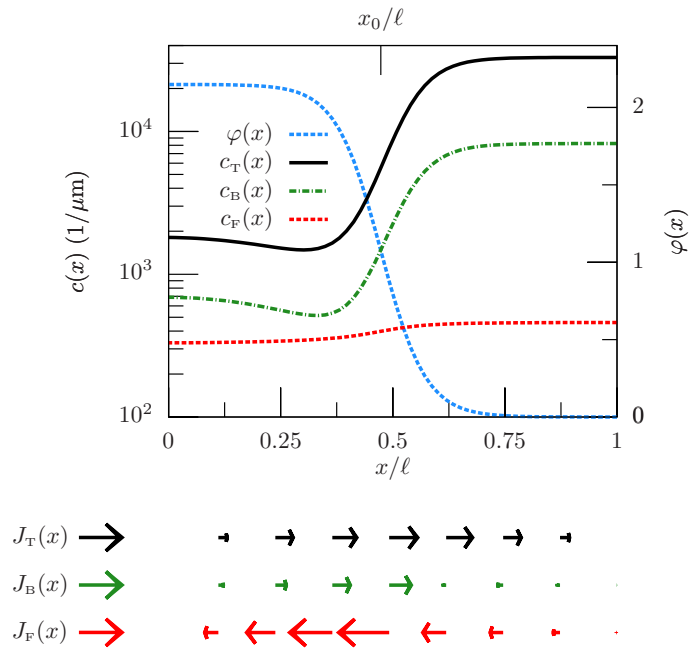

FIG. S24: Nucleoid profile, steady-state ribosome concentrations, and ribosome fluxes for the model including the force balance between nucleoid and polysomes. Top: DNA density $\varphi(x)$ normalized to unit area and concentrations $c_{\mathrm{T}}(x)$, $c_{\mathrm{B}}(x)$, and $c_{\mathrm{F}}(x)$ of translating (T), transiently bound (B), and free (F) ribosomes in the right half of the cell (compare Fig. 1). The nucleoid size $x_0/\ell$ is also marked. Bottom: Fluxes of T, B, and F ribosomes along the cell's long axis depicted in the top panel. The arrow length is proportional to local ribosome flux, and the arrows in the legends correspond to a flux of 20/s.

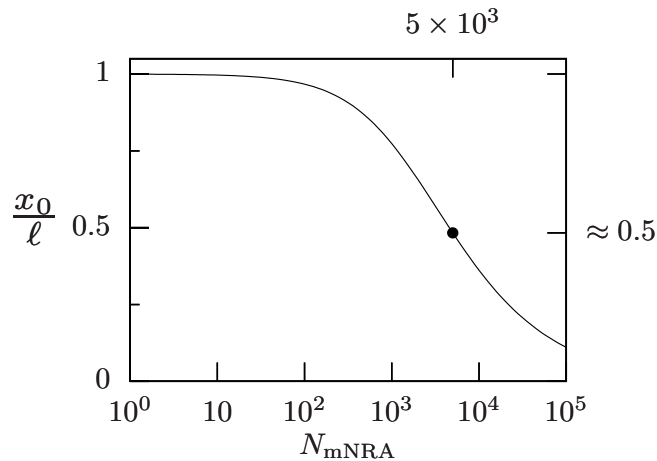

FIG. S25: Nucleoid size as a function of number of mRNAs, as predicted by force balance between nucleoid and polysomes. We plot the ratio $x_0/\ell$ between the nucleoid width and cell length as a function of the total number of mRNAs in the cell, from the solution of Eq. (3), which relies on the simplifying assumption that mRNAs cannot penetrate the nucleoid. The mRNA number $N_{\mathrm{mRNA}} = 5 \times 10^3$ used for our analysis in glucose minimal media and the predicted nucleoid size $x_0/\ell \approx 0.5$ are indicated with a dot.

the total mRNA concentration outside the nucleoid. Combining Eqs. (S124), (S125), (S126), and (S127) we obtain Eq. (3), which can be solved numerically for the nucleoid size $x_0$. The resulting values for the nucleoid size $x_0$ are shown in Fig. S25, where we plot $x_0/\ell$ as a function of the total mRNA number $N_{\mathrm{mRNA}}$ keeping all other parameters in Eq. (3) fixed.



\end{document}